\documentclass[preprint,aps,amsmath,pre,showpacs]{revtex4-2}
\usepackage{graphicx}
\usepackage{amsfonts}
\usepackage{xcolor}
\usepackage[english]{babel}

\newcommand*{\Scale}[2][4]{\scalebox{#1}{\ensuremath{#2}}}

\begin{document}

\title{The multi-allelic Moran process as a multi-zealot voter model: exact results and consequences for diversity thresholds}

\author{Dan Braha$^{1,3,}$} 

\author{Marcus A. M. de Aguiar$^{1,2}$\footnote{Corresponding author}}

\affiliation{$^1$New England Complex Systems Institute, Cambridge, Massachusetts, United States of America}

\affiliation{$^2$Instituto de F\'isica Gleb Wataghin, Universidade Estadual de Campinas, Unicamp 13083-970, Campinas, SP, Brazil}

\affiliation{$^3$University of Massachusetts Dartmouth, Dartmouth, Massachusetts, United States of America }

\begin{abstract}
The Moran process is a foundational model of genetic drift and mutation in finite populations. In its standard two-allele form with population size $n$, allele counts, and hence allele frequencies, change through stochastic replacement and mutation, yet converge to a stationary distribution. This distribution undergoes a qualitative transition at the \emph{critical mutation rate} $\mu_c=1/(2n)$: at $\mu=\mu_c$ it is exactly uniform, so that the probability of observing $k$ copies of allele~1 (and $n-k$ of allele~2) is $\pi(k)=1/(n+1)$ for $k=0,\dots,n$. For $\mu<\mu_c$ diversity is low: the stationary distribution places most of its mass near $k=0$ and $k=n$, and the population is therefore typically dominated by one allele. For $\mu>\mu_c$, on the other hand, diversity is high: the distribution concentrates around intermediate values, so that both alleles are commonly present at comparable frequencies. Recently, the two-allele Moran process was shown to be exactly equivalent to the voter model with two candidates and $\alpha_1$ and $\alpha_2$ committed voters (\emph{zealots}) in a population of $n+\alpha_1+\alpha_2$, where mutation is played by zealot influence. Here we extend this equivalence to multiple alleles and multiple candidates. Using the mapping, we derive the exact stationary distribution of allele counts for well-mixed populations with an arbitrary number $m$ of alleles, and obtain the critical mutation rate $\mu_c = 1/(m+2n-2)$, which depends explicitly on $m$. We then analyze the Moran process on randomly connected populations and show that both the stationary distribution and $\mu_c$ are invariant to network structure and coincide with the well-mixed results. Interestingly, this contrasts with the voter model, where zealot effects are strongly amplified on random networks yet do not depend on the number of candidates. Finally, simulations on general network topologies show that structural heterogeneity can substantially reshape the stationary allele distribution and, consequently, the level of genetic diversity.
\end{abstract}

\maketitle

\section{Introduction}
\label{intro}

Mutation and genetic drift are two unavoidable sources of stochasticity in population genetics. Mutation introduces novelty and can help maintain diversity, whereas drift erodes variation through random sampling. The Moran process, introduced by Patrick Moran in 1958 \cite{moran1958random}, is one of the simplest models that captures the interplay between drift and mutation in finite populations \cite{gillespie_population_2004,nowak_evolutionary_2006,ewens_mathematical_2004,burden2018population}. In its standard two-allele form, it tracks the allele count in a well-mixed population of fixed size $n$. At stationarity, the model yields, for example, the probability $\pi(k)$ of observing $k$ copies of allele~1 (and $n-k$ copies of allele~2), given mutation rates $\mu_{1\rightarrow 2}$ and $\mu_{2\rightarrow 1}$. In the absence of mutation, the process eventually fixates at one of the two absorbing states (all allele~1 or all allele~2), and the fixation probability of allele~1 equals $k_0/n$, where $k_0$ is the initial count of allele~1. When mutation is present, the dynamics instead admits a unique stationary distribution. Under symmetric mutation, $\mu_{1\rightarrow 2}=\mu_{2\rightarrow 1}\equiv \mu$, this stationary distribution undergoes a qualitative change at the critical mutation rate $\mu_c=1/(2n)$. For $\mu \ll \mu_c$, $\pi(k)$ is U-shaped and typically one allele dominates, whereas for $\mu \gg \mu_c$ it is unimodal and both alleles persist at substantial frequencies \cite{crow_introduction_1970,gillespie_population_2004}.

Despite its simplicity and pedagogical value, the two-allele Moran process does not reflect the prevalence of loci with many allelic variants. Beyond the well-known ABO system, which gives rise to the four blood types A, B, O, and AB, some genes can harbor dozens to hundreds of alleles. One example is the CFTR gene on chromosome~7, whose protein product helps regulate salt and water balance and is essential for maintaining healthy mucus in the lungs, digestive tract, and other organs \cite{brennan2016cystic}. Geneticists hypothesize that CFTR may admit on the order of one hundred alleles \cite{snustad2015principles}, contributing to wide variation in symptom severity across individuals. Another prominent example is the human leukocyte antigen (HLA) system on chromosome~6. Two major loci are HLA-A, with nearly twenty alleles, and HLA-B, with more than thirty \cite{choo2007hla}. Although HLA loci are subject to selection, it remains important to understand their baseline, neutral allele-frequency distribution as a function of mutation.

The Moran model, originally developed in population genetics \cite{watterson_markov_1961,cannings_latent_1974,gladstien_characteristic_1978,ewens_mathematical_2004}, has also been applied in other domains, including herding and animal decision making \cite{kirman_ants_1993}, cancer evolution \cite{durrett_spatial_2015}, the evolution of altruism \cite{debarre_social_2014}, and models of stock markets \cite{harmon_anticipating_2015} and voting behavior \cite{voting2017behavior}. In particular, the Moran process has been shown to be exactly equivalent to the voter model, a paradigmatic model of opinion dynamics in the social sciences \cite{de_aguiar_moran_2011}. The voter model and its variants have been widely studied in statistical physics, with applications to consensus formation \cite{ramirez2024ordering} and polarization \cite{jager2005uniformity,martins2010mass}. In the two-candidate voter model with zealots, the population consists of $n$ ``free voters,'' together with $\alpha_1$ zealots for candidate~1 and $\alpha_2$ zealots for candidate~2 \cite{mobilia_role_2007,chinellato_dynamical_2015}. Zealots hold fixed opinions and never change state, while free voters update their opinions through interactions with other free voters and with zealots. Under the Moran--voter correspondence, free voters map to alleles and zealots act as effective mutation sources. This mapping identifies not only the stationary probability of observing $k$ free voters supporting candidate~1 with the probability of observing $k$ copies of allele~1, but also the full dynamics of the two processes \cite{de_aguiar_moran_2011,schneider2016mutation,franco2021shannon}. The critical point of the two-candidate voter model occurs at $\alpha_1=\alpha_2=\alpha_c=1$, independent of $n$, in which case the stationary distribution is exactly uniform: the probability of finding $k$ voters for candidate~1 equals $1/(n+1)$ for all $k$.

In this paper, we generalize the fully mixed voter model with zealots from two candidates to $m$ candidates and derive closed-form expressions for the stationary distribution of vote counts. We then consider the Moran model with $m$ alleles in a well-mixed population and show that it maps exactly to the generalized voter model. This correspondence yields the stationary distribution of allele counts for finite populations and an arbitrary number of alleles. To obtain an analytically tractable multi-allele Moran model, we assume \emph{parent-independent mutations} (PIM): for $i\neq r$, the probability $\mu_{r\to i}$ that an allele $r$ mutates to allele $i$ equals $u_i$ and is independent of $r$. Under PIM, we show that the stationary probability that the population contains $k_i$ copies of allele~$i$ is a Dirichlet--multinomial distribution $\pi(k_1,\dots,k_m; \alpha_1,\dots,\alpha_m)$ (see Eq.~(\ref{DR_voter})), with parameters $\alpha_i = 2(n-1)u_i/(1-\Lambda)$ where $\Lambda = \sum_i u_i$. We also identify the critical mutation probability $u_c = 1/(2n-2+m)$ separating low- and high-diversity regimes.

Next, we study both the voter and Moran models on randomly connected populations modeled as Erd\H{o}s--R\'enyi networks. Here, each individual interacts with only a subset of the population, and all individuals have approximately the same number of interaction partners (degree) $d<n-1$. We show that, for the voter model, the stationary vote distribution can differ markedly from the fully mixed case, because the influence of zealots is strongly amplified when interactions are localized. For randomly connected populations, we further show that the Moran model can be mapped, under a mean-field approximation, to a corresponding voter model. Using this mapping, we obtain a Dirichlet--multinomial stationary distribution for allele counts that is, perhaps surprisingly, insensitive to the random connectivity.

Finally, we assess the robustness of the theoretical predictions by simulating the voter and Moran models on a range of network topologies, including scale-free, small-world, and ring networks. Our simulations show that, under the mean-field parameter mapping, the allele-count distribution in the Moran model matches the vote-count distribution in the voter model across all network topologies we considered. At the same time, structural heterogeneity can substantially reshape the Dirichlet--multinomial stationary distribution and, consequently, the level of genetic diversity.

This paper is organized as follows. In Sec.~\ref{voter} we review the two-candidate voter model and derive its stationary distribution on fully mixed populations. In the same section, we extend the voter model to $m$ candidates, obtain its stationary distribution, and analyze the transition from dominance by a single candidate to a more homogeneous vote distribution. We then consider the multi-candidate voter model on randomly connected populations modeled as Erd\H{o}s--R\'enyi networks. In Sec.~\ref{moran} we study the multi-allele Moran model on fully mixed populations, establish its exact mapping to the generalized voter model, and derive the stationary allele distribution and the critical mutation rate separating low- and high-diversity regimes. Extending the Moran model to randomly connected populations in Sec.~\ref{moranrandom}, we show that the stationary allele distribution is independent of the network's average degree. A summary of the main results is presented in Sec.~\ref{summary} before we report simulation results in Sec.~\ref{simulatons}. Finally, in Sec.~\ref{discussion} we discuss implications for models in population genetics and related applications. Because some sections contain technical details that may not interest all readers, we state the main results at the beginning of each section. Proofs follow immediately thereafter and may be skipped on a first reading.

\raggedbottom
\section{The voter model with zealots on fully mixed populations}
\label{voter}

\subsection{Two candidates}
\label{voter-two}

Consider a population of $n+\alpha_1+\alpha_2$ individuals choosing between candidates 1 and 2 in an election \cite{liggett_interacting_2012}. Only $n$ individuals are \emph{free voters} (undecided), while the remaining $\alpha_1$ and $\alpha_2$ individuals are \emph{zealots} with fixed opinions in favor of candidates 1 and 2, respectively. Zealots never change their states, whereas free voters may update their opinions through social interaction. In network language, we refer to free voters as \emph{free nodes} and zealots as \emph{frozen nodes}. Each node has a \emph{state} in $\{1,2\}$. Because the population is fully mixed, the system state at time $t$, denoted $X_t$, is completely determined by the number of free nodes in state 1. Thus $X_t$ takes values in $\{0,1,\dots,n\}$.

The dynamics proceeds as follows. At each time step, a free node is selected uniformly at random. With probability $\ell$ (retention) it keeps its current state, and with probability $1-\ell$ (imitation) it copies the state of a second node chosen uniformly from the remaining $n-1+\alpha_1+\alpha_2$ individuals (free or frozen). Since only one free node is updated at each step, if $X_t=k$ then $X_{t+1}\in\{k-1,k,k+1\}$. The one-step transition probabilities are
\begin{eqnarray}
P_{k,k+1} & = & (1-\ell)\,\frac{n-k}{n}\cdot\frac{k+\alpha_1}{\,n-1+\alpha_1+\alpha_2\,},\nonumber \\
P_{k,k-1} & = & (1-\ell)\,\frac{k}{n}\cdot\frac{n-k+\alpha_2}{\,n-1+\alpha_1+\alpha_2\,},\nonumber
\end{eqnarray}
and $P_{k,k}=1-P_{k,k+1}-P_{k,k-1}$. For example, $P_{k,k+1}$ is the probability that the updated free node is currently in state 2, $(n-k)/n$, times the probability it imitates, $(1-\ell)$, times the probability it copies a node in state 1, $(k+\alpha_1)/(n-1+\alpha_1+\alpha_2)$.

For $\alpha_1,\alpha_2>0$ and $\ell<1$, we have $P_{k,k+1}>0$ for $k<n$ and $P_{k,k-1}>0$ for $k>0$, so the dynamics has no absorbing states. Since $P_{k,k}>0$ for all $k=0,1,\dots,n$, the chain is aperiodic. Moreover, every state communicates with every other through $\pm1$ moves (that is, the chain is irreducible), and therefore the chain admits a unique stationary distribution. We now establish this stationary distribution. Let $\pi_t(k)=\mathbb{P}(X_t=k)$ and write $\pi_t$ for the row vector with components $\pi_t(0),\pi_t(1),\dots,\pi_t(n)$. From the update rule,
\begin{equation*}
\pi_{t+1} = \ell \pi_t  + (1-\ell) \pi_t Q = \pi_t [\ell I +(1-\ell)Q] \equiv \pi_t P,
\end{equation*}
where $Q$ is the \emph{copy-step} kernel (the dynamics with $\ell=0$), given by
\begin{eqnarray}
Q_{k,k+1} & = & \frac{n-k}{n}\cdot\frac{k+\alpha_1}{\,n-1+\alpha_1+\alpha_2\,}, \nonumber \\ 
Q_{k,k-1} & = & \frac{k}{n}\cdot\frac{n-k+\alpha_2}{\,n-1+\alpha_1+\alpha_2\,},\\
Q_{k,k} & = & 1-Q_{k,k+1}-Q_{k,k-1}. \nonumber
\label{q2a}
\end{eqnarray}
Stationarity requires $\pi_{t+1}=\pi_t$, so $\pi$ is a left eigenvector of $P$ with eigenvalue 1. Since $P=\ell I+(1-\ell)Q$, any stationary distribution of $Q$ is also stationary for $P$. Therefore, the stationary law does not depend on $\ell$, and it suffices to determine the stationary distribution of $Q$.

We will show that the stationary distribution for the two-state zealot voter model on a fully mixed population is
\begin{equation}
\pi(k;\alpha_1,\alpha_2)=\binom{n}{k}\,\frac{(\alpha_1)_k\,(\alpha_2)_{n-k}}{(\alpha_1+\alpha_2)_n}
=\frac{\displaystyle\binom{\alpha_1+k-1}{k}\,\binom{n+\alpha_2-k-1}{\,n-k\,}}
{\displaystyle\binom{n+\alpha_1+\alpha_2-1}{\,n\,}},
\end{equation}
for $k=0,1,\dots,n$, where $(a)_m:=a(a+1)\cdots(a+m-1)$ is the Pochhammer (rising) factorial, with $(a)_0:=1$. The generalized binomial form extends to all real $\alpha_1,\alpha_2>0$ via the gamma function. For $\alpha_1=0$ (resp.\ $\alpha_2=0$), the unique stationary distribution is the point mass at $k=0$ (resp.\ at $k=n$). Importantly, when $\alpha_1=\alpha_2=1$ the distribution is uniform, with $\pi(k; 1, 1)=1/(n+1)$. We include $(\alpha_1,\alpha_2)$ explicitly in $\pi(k;\alpha_1,\alpha_2)$ to facilitate comparison with the multi-candidate and multi-allele results in later sections.

\medskip

\noindent {\bf Proof --} Although the stationary distribution can be obtained via hypergeometric functions \cite{chinellato_dynamical_2015}, the most direct route uses detailed balance. In equilibrium, the probability flux from $k$ to $k+1$ equals the flux from $k+1$ to $k$:
\[
\pi(k)\,Q_{k,k+1}=\pi(k+1)\,Q_{k+1,k}\qquad (k=0,1,\dots,n-1).
\]
Therefore,
\[
\frac{\pi(k+1)}{\pi(k)}=\frac{Q_{k,k+1}}{Q_{k+1,k}}
=\frac{(n-k)(k+\alpha_1)}{(k+1)(n-k-1+\alpha_2)}.
\]
Solving this first-order recurrence yields
\[
\pi(k)=\pi(0)\,\prod_{i=0}^{k-1}\frac{(n-i)(i+\alpha_1)}{(i+1)(n-i-1+\alpha_2)}
=\pi(0)\,\binom{n}{k}\,(\alpha_1)_k\,\frac{(\alpha_2)_{n-k}}{(\alpha_2)_n}.
\]
To identify $\pi(0)$, use Vandermonde's identity in Pochhammer form,
\[
\sum_{k=0}^{n}\binom{n}{k}\,(\alpha_1)_k\,(\alpha_2)_{n-k}=(\alpha_1+\alpha_2)_n,
\]
which follows by multiplying the binomial series $(1-z)^{-\alpha_1}=\sum_{k\ge0}(\alpha_1)_k z^k/k!$ and $(1-z)^{-\alpha_2}=\sum_{m\ge0}(\alpha_2)_m z^m/m!$ and equating the $z^n$ coefficients after multiplying by $n!$. Hence
\[
\pi(k)=\binom{n}{k}\,\frac{(\alpha_1)_k\,(\alpha_2)_{n-k}}{(\alpha_1+\alpha_2)_n}.
\]
Finally, using $(a)_m=\Gamma(a+m)/\Gamma(a)=m!\,\binom{a+m-1}{m}$ gives the equivalent generalized binomial form displayed above.

Before closing this section, we rewrite the transition probabilities in a form that will be convenient for the multi-candidate voter model and for the multi-allele Moran model, where the state is a count vector $k=(k_1,\dots,k_m)$. Although in the present two-candidate case the chain is fully specified by the scalar $k=X_t$ (the number of free nodes in state~1), it is convenient to represent the state as the vector $k=(k_1,k_2)$ with $k_1+k_2=n$ (so $k_1=k$ and $k_2=n-k$). Since each update changes only one individual's opinion, the updated state can be written as $k' = k + e_i - e_j$, where $i,j \in \{1,2\}$ and $e_r$ denotes the $r$th standard basis vector, i.e.\ $e_1=(1,0)$ and $e_2=(0,1)$. This corresponds to a single voter switching from opinion $j$ to opinion $i$. It then follows directly from Eq.~(\ref{q2a}) that
\begin{equation}
Q\big(k \to k' \big)=\frac{k_j}{n}\cdot \frac{k_i+\alpha_i}{\,n-1+ \alpha_1 + \alpha_2\,},
\label{qqp}
\end{equation}
since we choose a free $j$-node with probability $k_j/n$ and it copies an $i$-neighbor (free or zealot) with probability $(k_i+\alpha_i)/(n-1+\alpha_1+\alpha_2)$.

\subsection{Multiple candidates}\label{subsec:multiple-candidates}

We next consider the multi-candidate extension of the zealot voter model. The population consists of $n+\alpha_0$ individuals who must choose among candidates $1,2,\dots,m$, where $\alpha_i$ denotes the number of zealots committed to candidate $i$ and 
\[ 
\alpha_0  \equiv \sum_{i=1}^m \alpha_i
\]
is the total number of zealots. As before, only $n$ individuals are undecided (free voters), while zealots keep their opinions fixed throughout the dynamics. At each time step, a free voter may either retain its current opinion or copy the opinion of another individual (free voter or zealot) in the fully mixed population.

Because the population is fully mixed, the state of the system at time $t$ is specified by the vector of free-voter counts
\[
X_t=(X_{t,1},\dots,X_{t,m}) \in \mathcal{S}_n
:=\Big\{k\in\mathbb{Z}_{\ge0}^m:\ \sum_{i=1}^m k_i=n\Big\},
\]
where $X_{t,i}$ is the number of free voters in state $i$ at time $t$. The dynamics parallels the two-candidate case: at each time step a free voter is selected uniformly at random. With probability $\ell$ it retains its state, and with probability $1-\ell$ it imitates the state of another individual chosen uniformly from the remaining $n-1+\alpha_0$ individuals.

Let $k=(k_1,\dots,k_m)\in\mathcal{S}_n$ denote the current state, and let $e_r$ be the $r$th standard basis vector. An elementary update corresponds to a single voter switching from $j$ to $i$, so the next state is $k' = k + e_i - e_j$, where $i,j \in \{1,2,\dots,m\}$ and $i \neq j$. The copy-step transition kernel $Q$ therefore satisfies 
\[
Q\big(k \to k' \big)=\frac{k_j}{n}\cdot \frac{k_i+\alpha_i}{\,n-1+\alpha_0\,},
\]
in agreement with the two-candidate expression in Eq.~(\ref{qqp}). The one-step transition kernel for the full dynamics is
\begin{equation}
	P = \ell I+(1-\ell)Q.
	\label{kernelmv}
\end{equation}
For $\ell<1$ and $\alpha_i>0$ for all $i$, all unit moves $e_i-e_j$ occur with positive probability whenever $k_j>0$, and the chain can move throughout $\mathcal{S}_n$. Moreover, there is always a positive probability of remaining in the same state (by retaining, or by copying a neighbor with the same opinion), so the chain is aperiodic. Consequently, the chain admits a unique stationary distribution. As in the two-candidate case, the stationary distribution does not depend on $\ell$: if $\pi Q=\pi$, then $\pi P=\ell\pi+(1-\ell)\pi=\pi$.

The stationary distribution is
\begin{equation}
	\pi(k_1,\dots,k_m; \alpha_1,\dots,\alpha_m)\;=\;
	\frac{\displaystyle\prod_{i=1}^m \binom{\alpha_i+k_i-1}{\,k_i\,}}
	{\displaystyle \binom{\alpha_0+n-1}{\,n\,}},
	\label{DR_voter}
\end{equation}
where generalized binomial coefficients are understood in the Gamma-function sense; we include $(\alpha_1,\dots,\alpha_m)$ explicitly in $\pi$ to facilitate comparison with later sections. Equivalently,
\[
\begin{aligned}
	\pi(k_1,\dots,k_m; \alpha_1,\dots,\alpha_m)
	&= \frac{n!}{k_1!\cdots k_m!}\,
	\frac{(\alpha_1)_{k_1}\cdots(\alpha_m)_{k_m}}{(\alpha_1+\cdots+\alpha_m)_{n}} \\
	&= \frac{n!}{k_1!\cdots k_m!}\,
	\frac{\Gamma(\alpha_0)}{\Gamma(\alpha_0+n)}
	\prod_{i=1}^m \frac{\Gamma(\alpha_i+k_i)}{\Gamma(\alpha_i)},
\end{aligned}
\]
which is the Dirichlet--multinomial distribution. Moreover, when $\alpha_1=\cdots=\alpha_m=1$, we have $(1)_{k_i}=k_i!$, so the factorials cancel in the Dirichlet--multinomial pmf and $\pi(k_1,\dots,k_m; 1, \dots, 1)$ becomes constant over $\mathcal{S}_n$. Since $|\mathcal{S}_n|=\binom{n+m-1}{m-1}$, the stationary distribution is uniform:
\[
\pi(k_1,\dots,k_m; 1, \dots, 1)=\frac{1}{\binom{n+m-1}{m-1}}
=\frac{(m-1)!\,n!}{(n+m-1)!}.
\]

\medskip

\noindent {\bf Proof --} We derive Eq.~(\ref{DR_voter}) via detailed balance. Define the weight function
\[
w(k)=\prod_{i=1}^m \binom{\alpha_i+k_i-1}{\,k_i\,}.
\]
Fix $i\neq j$ and consider adjacent states $k' = k+e_i-e_j$ (so $k_j\ge1$). Using the elementary identities
\[
\frac{\binom{a+r}{\,r+1\,}}{\binom{a+r-1}{\,r\,}}=\frac{a+r}{r+1},
\qquad
\frac{\binom{a+(r-1)-1}{\,r-1\,}}{\binom{a+r-1}{\,r\,}}=\frac{r}{a+r-1},
\]
valid for all real $a>0$ and integers $r\ge1$, we obtain
\[
\frac{w(k')}{w(k)}
=\frac{\binom{\alpha_i+k_i}{\,k_i+1\,}}{\binom{\alpha_i+k_i-1}{\,k_i\,}}
\cdot
\frac{\binom{\alpha_j+(k_j-1)-1}{\,k_j-1\,}}{\binom{\alpha_j+k_j-1}{\,k_j\,}}
=\frac{\alpha_i+k_i}{k_i+1}\cdot \frac{k_j}{\alpha_j+k_j-1}.
\]
On the other hand, the reverse move $k'\to k$ has probability
\[
Q(k'\to k)=\frac{k_i+1}{n}\cdot \frac{k_j-1+\alpha_j}{\,n-1+\alpha_0\,},
\]
so the transition ratio is
\[
\frac{Q(k\to k')}{Q(k'\to k)}
=\frac{\frac{k_j}{n}\cdot \frac{k_i+\alpha_i}{n-1+\alpha_0}}{\frac{k_i+1}{n}\cdot \frac{k_j-1+\alpha_j}{n-1+\alpha_0}}
=\frac{k_j}{k_i+1}\cdot \frac{\alpha_i+k_i}{\alpha_j+k_j-1}
=\frac{w(k')}{w(k)}.
\]
Thus $w(k)\,Q(k\to k')=w(k')\,Q(k'\to k)$ for all adjacent pairs, and detailed balance holds for $Q$. Hence the stationary distribution satisfies $\pi\propto w$. Moreover, if $\pi Q=\pi$ then $\pi P=\ell\pi+(1-\ell)\pi=\pi$, so $\pi$ is also stationary for $P$.

It remains to compute the normalizing constant. Using the generating function identity
\[
\sum_{r\ge0} \binom{\alpha_i+r-1}{\,r\,} z^r=(1-z)^{-\alpha_i}\qquad (|z|<1),
\]
multiplying over $i=1,\dots,m$, and extracting the coefficient of $z^n$ yields
\[
\sum_{k\in\mathcal{S}_n}\prod_{i=1}^m \binom{\alpha_i+k_i-1}{\,k_i\,}
=\binom{\alpha_0+n-1}{\,n\,}.
\]
Therefore $C=\binom{\alpha_0+n-1}{\,n\,}^{-1}$ and Eq.~(\ref{DR_voter}) follows.

As a final observation, in stationarity the frequency vector $V_n:=X/n=(X_1/n,\dots,X_m/n)$ converges in distribution, as $n\to\infty$, to $\mathrm{Dirichlet}(\alpha_1,\dots,\alpha_m)$ with density
\begin{equation}
	f(v_1,\dots,v_m; \alpha_1,\dots,\alpha_m) = \frac{\Gamma(\alpha_0)}{\displaystyle\prod_{i=1}^m \Gamma(\alpha_i)}
	\prod_{i=1}^m v_i^{\alpha_i-1}.
\end{equation}
For $m=2$, this reduces to the $\mathrm{Beta}(\alpha_1,\alpha_2)$ distribution. This result allows us to focus on vote frequencies (with a Dirichlet limit) rather than vote counts (with a Dirichlet-- multinomial law) in the large-population regime.

\subsection{Multiple candidates on random populations under global zealot influence}
\label{voter-random}
All results so far assumed a well-mixed population. In this setting, each free voter interacts uniformly with every other individual, corresponding to a complete graph. We now relax this assumption by allowing interactions among free voters to be \emph{local}, while keeping zealot influence \emph{global}. Specifically, the free--free connections form an Erd\H{o}s--R\'enyi (ER) graph $G(n,p)$ \cite{erdds1959random,newman2018networks}, in which each pair of free voters is linked independently with probability $p$, so the mean free-degree is $d=p(n-1)$. We treat the network as fixed, so the edge set is held constant throughout the dynamics.

Fix $m\ge2$ and zealot counts $\alpha^{\mathrm{ER}}_i>0$ for $i=1,\dots,m$, and write $\alpha^{\mathrm{ER}}_0:=\sum_{i=1}^m \alpha^{\mathrm{ER}}_i$. Consider a population with $n$ free nodes and $\alpha^{\mathrm{ER}}_0$ zealots, where \emph{every free node is connected to all zealots}, and the free--free edges are as above (ER with mean free-degree $d$). At each time step, a free node is selected uniformly; with probability $\ell\in[0,1)$ it keeps its current state, and with probability $1-\ell$ it copies the state of a uniformly chosen neighbor (free or zealot).

As before, let $X_t=(X_{t,1},\dots,X_{t,m})$ denote the free-voter counts. Then, under the ER mean-field (random-neighborhood) approximation, the stationary distribution of the count vector is Dirichlet--multinomial with \emph{effective} zealot counts:
\begin{eqnarray}
\pi^{\mathrm{ER}}(k_1,\dots,k_m; \alpha^{\mathrm{ER}}_1,\dots,\alpha^{\mathrm{ER}}_m) &=&
\pi(k_1,\dots,k_m; \alpha_1^{\text{eff}},\dots,\alpha_m^{\text{eff}}) \\
&=&\frac{\displaystyle\prod_{i=1}^m \binom{\alpha_i^{\text{eff}}+k_i-1}{\,k_i\,}}
{\displaystyle \binom{\alpha_0^{\text{eff}}+n-1}{\,n\,}}.
\label{voterer1}
\end{eqnarray}

Here
\begin{equation}
\alpha_i^{\text{eff}}=\frac{n-1}{d}\,\alpha^{\mathrm{ER}}_i,\quad
\alpha_0^{\text{eff}}=\frac{n-1}{d}\,\alpha^{\mathrm{ER}}_0,
\label{voterer2}
\end{equation}
so zealot influence is amplified by the factor $(n-1)/d$ relative to the fully mixed case (recovering the complete-graph result $\alpha_i^{\text{eff}}=\alpha^{\mathrm{ER}}_i$ when $d=n-1$). Appendix~\ref{app:second-order-degree} develops a second-order correction to this ER mean-field reduction that accounts for degree fluctuations, and we use it in Section~\ref{simulatons} when comparing simulations across network topologies.

\medskip

\noindent {\bf Mean-field derivation --} Fix a configuration $k=(k_1,\dots,k_m)$. The probability that the updated node is a free voter currently in state $j$ is $k_j/n$. Let $d_j$ denote its free-degree (the number of free neighbors). Since every free voter is connected to all zealots, the probability that the updated node chooses a neighbor in state $i$ is
\[
\frac{(\text{\# free $i$-neighbors of the updated node})+\alpha^{\mathrm{ER}}_i}{d_j+\alpha^{\mathrm{ER}}_0}.
\]
In the mean-field approximation for ER graphs, we replace $d_j$ by $d$ and the number of free $i$-neighbors by its expectation $d\,k_i/(n-1)$ (excluding the updated node), obtaining
\[
\text{Prob}(\text{$j$ copies $i$})
=\frac{\tfrac{d}{n-1}k_i+\alpha^{\mathrm{ER}}_i}{d+\alpha^{\mathrm{ER}}_0}
=\frac{k_i+\tfrac{n-1}{d}\alpha^{\mathrm{ER}}_i}{(n-1)+\tfrac{n-1}{d}\alpha^{\mathrm{ER}}_0}
=\frac{k_i+\alpha_i^{\text{eff}}}{(n-1)+\alpha_0^{\text{eff}}}.
\]
Hence the copy-step kernel $Q^{\mathrm{ER}}$ satisfies
\[
Q^{\mathrm{ER}}\big(k\to k+e_i-e_j\big)
=\frac{k_j}{n}\cdot \frac{k_i+\alpha_i^{\text{eff}}}{(n-1)+\alpha_0^{\text{eff}}},
\]
which coincides with the complete-graph kernel with zealot counts $\alpha^{\text{eff}}$. Therefore the detailed-balance weights
\[
w(k):=\prod_{i=1}^m \binom{\alpha_i^{\text{eff}}+k_i-1}{\,k_i\,}
\]
satisfy $w(k)\,Q^{\mathrm{ER}}(k\to k')=w(k')\,Q^{\mathrm{ER}}(k'\to k)$ for all adjacent pairs $k\leftrightarrow k'$. Normalizing $\pi^{\mathrm{ER}}\propto w$ yields stationarity for $Q^{\mathrm{ER}}$, and therefore also for $P^{\mathrm{ER}}=\ell I+(1-\ell)Q^{\mathrm{ER}}$ (independently of $\ell$).

Finally, it is worth emphasizing what this reduction does and does not capture. On ER graphs, the random-neighborhood approximation makes the population effectively exchangeable at the macroscopic level, so the count vector $k$ provides an accurate state description. By contrast, on structured graphs the same counts can correspond to inequivalent microscopic configurations. For example, on a ring (degree $d=2$), the state with $n/2$ voters in candidate~1 and $n/2$ in candidate~2 may consist of a single contiguous block of 1's followed by a block of 2's, or it may alternate $1,2,1,2,\dots$ around the circle. These configurations are not dynamically equivalent, and tracking only $k$ can miss important spatial effects. A detailed two-candidate analysis of this phenomenon appears in \cite{schneider2016mutation}, and we return to related issues in Section~\ref{simulatons}.

\section{The Moran model on fully mixed populations}
\label{moran}

The Moran model describes the evolution of allele counts in a well-mixed population of haploid individuals carrying a single biallelic gene. In its standard formulation, each discrete-time update replaces a randomly chosen individual by a copy of another randomly chosen individual.

Following \cite{de_aguiar_moran_2011}, we use an equivalent formulation that is convenient for comparison with the voter model. At each step, an individual is chosen uniformly to be replaced (the \emph{focal} individual), and a mating partner is chosen uniformly from the remaining $n-1$ individuals. The offspring inherits the allele of either parent with probability $1/2$. If it inherits the allele of the focal individual, the population state is unchanged; if it inherits the partner’s allele, the net effect is that the focal individual is replaced by a copy of the partner. Thus, relative to the standard Moran process, this dynamics has the same transitions but occurs at half the rate. We incorporate mutation by allowing the inherited allele to change during reproduction: allele~1 mutates to~2 with probability $\mu_{1\to 2}$, and allele~2 mutates to~1 with probability $\mu_{2\to 1}$.

Before proving the equivalence between the Moran and voter models, we define the multi-allele Moran process. Consider a well-mixed haploid population of size $n$ with a single gene having $m$ alleles $1,2,\dots,m$. Each update proceeds as follows: choose a focal individual uniformly and choose a mating partner uniformly from the remaining $n-1$ individuals. The offspring copies one of the two parental alleles with probability $1/2$. If the copied allele is $r$, the offspring mutates to allele~$i$ with probability $\mu_{r\to i}$, and the focal individual is then replaced by the offspring. Throughout we assume a parent-independent mutation (PIM; equal-input) structure, meaning that the mutation probability depends only on the destination allele:
\[
\mu_{r\to i}=u_i \quad \text{for } r\neq i.
\]
By normalization, $\sum_{i=1}^m \mu_{r\to i}=1$ for each $r$, which implies
\[
\mu_{r\to r}=1-\sum_{i\neq r}u_i=1-\Lambda+u_r,
\qquad
\Lambda:=\sum_{q=1}^m u_q<1.
\]

We shall show that the $m$-allele Moran model with PIM parameters $u_i$ maps exactly to the voter model with $m$ zealot types and retention probability $\ell$, with parameters related by
\begin{equation}
\begin{aligned}
\alpha_i &= \frac{2(n-1)u_i}{1-\Lambda}, \qquad
\ell = \frac{1-\Lambda}{2}, \qquad
\alpha_0 = \sum_{i=1}^m \alpha_i = \frac{2(n-1)\Lambda}{1-\Lambda}.
\end{aligned}
\label{alpha_map}
\end{equation}

Conversely,
\begin{equation}
\begin{aligned}
u_i &= \frac{\alpha_i}{\alpha_0+2(n-1)}, \qquad
\Lambda = \sum_{i=1}^m u_i = \frac{\alpha_0}{\alpha_0+2(n-1)}, \\
\mu_{r\to i} &=
\begin{cases}
\dfrac{\alpha_i}{\alpha_0+2(n-1)}, & r\neq i,\\[4pt]
\dfrac{2(n-1)+\alpha_i}{\alpha_0+2(n-1)}, & r=i.
\end{cases}
\end{aligned}
\label{eq:inv}
\end{equation}

We can summarize this correspondence as
\begin{equation}
\pi^{\mathrm{M}}(k_1,\dots,k_m; u_1,\dots,u_m)
=
\pi(k_1,\dots,k_m; \alpha_1,\dots,\alpha_m),
\label{Moran_voter_equality}
\end{equation}
where $\pi^{\mathrm{M}}(k_1,\dots,k_m;u_1,\dots,u_m)$ denotes the stationary distribution of allele counts in the Moran model (i.e., the probability of observing $k_i$ copies of allele~$i$, $i=1,\dots,m$). The right-hand side is the voter-model stationary distribution (see Eq.~(\ref{DR_voter})). Hence the stationary distribution of the Moran model is Dirichlet--multinomial with parameters $\alpha_i$ given by Eq.~(\ref{alpha_map}).

The critical mutation rate marking the transition between low- and high-diversity regimes (where the stationary distribution is uniform over the count simplex; see Section~\ref{subsec:multiple-candidates}) is obtained by setting $\alpha_i=1$ for all $i$ in Eq.~(\ref{eq:inv}), which forces $u_i=u_c$ for all $i$. This yields the critical mutation rate for the multi-allele Moran process,
\begin{equation}
u_c=\frac{1}{2n-2+m},
\label{criticalmu}
\end{equation}
which reduces to $u_c=1/(2n)$ when $m=2$.


{\bf Proof of equivalence with the voter model} -- To establish the equivalence between the Moran and voter models, we compute the one-step probability that a free individual of type $j$ is replaced by one of type $i\neq j$. In the allele-count notation, the move $k\to k'=k+e_i-e_j$ occurs when the focal individual has type $j$ and the offspring ends in type $i$. Writing $T_{i\mid j}(k)$ for this conditional probability, we have
\[
Q^{\mathrm{M}}\big(k\to k+e_i-e_j\big)=\frac{k_j}{n}\,T_{i\mid j}(k).
\]

There are three mutually exclusive ways the offspring can end in type $i$:
(i) it inherits the focal allele $j$ and mutates to $i$;
(ii) it inherits the partner allele, the partner is also type $j$, and the inherited allele mutates to $i$; or
(iii) it inherits the partner allele from a partner of type $r\neq j$ and mutates to $i$.
Therefore,
\[
T_{i\mid j}(k)
=\frac{1}{2}\mu_{j\to i}
+\frac{1}{2}\frac{k_j-1}{n-1}\mu_{j\to i}
+\frac{1}{2}\sum_{r\neq j}\frac{k_r}{n-1}\mu_{r\to i}.
\]
Adding and subtracting the $r=j$ term in the last sum and rearranging yields
\begin{equation}
T_{i\mid j}(k)
=\frac{1}{2}\left(1-\frac{1}{n-1}\right)\mu_{j\to i}
+\frac{1}{2(n-1)}\sum_{r=1}^m k_r\,\mu_{r\to i}.
\label{tij}
\end{equation}
This is the only point where the details of the reproduction mechanism enter.

Recall the parent-independent mutation (PIM) assumption: $\mu_{r\to i}=u_i$ for $r\neq i$, and $\mu_{i\to i}=1-\Lambda+u_i$, where $\Lambda=\sum_{q=1}^m u_q<1$. Using $\sum_{r=1}^m k_r=n$, we obtain
\[
\sum_{r=1}^m k_r\,\mu_{r\to i}
=(1-\Lambda+u_i)k_i+u_i\sum_{r\neq i}k_r
=u_i n+(1-\Lambda)k_i.
\]
Substituting into \eqref{tij} (and using $\mu_{j\to i}=u_i$ for $i\neq j$) gives
\begin{equation}
T_{i\mid j}(k)=u_i+\frac{1-\Lambda}{2(n-1)}\,k_i \;=: \; u_i+b\,k_i,
\label{tijm}
\end{equation}
where
\begin{equation}
b=\frac{1-\Lambda}{2(n-1)}.
\end{equation}
Hence, for $i\neq j$,
\[
Q^{\mathrm{M}}\big(k\to k+e_i-e_j\big)
=\frac{k_j}{n}\big(u_i+b\,k_i\big)
=\frac{k_j}{n}\,b\,(k_i+\alpha_i),
\qquad
\alpha_i:=\frac{u_i}{b}=\frac{2(n-1)u_i}{1-\Lambda}.
\]

For the voter model with zealots we previously obtained
\[
Q\big(k\to k+e_i-e_j\big)
=\frac{k_j}{n}\cdot\frac{k_i+\alpha_i}{n-1+\alpha_0},
\qquad
\alpha_0:=\sum_{i=1}^m \alpha_i.
\]
Therefore,
\[
Q^{\mathrm{M}}\big(k\to k+e_i-e_j\big)=c\,Q\big(k\to k+e_i-e_j\big),
\]
with
\[
c=b\,(n-1+\alpha_0)
=\frac{1-\Lambda}{2(n-1)}\left(n-1+\frac{2(n-1)\Lambda}{1-\Lambda}\right)
=\frac{1+\Lambda}{2},
\]
since $\alpha_0=\sum_{i=1}^m \alpha_i=\frac{2(n-1)\Lambda}{1-\Lambda}$.
Because this proportionality holds for all off-diagonal moves, the full kernels satisfy (compare with \ Eq.~(\ref{kernelmv}))
\[
P^{\mathrm{M}}=(1-c)\,I+c\,Q,
\]
with retention probability $\ell:=1-c=\frac{1-\Lambda}{2}\in[0,1)$. Thus the one-step transition kernel of the Moran model has exactly the voter-model form, and since the voter model has a Dirichlet--multinomial stationary distribution, the Moran model does as well.

\section{Moran model on random populations}
\label{moranrandom}

We now study the Moran model on a population whose interaction structure is an ER graph $G(n,p)$ with mean degree $d=p(n-1)$. In this setting interactions are local: mating partners are chosen only among the neighbors of the focal individual. The dynamics is the same as in the fully mixed case except for this restriction. At each step: (i) choose a focal individual $j$ uniformly; (ii) choose a partner uniformly among the neighbors of $j$ in $G(n,p)$; (iii) the offspring copies one of the two parental alleles with probability $1/2$; (iv) conditional on the pre-mutation allele $r$, it mutates to $i$ with probability $\mu_{r\to i}$; and (v) the focal individual is replaced by the offspring.

Within the ER mean-field (random-neighborhood) approximation, the induced dynamics of the allele-count vector in the $m$-allele Moran model with parent-independent mutation (PIM) rates $u_i$ (with $\Lambda:=\sum_{q=1}^m u_q<1$) coincides with that of the $m$-candidate voter model on the same ER network. The corresponding retention probability and zealot counts are
\[
\ell=\frac{1-\Lambda}{2},\qquad
\alpha^{\mathrm{ER}}_i=\frac{2d\,u_i}{1-\Lambda},\qquad
\alpha^{\mathrm{ER}}_0:=\sum_{i=1}^m \alpha^{\mathrm{ER}}_i=\frac{2d\,\Lambda}{1-\Lambda}.
\]
Moreover, the stationary distribution of allele counts is Dirichlet--multinomial with parameters
\[
\alpha_i=\frac{n-1}{d}\,\alpha^{\mathrm{ER}}_i=\frac{2(n-1)u_i}{1-\Lambda},
\]
which are independent of the mean degree $d$. (This expression for $\alpha_i$ follows directly from the ER--to--fully mixed reduction for the voter model via the effective zealot counts in Eq.~(\ref{voterer2}).) Consequently, the critical mutation rate is still given by Eq.~(\ref{criticalmu}) and does not depend on $d$.

We summarize these correspondences as
\begin{equation}
\pi^{\mathrm{M},\mathrm{ER}}(k_1,\dots,k_m;u_1,\dots,u_m)
=
\pi^{\mathrm{ER}}\Bigl(k_1,\dots,k_m;\,
\overbrace{\alpha^{\mathrm{ER}}_1}^{\frac{2d\,u_1}{1-\Lambda}},\dots,
\overbrace{\alpha^{\mathrm{ER}}_m}^{\frac{2d\,u_m}{1-\Lambda}}
\Bigr)
=
\pi(k_1,\dots,k_m;\alpha_1,\dots,\alpha_m).
\label{moran-random-voter-corr}
\end{equation}

Despite the locality of interactions, the stationary allele-count distribution coincides with that of the fully mixed voter model under the parameter map above.

In addition, combining \eqref{moran-random-voter-corr} with the fully mixed Moran--voter identity
\eqref{Moran_voter_equality}, we expect that the stationary allele-count distribution of the Moran model is
insensitive to ER sparsity. That is,
\begin{equation}
\pi^{\mathrm{M},\mathrm{ER}}(k_1,\dots,k_m;u_1,\dots,u_m)
=
\pi^{\mathrm{M}}(k_1,\dots,k_m;u_1,\dots,u_m).
\label{moran-random-moran-corr}
\end{equation}

\medskip
\noindent {\bf Mean-field derivation--}
The probability that a type-$j$ individual is chosen as the focal parent is $k_j/n$. Let $d_j$ denote the degree of this focal node, and let $N_{j,r}$ be the number of its neighbors of type $r$. Conditional on choosing $j$ as focal, the probability that the mating partner has allele $r$ is $N_{j,r}/d_j$.

Under the ER random-neighborhood approximation, conditional on the macroscopic counts $k$ and on $d_j>0$, the neighbors of the focal node are effectively a uniform sample from the other $n-1$ free nodes. Consequently, the expected fraction of type-$r$ neighbors matches the global frequency:
\[
\mathbb{E}\!\left[\frac{N_{j,r}}{d_j}\,\Big|\,k,d_j>0\right]=\frac{k_r}{n-1}\quad (r\neq j),
\qquad
\mathbb{E}\!\left[\frac{N_{j,j}}{d_j}\,\Big|\,k,d_j>0\right]=\frac{k_j-1}{n-1}.
\]
In the mean-field closure we therefore replace $N_{j,r}/d_j$ by these expressions, so the partner’s type is (approximately) distributed as a uniform draw from the remaining $n-1$ individuals. This step hinges on the randomness of ER neighborhoods and can fail on structured graphs (e.g., rings or other regular lattices); see Section~\ref{voter-random}.

Next, compute $T_{i\mid j}(k)$, the conditional probability that the offspring ends in allele $i\neq j$ given that the focal parent is type $j$. As in Section~\ref{moran}, there are three contributions:
(i) the offspring inherits the focal allele $j$ and mutates to $i$;
(ii) it inherits the partner allele, the partner is also type $j$, and it mutates to $i$; or
(iii) it inherits the partner allele from a type-$r\neq j$ partner and mutates to $i$.
Hence,
\[
T_{i\mid j}(k)
=\frac{1}{2}\mu_{j\to i}
+\frac{1}{2}\frac{N_{j,j}}{d_j}\mu_{j\to i}
+\frac{1}{2}\sum_{r\neq j}\frac{N_{j,r}}{d_j}\mu_{r\to i}.
\]
Using the mean-field substitutions above gives
\[
T_{i\mid j}(k)
=\frac{1}{2}\mu_{j\to i}
+\frac{1}{2}\frac{k_j-1}{n-1}\mu_{j\to i}
+\frac{1}{2}\sum_{r\neq j}\frac{k_r}{n-1}\mu_{r\to i}
=\frac{1}{2}\left(1-\frac{1}{n-1}\right)\mu_{j\to i}
+\frac{1}{2(n-1)}\sum_{r=1}^m k_r\,\mu_{r\to i},
\]
which coincides with Eq.~(\ref{tij}) for fully mixed populations. The remaining steps therefore follow exactly as in the fully mixed case, yielding
\[
P^{\mathrm{M},\mathrm{ER}}=\ell I+(1-\ell)Q,
\]
where $P^{\mathrm{M},\mathrm{ER}}$ is the one-step transition kernel of the Moran model on the ER network, and $Q$ and $\ell$ are the retention probability and copy-step kernel of the corresponding voter model on a fully mixed population (see Section~\ref{subsec:multiple-candidates}):
\[
Q\big(k\to k+e_i-e_j\big)
=\frac{k_j}{n}\cdot\frac{k_i+\alpha_i}{n-1+\alpha_0},
\qquad
\alpha_i=\frac{2(n-1)u_i}{1-\Lambda},
\qquad
\alpha_0=\sum_{i=1}^m \alpha_i,
\qquad
\ell=\frac{1-\Lambda}{2}.
\]
We therefore obtain, within the ER mean-field approximation, that the stationary distribution of the Moran model on ER networks coincides with the Dirichlet--multinomial stationary law of the voter model on fully mixed populations, establishing the second equality in Eq.~(\ref{moran-random-voter-corr}) at the mean-field level. In particular, the stationary allele-count distribution and the critical mutation rate are independent of the mean degree $d$ within this approximation. The first equality in Eq.~(\ref{moran-random-voter-corr}) follows by solving for $\alpha^{\mathrm{ER}}_i$ using the effective-zealot relation in Eq.~(\ref{voterer2}).

\section{Summary of Results}
\label{summary}

Before presenting simulations, we summarize our main analytical results. For the voter model on fully mixed populations (complete graphs) with $m$ candidates and $\alpha_i$ zealots committed to candidate~$i$, the stationary distribution of the free-voter count vector $k=(k_1,\dots,k_m)$ is Dirichlet--multinomial,
\[
\pi(k_1,\dots,k_m; \alpha_1,\dots,\alpha_m),
\]
given explicitly in Eq.~\eqref{DR_voter}. The critical (uniform) state is obtained by setting $\alpha_i=1$ for all $i=1,\dots,m$.

For the voter model on Erd\H{o}s--R\'enyi graphs with mean free-degree $d=p(n-1)$, our ER mean-field approximation yields the same Dirichlet--multinomial form, but with rescaled zealot parameters. Equivalently, relative to the fully mixed case, zealot influence is amplified by the factor $(n-1)/d$ (Eqs.~\eqref{voterer1} and \eqref{voterer2}). The mapping can be written as
\begin{equation}
\pi^{\mathrm{ER}}\left(k_1,\dots,k_m;\,
\overbrace{\alpha_1 \frac{d}{n-1}}^{\alpha^{\mathrm{ER}}_1},\dots,
\overbrace{\alpha_m \frac{d}{n-1}}^{\alpha^{\mathrm{ER}}_m}
\right)
=
\pi(k_1,\dots,k_m; \alpha_1,\dots,\alpha_m),
\label{summary_1}
\end{equation}
which shows that, when $d\ll n$, proportionally fewer zealots suffice to reproduce the fully mixed stationary distribution. In particular, the critical state on ER graphs corresponds to $\alpha^{\mathrm{ER}}_i=d/(n-1)$ for all $i$ (i.e., $\alpha_i=1$ in \eqref{summary_1}).

For the $m$-allele Moran process on both fully mixed populations and Erd\H{o}s--R\'enyi graphs, Eqs.~\eqref{moran-random-voter-corr} and \eqref{moran-random-moran-corr} imply that the stationary allele-count distribution is again Dirichlet--multinomial:
\begin{equation}
\pi^{\mathrm{M},\mathrm{ER}}(k_1,\dots,k_m;u_1,\dots,u_m)
=
\pi^{\mathrm{M}}(k_1,\dots,k_m;u_1,\dots,u_m)
=
\pi(k_1,\dots,k_m; \alpha_1,\dots,\alpha_m),
\label{summary_2}
\end{equation}
with parameters $\alpha_i=\frac{2(n-1)u_i}{1-\Lambda}$, where $\Lambda=\sum_{i=1}^m u_i$. In particular, within this approximation the stationary law is independent of $d$. At criticality the stationary distribution is uniform: setting $\alpha_i=1$ for all $i$ forces $u_i=u_c$ for all $i$, yielding the \emph{critical mutation}
\[
u_c=\frac{1}{2n-2+m}
\]
(Eq.~\eqref{criticalmu}).

Finally, Eq.~\eqref{moran-random-voter-corr} gives the induced correspondence between the Moran process and the voter model on ER networks:
\begin{equation}
\pi^{\mathrm{M},\mathrm{ER}}(k_1,\dots,k_m;u_1,\dots,u_m)
=
\pi^{\mathrm{ER}}\Bigl(k_1,\dots,k_m;\,
\overbrace{\alpha^{\mathrm{ER}}_1}^{\frac{2d\,u_1}{1-\Lambda}},\dots,
\overbrace{\alpha^{\mathrm{ER}}_m}^{\frac{2d\,u_m}{1-\Lambda}}
\Bigr).
\label{summary_3}
\end{equation}
In particular, when $d=n-1$ (the complete-graph limit), we recover the exact fully mixed Moran--voter correspondence, namely the second equality in Eq.~\eqref{summary_2}.

\section{Simulations}
\label{simulatons}

In this section we present simulations of the Moran and voter models with three alleles/candidates on several network topologies. Our goal is twofold: (i) to verify the theoretical predictions derived in the previous sections (and Appendix~\ref{app:second-order-degree}), and (ii) to probe their robustness and limitations beyond the settings where the mean-field assumptions are expected to hold.

For a fixed network and parameter set (zealot counts $\alpha_i$ for the voter model, and mutation rates $u_i$ for the Moran model), we initialize the population and run the dynamics for $N_t$ update steps. Each run ends in a state $(k_1,k_2,k_3)$, where $k_i$ is the number of individuals of allele~$i$ (Moran) or supporting candidate~$i$ (voter). Repeating this procedure $N_s$ times yields an empirical distribution: we count how many times each terminal state $(k_1,k_2,k_3)$ occurs and divide by $N_s$ to estimate its stationary probability.

\begin{figure}[htb!]
	\centering
	\includegraphics[width=0.35\textwidth]{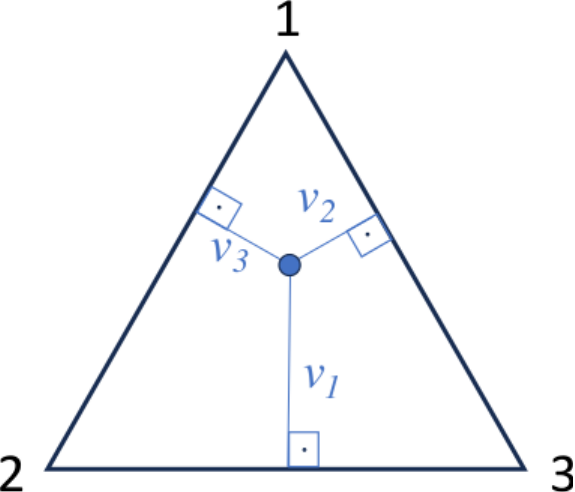} 
	\caption{Simplex depicting the set of states $(k_1,k_2,k_3)$ with $k_1+k_2+k_3=n$. Each point in the simplex corresponds to frequencies $(v_1,v_2,v_3) = (k_1/n,k_2/n,k_3/n)$. The corners represent all individuals of type 1 (upper corner), or type 2 (left corner) or type 3 (right corner). An internal point has frequencies corresponding to the length of the line drawn from the point to the sides as indicated.}
	\label{fig:simplex}
\end{figure}

\begin{figure}[htb!]
	\centering
	$\Scale[1.5]{\qquad \qquad {\rm Exact} \qquad \qquad \qquad {\rm Sampled} }$ \\
	\vspace{0.2cm}
	\raisebox{1.5cm}{$\Scale[1.5]{u < u_c}$} \quad
	\includegraphics[width=0.25\textwidth]{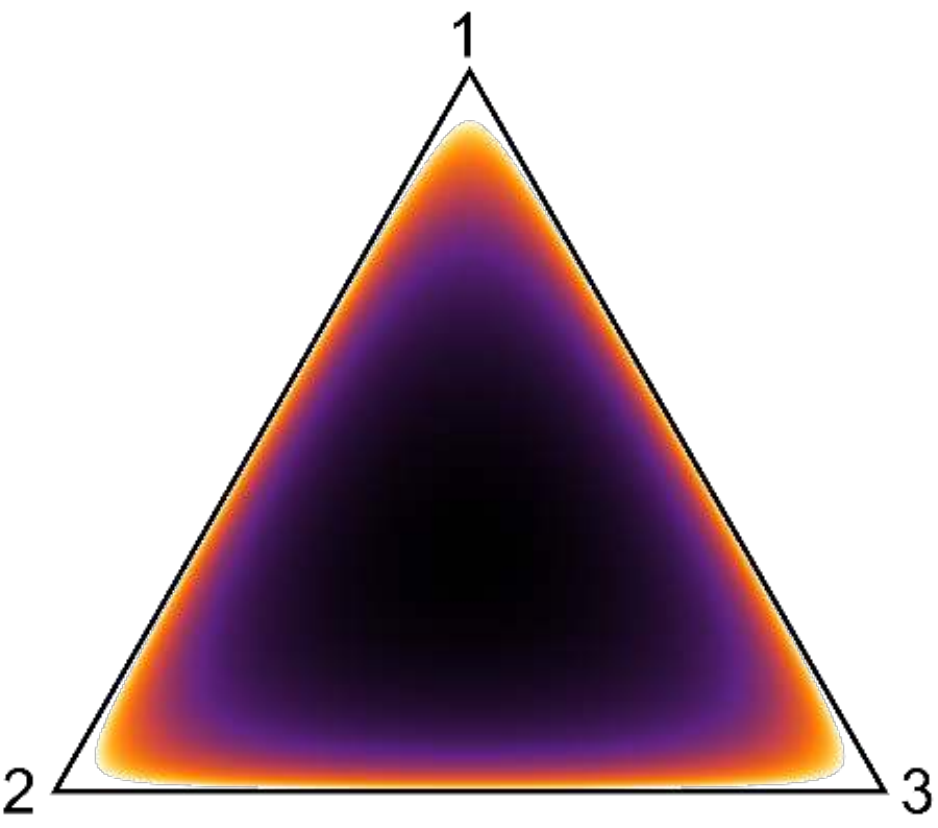}  \qquad
	\includegraphics[width=0.25\textwidth]{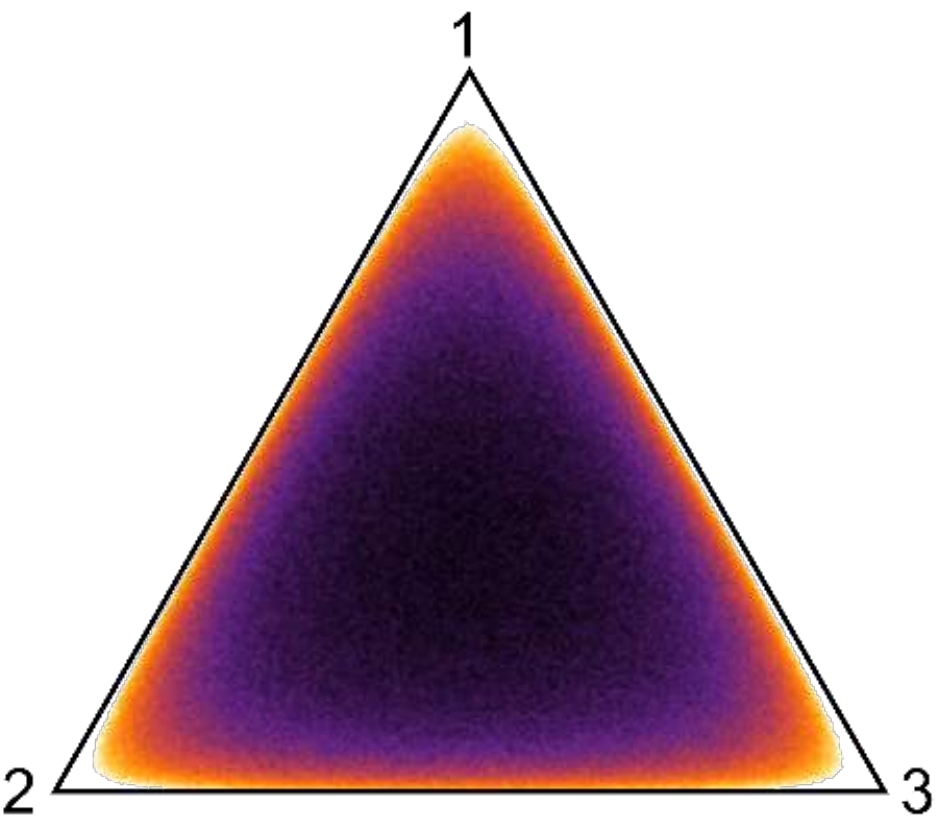} \\
	\raisebox{1.5cm}{$\Scale[1.5]{u = u_c}$} \quad
	\includegraphics[width=0.25\textwidth]{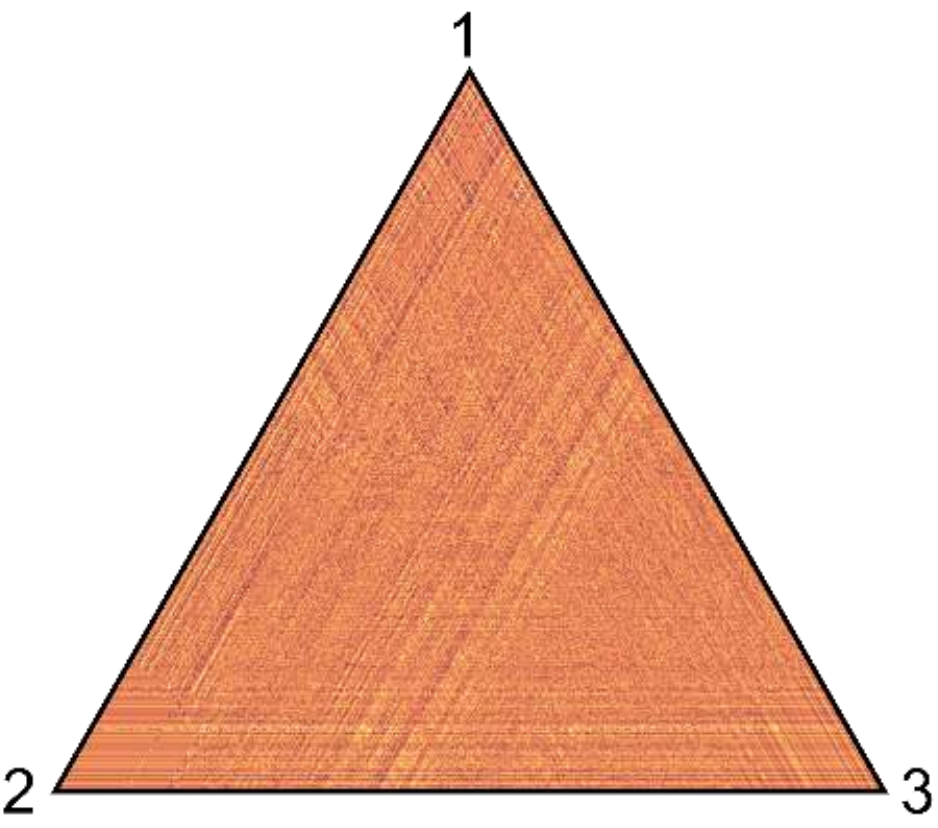} \qquad
	\includegraphics[width=0.25\textwidth]{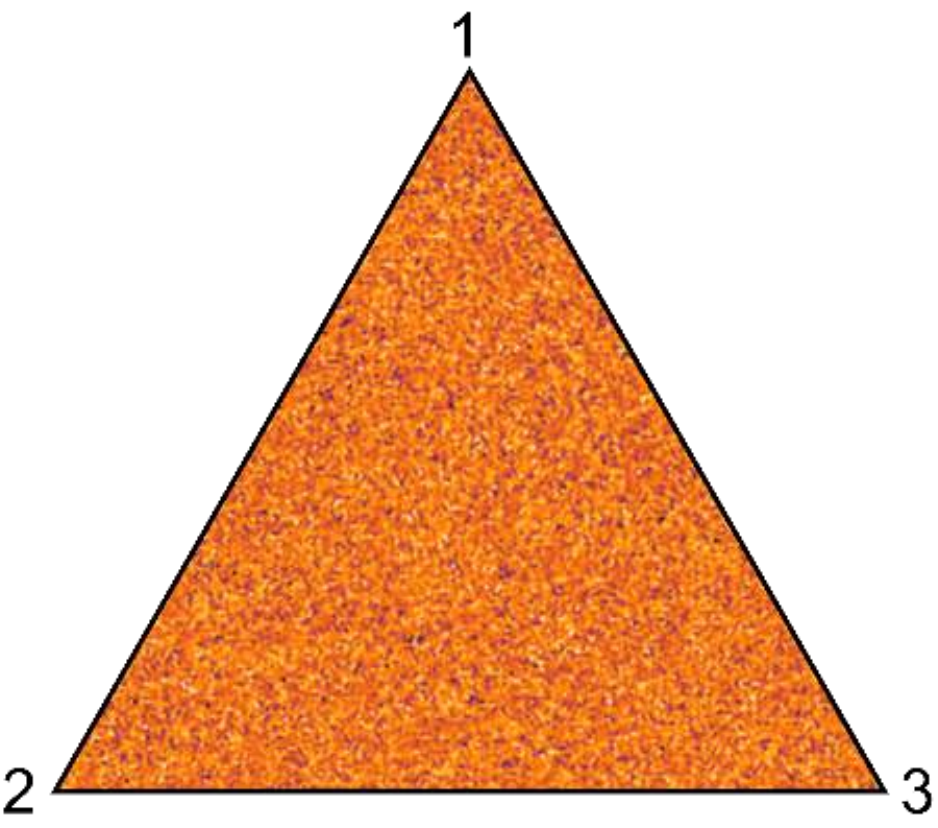} \\
	\raisebox{1.5cm}{$\Scale[1.5]{u > u_c}$} \quad
	\includegraphics[width=0.25\textwidth]{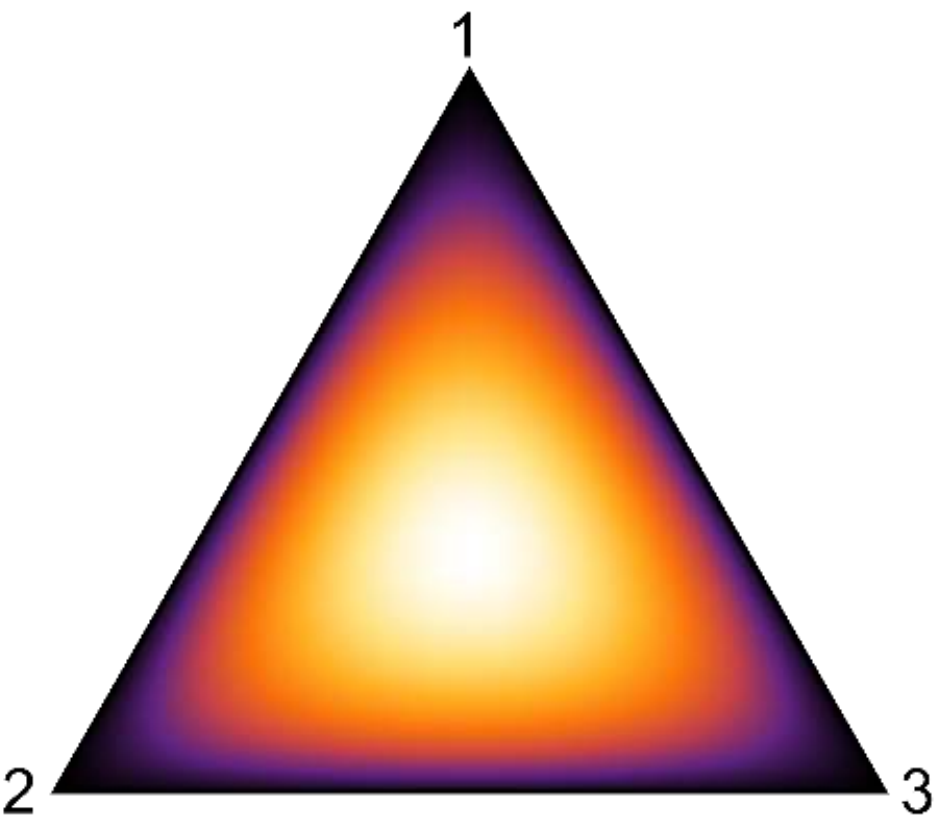} \qquad
	\includegraphics[width=0.25\textwidth]{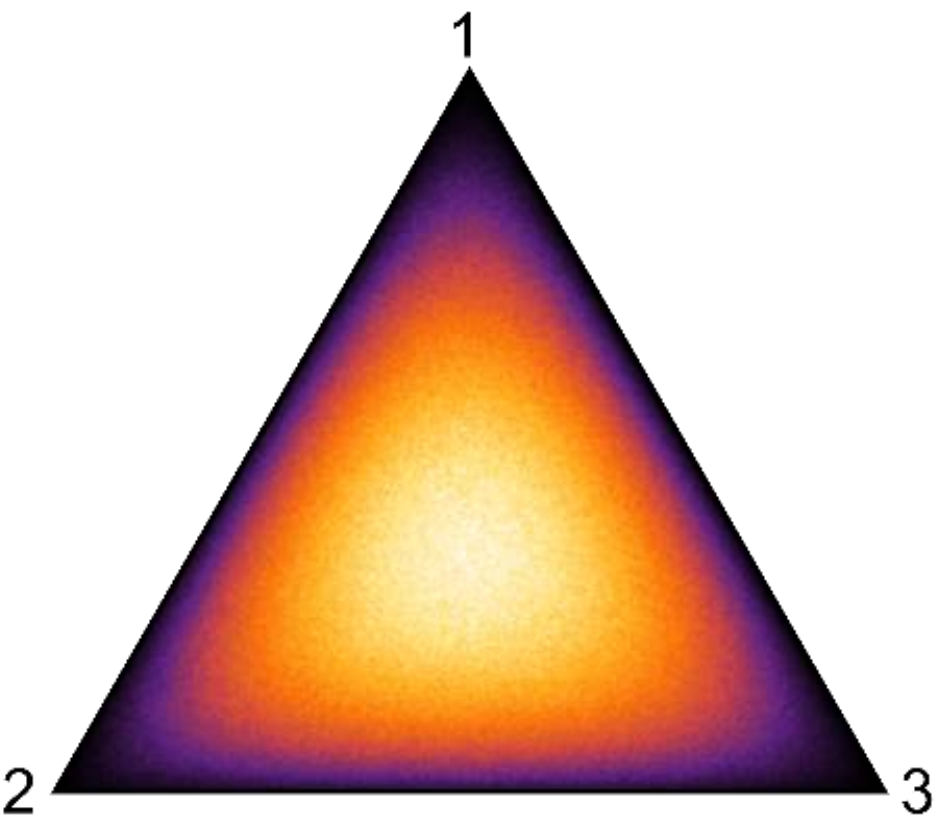} \\ 
	\vspace{0.25cm}
	\qquad \qquad \includegraphics[width=0.6\textwidth]{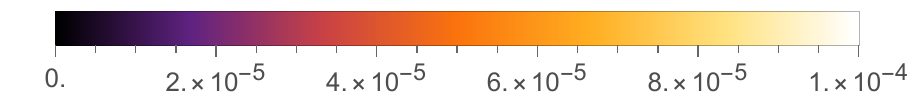} 
	\caption{Heat-maps comparing the exact Dirichlet--multinomial distribution with the distribution obtained from $10^7$ samples generated from the exact distribution for 3 alleles and $n=201$. The critical mutation probability (used in the corresponding Moran model) is $u_c=1/(2n+1)$ (middle row), corresponding to $\alpha_c=1$. Top row: $u = 1/(4n-1) < u_c$, corresponding to $\alpha=0.5$; Bottom row: $u = 1/(n+2) > u_c$, corresponding to $\alpha=2.0$.}
	\label{fig:exact}
\end{figure}

Because $k_1+k_2+k_3=n$, only two coordinates are independent, so we visualize the distribution as a heat-map on the simplex (Fig.~\ref{fig:simplex}). The probability assigned to a configuration is encoded by the color intensity in the heat map.
High-diversity regimes concentrate near the center of the simplex, whereas low-diversity regimes concentrate near the corners (fixation by a single type). Along an edge of the simplex, the allele indicated at the opposite corner has zero frequency.

Throughout, we consider the symmetric case $\alpha_1=\alpha_2=\alpha_3\equiv \alpha$ and $u_1=u_2=u_3\equiv u$. Each realization starts from i.i.d.\ initial states with $\Pr(\text{type }i)=1/3$, so the initial condition is close to $(1/3,1/3,1/3)$. We fix $n=201$. Unless stated otherwise, we use $N_s=10^7$ independent realizations and run each realization for $N_t=10^5$ update steps for the voter model and $N_t=2\times 10^5$ update steps for the Moran model (which evolves at half the rate in our equivalent formulation; see Sec.~\ref{moran}). We fix the population size to $n=201$ to keep the computational cost manageable while still resolving the stationary distribution on the simplex.

\medskip
\noindent\textbf{Fully mixed baseline.}
As noted in Sec.~\ref{summary}, the Dirichlet--multinomial distribution underlies our theory for both the voter and Moran models. Figure~\ref{fig:exact} (middle column, ``Exact'') shows the Dirichlet--multinomial stationary distribution (Eq.~\eqref{DR_voter}) for $\alpha=0.5$, $1$, and $2$. Using the Moran--voter correspondence (Eq.~\eqref{eq:inv}) in the symmetric three-allele case gives
\[
u=\frac{\alpha}{2n-2+3\alpha}.
\]
Thus $\alpha=0.5,1,2$ correspond to $u=1/(4n-1)<u_c$, $u=1/(2n+1)=u_c$, and $u=1/(n+2)>u_c$, respectively. The ``Sampled'' column in Fig.~\ref{fig:exact} shows the distribution obtained by drawing $N_s=10^7$ samples from the exact Dirichlet--multinomial distribution. Because this sampled distribution is visually indistinguishable from the exact one at this resolution, we use the same sampling-based representation when comparing theory to simulation outputs generated with the same $N_s$.

\begin{figure}[htb!]
	\centering
	$\Scale[1.5]{\qquad {\rm Mean} \; {\rm Field} \qquad \quad \; {\rm Moran} \qquad  \qquad \quad {\rm Voter}}$ \\
	\vspace{0.2cm}
	\raisebox{1.5cm}{$\Scale[1.5]{u < u_c}$} \quad
	\includegraphics[width=0.25\textwidth]{th-sample-equi-alpha-05-r}
	\includegraphics[width=0.25\textwidth]{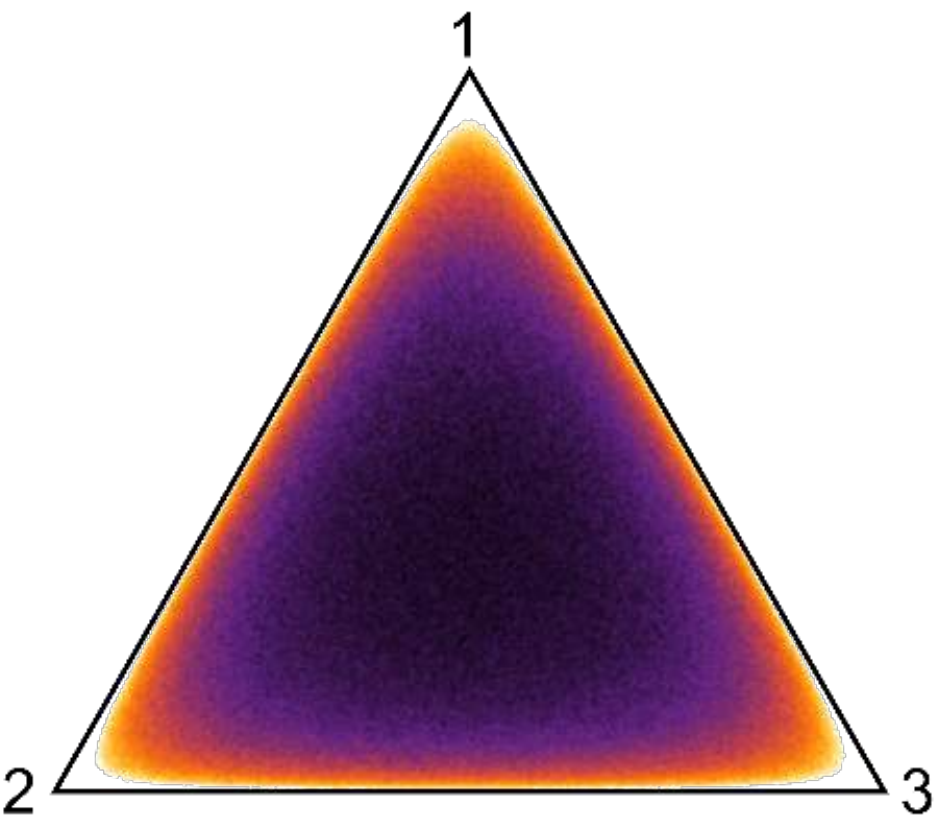}
	\includegraphics[width=0.25\textwidth]{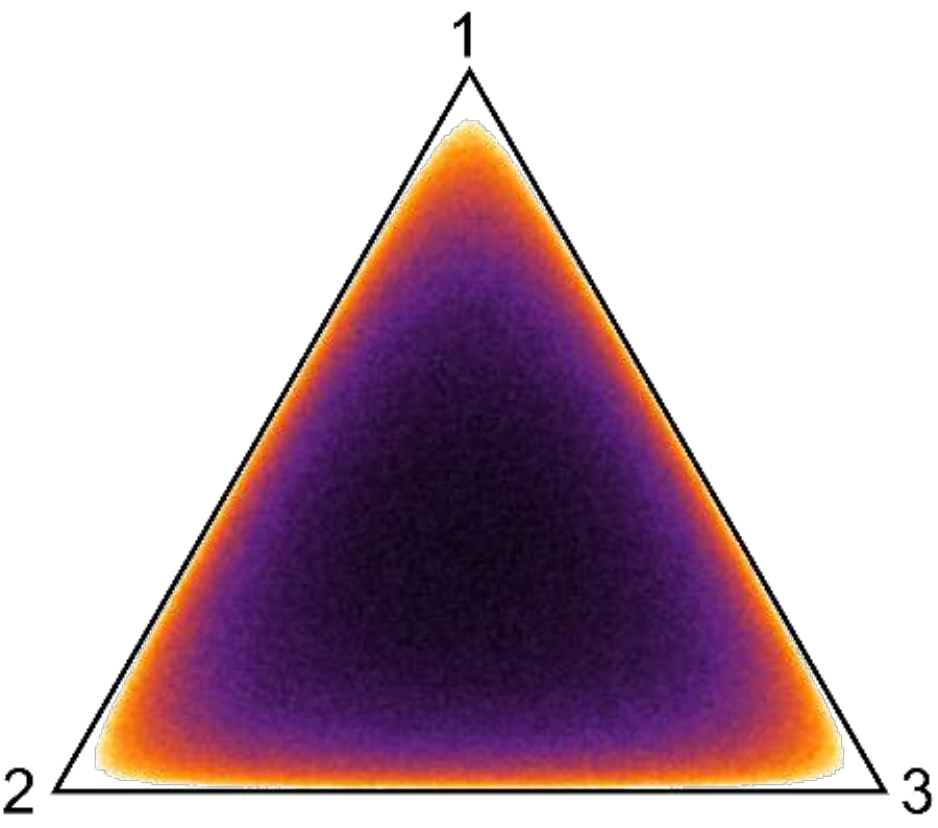}\\
	\raisebox{1.5cm}{$\Scale[1.5]{u = u_c}$} \quad
	\includegraphics[width=0.25\textwidth]{th-sample-equi-alpha-10-r}
	\includegraphics[width=0.25\textwidth]{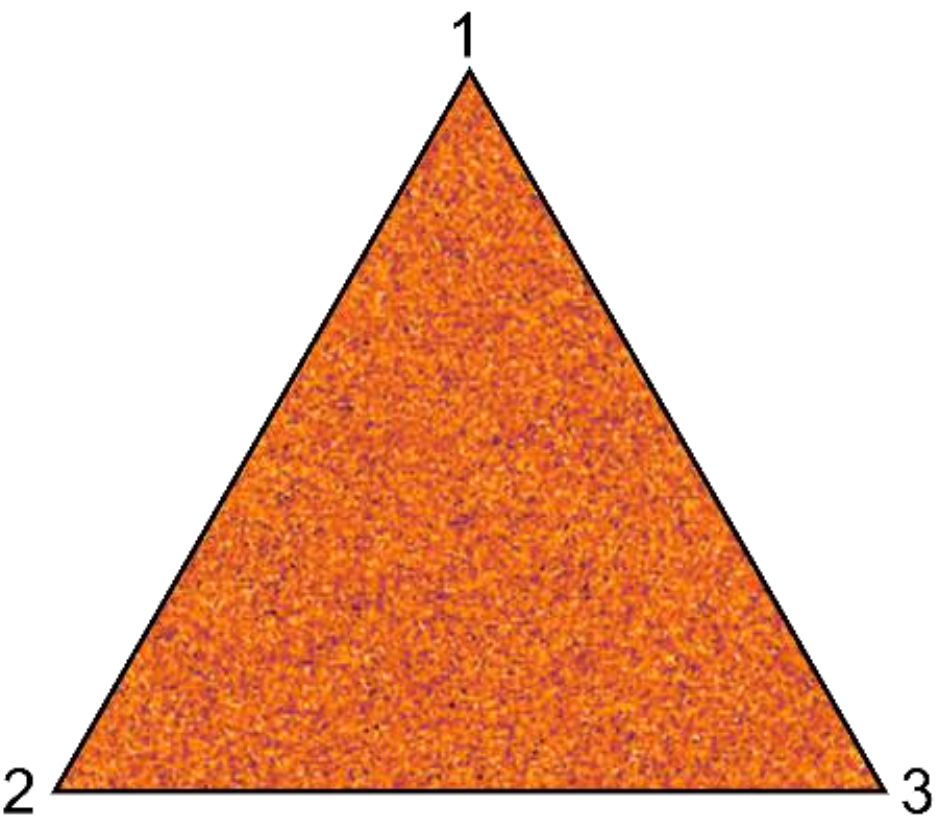}
	\includegraphics[width=0.25\textwidth]{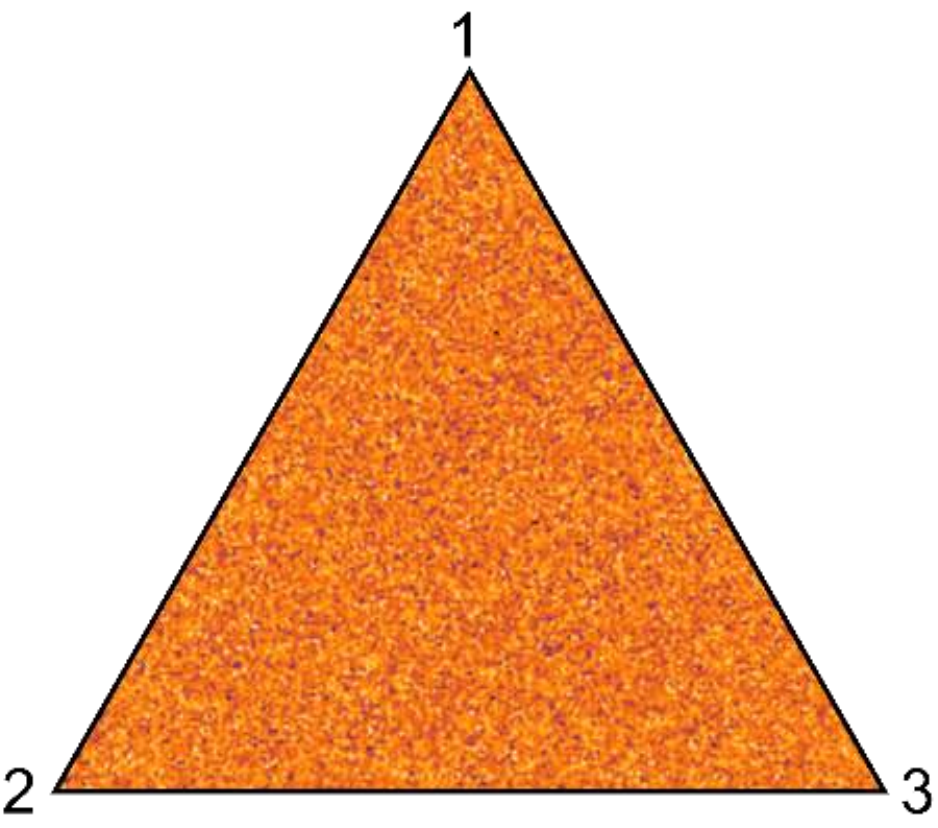}\\
	\raisebox{1.5cm}{$\Scale[1.5]{u > u_c}$} \quad
	\includegraphics[width=0.25\textwidth]{th-sample-equi-alpha-20-r}
	\includegraphics[width=0.25\textwidth]{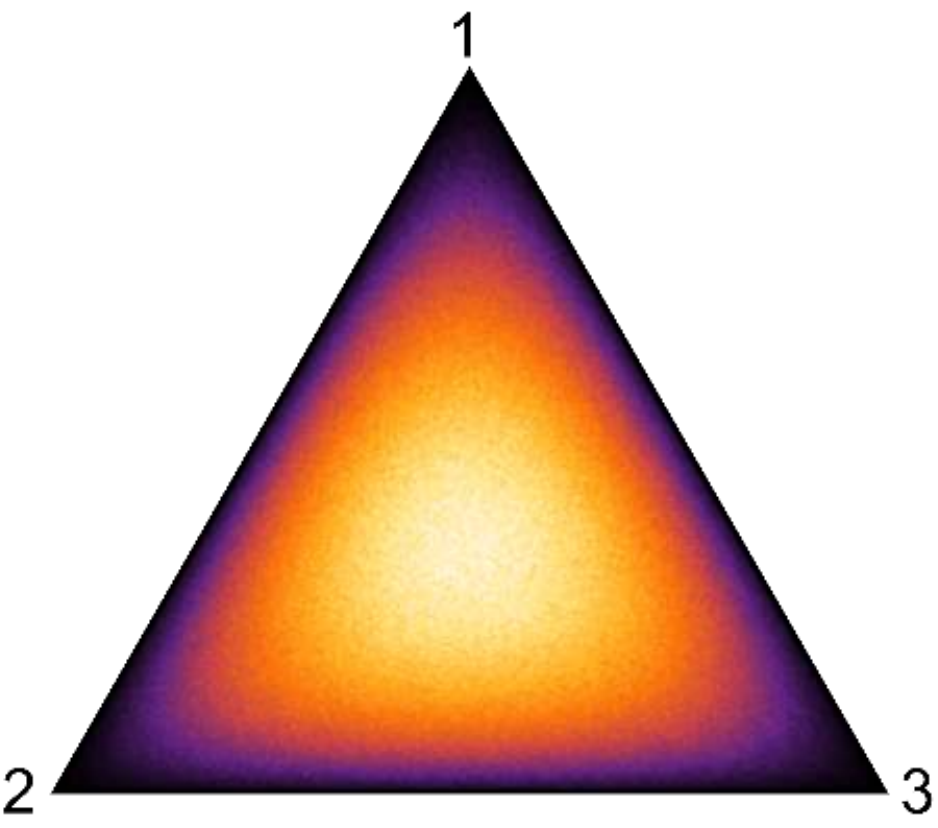}
	\includegraphics[width=0.25\textwidth]{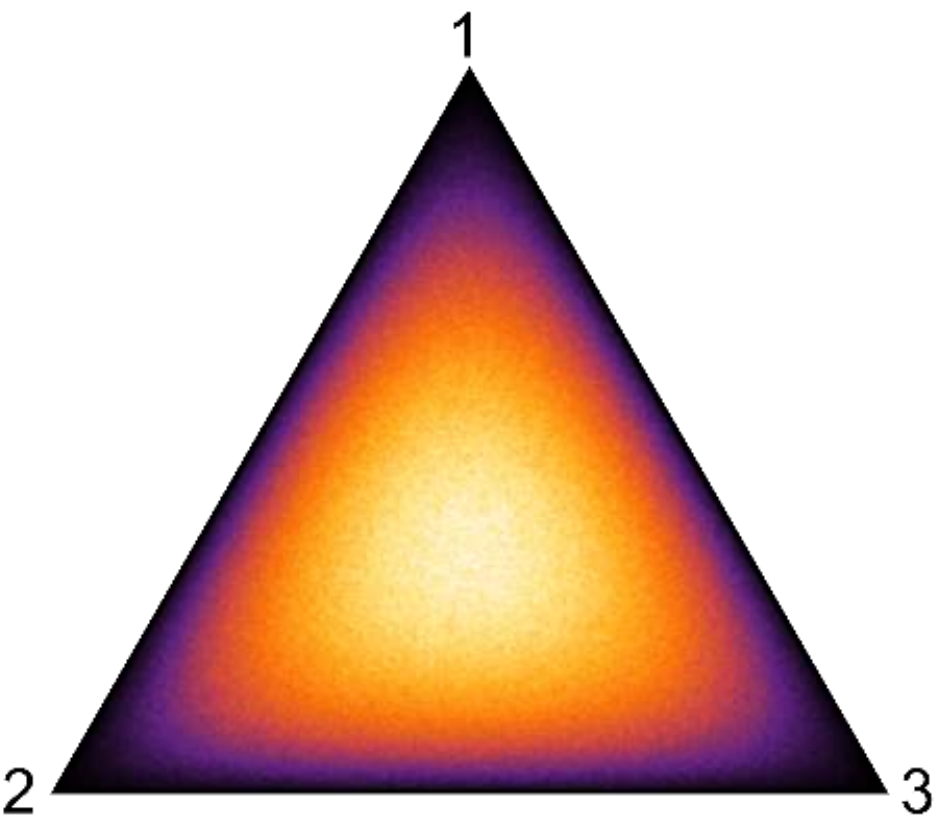} 
	\caption{Simulations on Erd\H{o}s--R\'enyi graphs with connection probability $p=0.3$. The critical
		mutation probability is $u_c=1/(2n+1)$ (middle row), corresponding to $\alpha_c=1$. Top row: $u = 1/(4n-1) < u_c$ and
		$\alpha=0.5$; Bottom row: $u = 1/(n+2) > u_c$ and $\alpha=2.0$. Columns show the mean-field prediction, and the simulated stationary distributions for the Moran and voter models. Color bar is the same as in Fig.~\ref{fig:exact}.}
	\label{fig:er}
\end{figure}

\medskip
\noindent\textbf{Erd\H{o}s--R\'enyi networks.}
Figure~\ref{fig:er} shows simulations on ER graphs with connection probability $p=0.3$, $n=201$ nodes, and mean degree $d=p(n-1)=60$. To compare simulations with the ER mean-field predictions summarized in Sec.~\ref{summary}, we use the same values of $\alpha$ as in Fig.~\ref{fig:exact}, namely $\alpha=0.5,1,2$, corresponding to $u=1/(4n-1)$, $u=1/(2n+1)$, and $u=1/(n+2)$, respectively. The ``Mean Field'' column shows the predicted Dirichlet--multinomial distribution generated by sampling, as in Fig.~\ref{fig:exact}. The ``Moran'' column uses the corresponding values of $u$. For the voter model, we rescale $\alpha$ following Eq.~\eqref{summary_1} by defining $\alpha^{\mathrm{ER}} := \alpha\, d/(n-1)$, which is also consistent with Eq.~\eqref{summary_3}. As predicted, the stationary distributions for the Moran and voter models are virtually identical and match the mean-field prediction.

\begin{figure}[htb!]
\centering
$\Scale[1.5]{\qquad {\rm Mean} \; {\rm Field} \qquad {\rm Corrected} \qquad  {\rm Moran} \qquad  \qquad {\rm Voter}}$ \\
\vspace{0.2cm}
\raisebox{1.5cm}{$\Scale[1.5]{u < u_c}$} \quad
\includegraphics[width=0.20\textwidth]{th-sample-equi-alpha-05-r}
\includegraphics[width=0.20\textwidth]{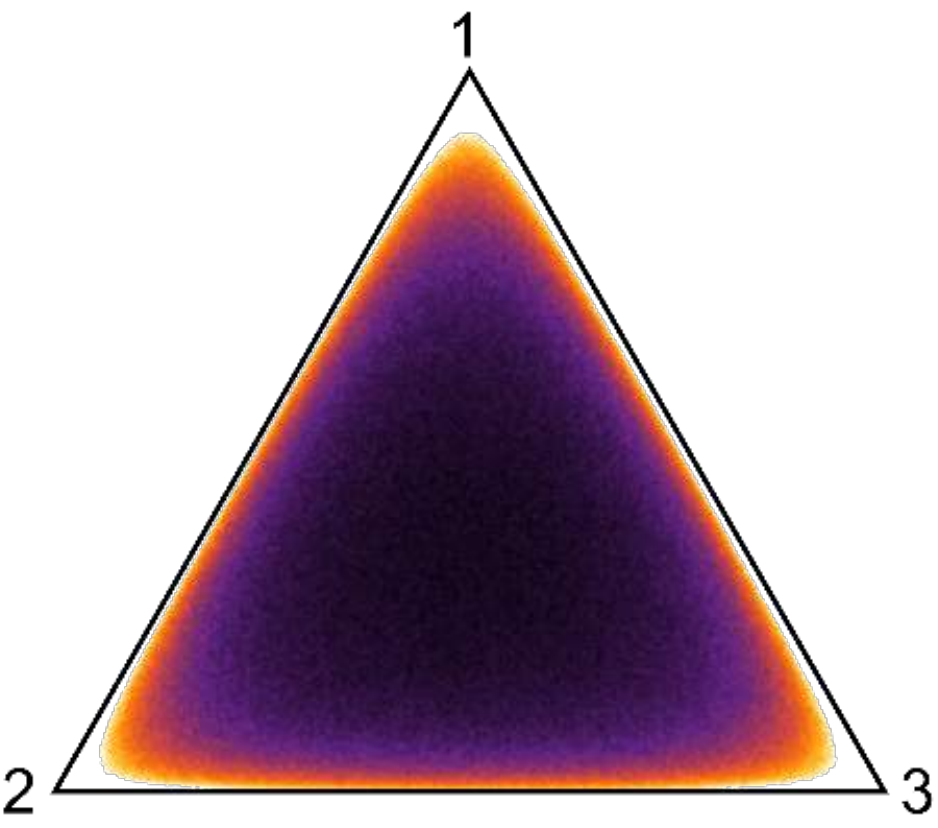}
\includegraphics[width=0.20\textwidth]{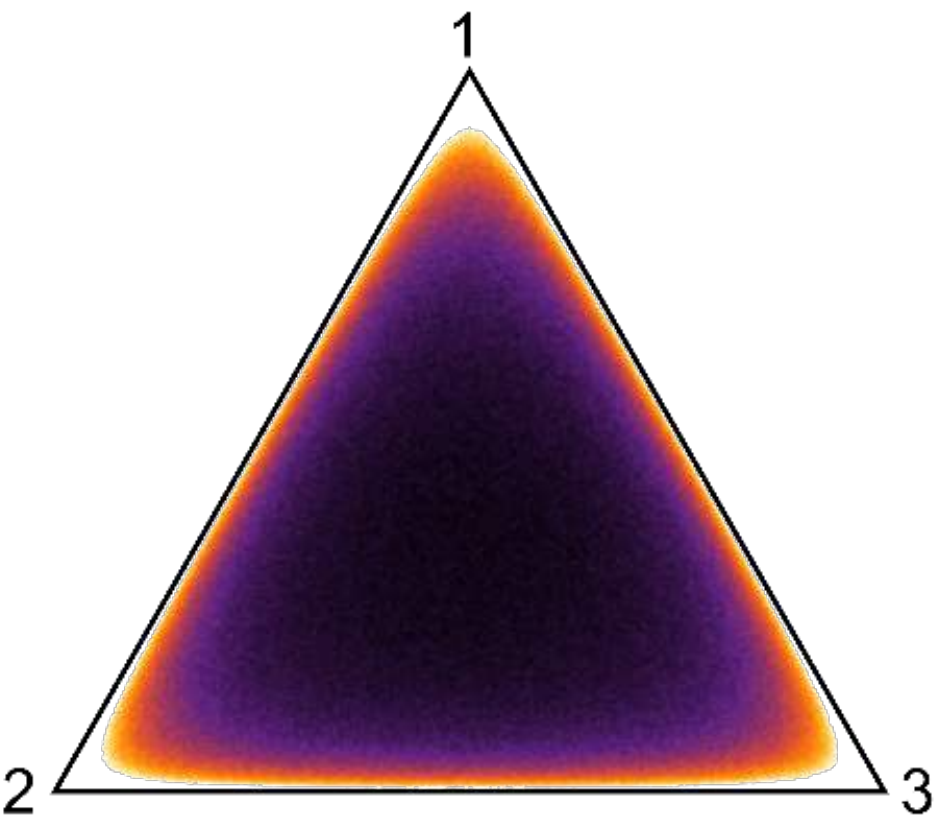}
\includegraphics[width=0.20\textwidth]{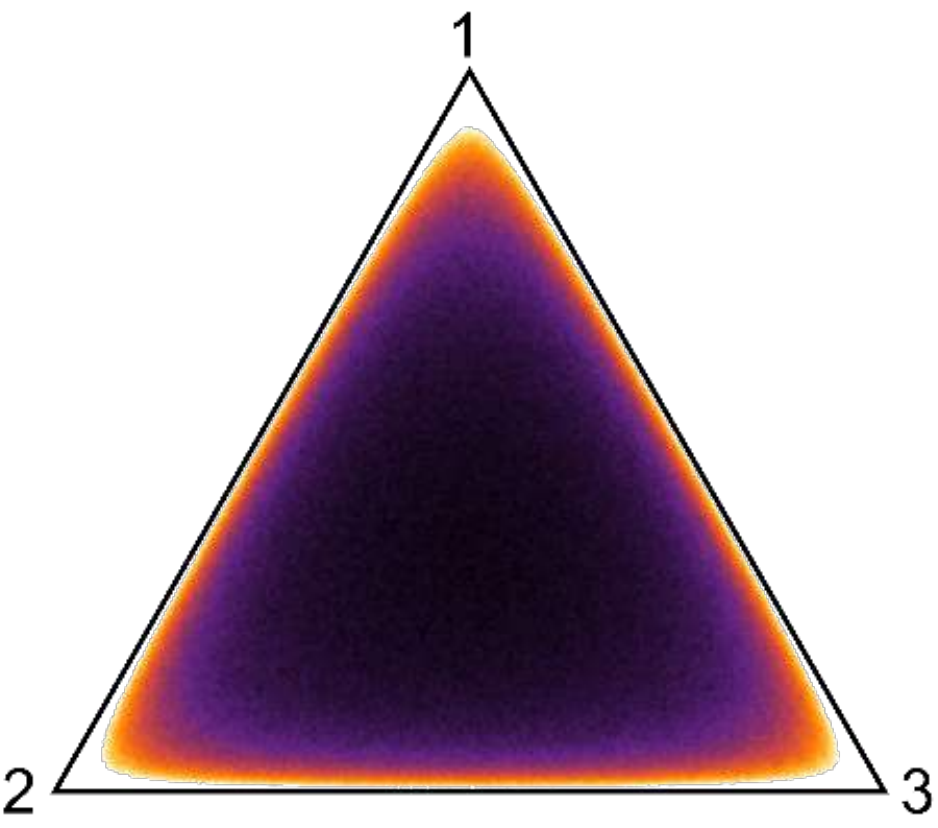}\\
\raisebox{1.5cm}{$\Scale[1.5]{u = u_c}$} \quad
\includegraphics[width=0.20\textwidth]{th-sample-equi-alpha-10-r}
\includegraphics[width=0.20\textwidth]{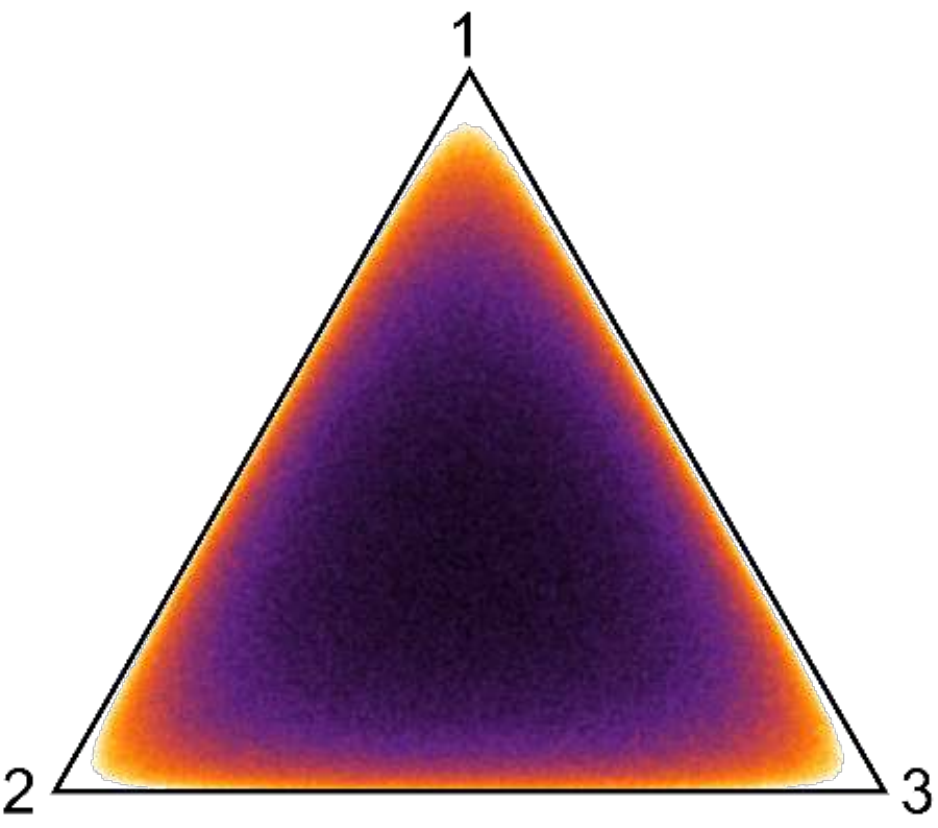}
\includegraphics[width=0.20\textwidth]{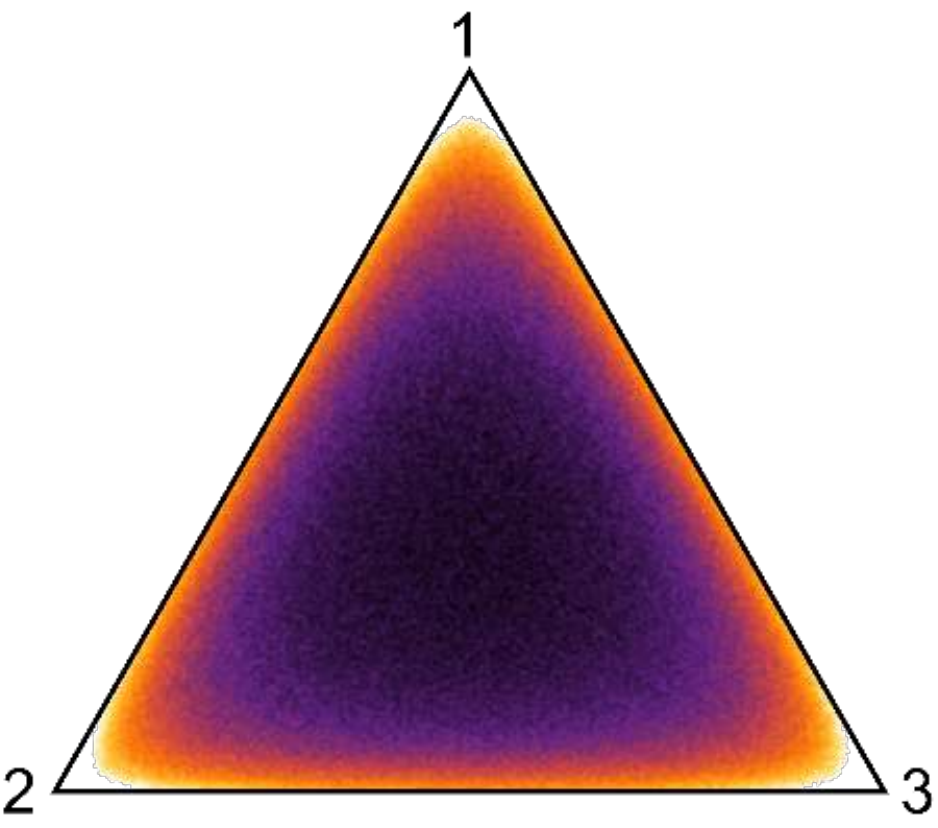}
\includegraphics[width=0.20\textwidth]{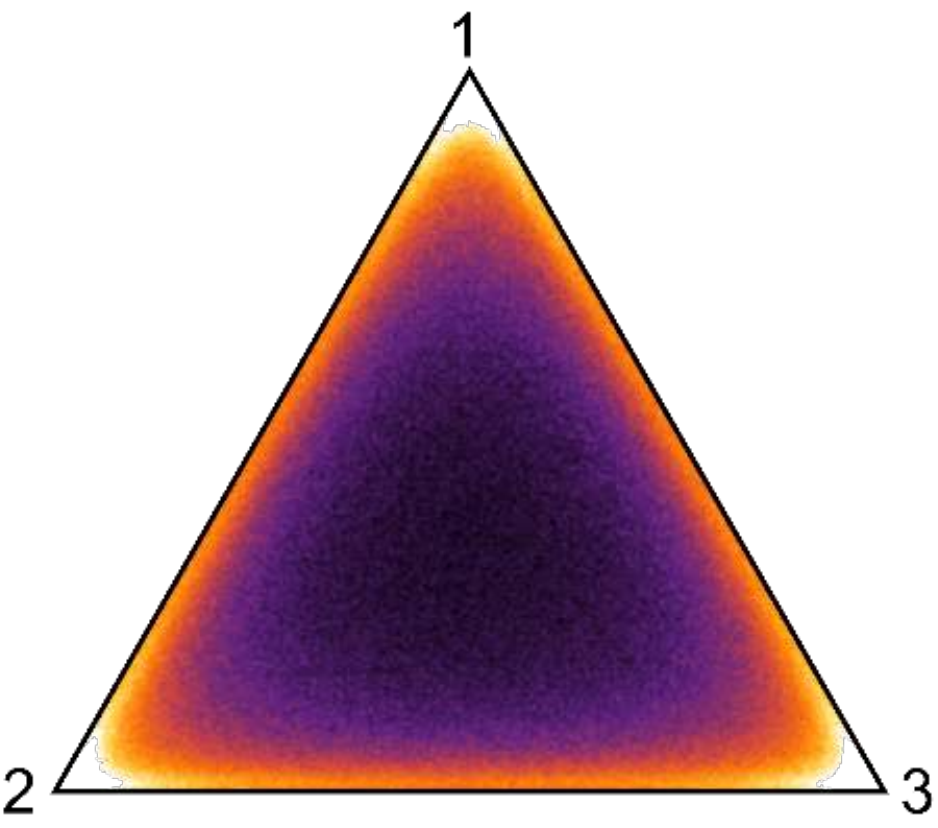}\\
\raisebox{1.5cm}{$\Scale[1.5]{u > u_c}$} \quad
\includegraphics[width=0.20\textwidth]{th-sample-equi-alpha-20-r}
\includegraphics[width=0.20\textwidth]{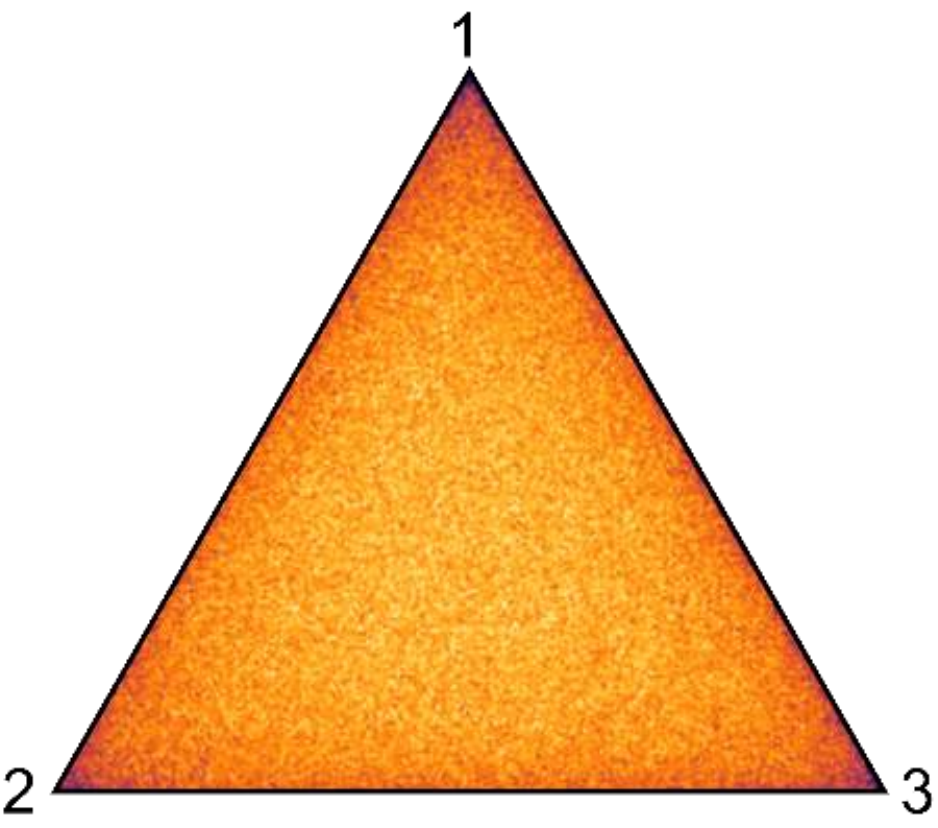}
\includegraphics[width=0.20\textwidth]{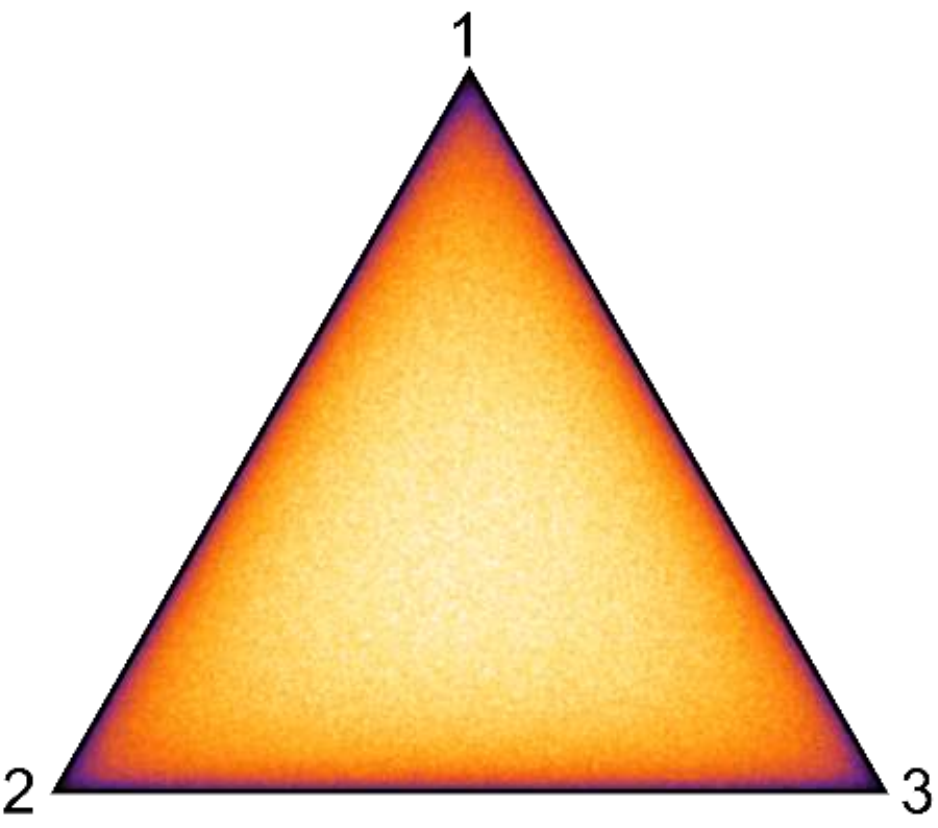}
\includegraphics[width=0.20\textwidth]{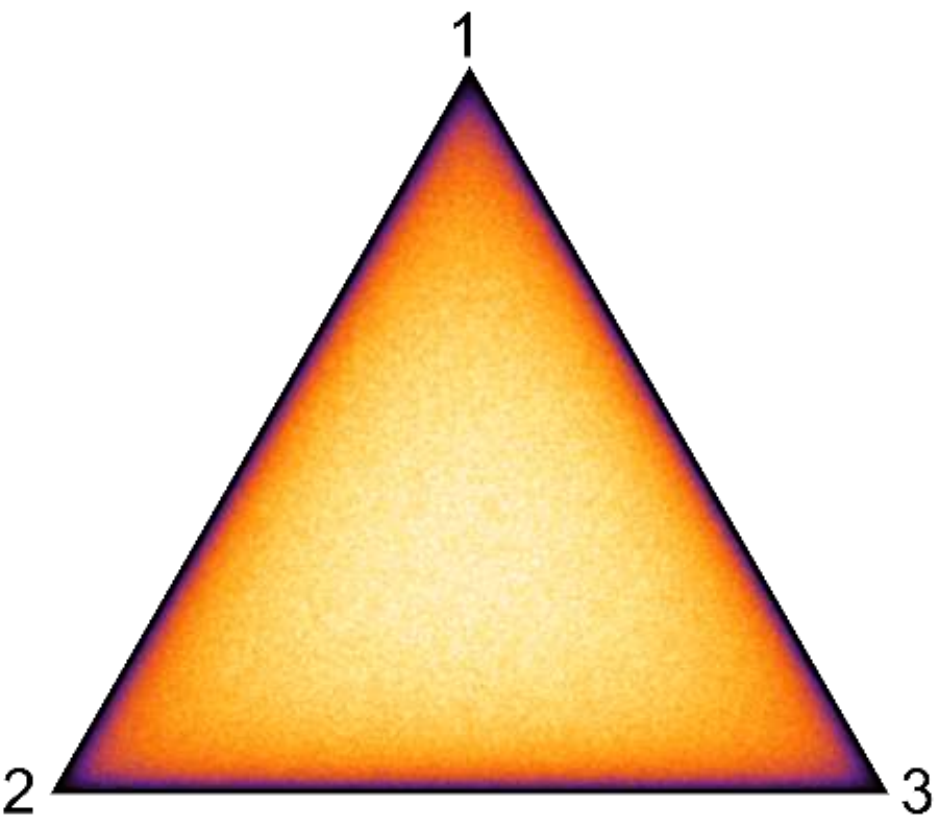}
\caption{Simulations on Barabasi--Albert scale-free graphs with parameters $m_0=m=3$. The critical
	mutation probability is $u_c=1/(2n+1)$ (middle row), corresponding to $\alpha_c=1$. Top row: $u = 1/(4n-1) < u_c$ and
	$\alpha=0.5$; Bottom row: $u = 1/(n+2) > u_c$ and $\alpha=2.0$. Columns show the mean-field prediction, the corrected mean-field approximation (Appendix~\ref{app:second-order-degree}), and simulations for the Moran and voter models, respectively. For the voter model the values of $\alpha$ were rescaled by $d/(n-1)$. Color bar is the same as in Fig.~\ref{fig:exact}.}
\label{fig:sf}
\end{figure}

\medskip
\noindent\textbf{Scale-free networks.}
To assess the role of degree heterogeneity and the accuracy of the mean-field approximations, we also ran simulations on Barabasi--Albert scale-free networks \cite{albert2002statistical}. Starting from $m_0=3$ fully connected nodes, new nodes are added sequentially until a total of $n$ nodes is reached. Each new node forms $m=3$ links with pre-existing nodes via preferential attachment, meaning that the probability of attaching to a given node is proportional to its degree, producing a heterogeneous degree distribution with a few high-degree hubs and many low-degree nodes. The resulting mean degree is $d\approx 2m=6$.

Figure~\ref{fig:sf} shows that, on the same scale-free network, the Moran and voter models produce nearly identical stationary distributions (``Moran'' and ``Voter'' columns), suggesting that the Moran--voter correspondence extends beyond ER graphs to degree-heterogeneous topologies. To match parameters, we run the Moran model with mutation rate $u$ and run the voter model with zealot strength $\alpha^{\mathrm{SF}}=\alpha\,\frac{d}{n-1}$ (with $d\approx 6$), in direct analogy with the ER mapping in Eq.~\eqref{summary_1} and consistent with Eq.~\eqref{summary_3}.  However, the mean-field prediction (``Mean Field'' column) misplaces the apparent critical regime: relative to ER and fully mixed populations, the scale-free topology shifts the distribution toward lower diversity, and simulations suggest a critical mutation larger than the mean-field value $u_c=1/(2n+1)$.

To account for degree heterogeneity, we apply the second-order correction developed in Appendix~\ref{app:second-order-degree}, which incorporates fluctuations of the degree about its mean. The ``Corrected'' column shows the resulting Dirichlet--multinomial approximation with effective parameters given explicitly in Eq.~\eqref{app:eq:alpha-eff-2}, and it agrees closely with the simulations.

We note that even ER graphs can deviate from the leading-order mean-field description when $p$ is small. For instance, for $n=201$ and $p=0.02$ we have $d=4$ and degree standard deviation $\sigma_d \approx 2$. The second-order correction in Appendix~\ref{app:second-order-degree} predicts a nontrivial parameter re-parameterization in this regime; empirically, matching simulations requires effective values of $\alpha$ about $1.25$ times smaller than those predicted by the leading-order mean-field theory (not shown). Such close agreement is expected when neighborhoods are approximately well mixed (as in ER graphs) and is typically weaker in highly degree-heterogeneous networks where links concentrate around hubs (as in scale-free networks).

\begin{figure}[htb!]
	\centering
	$\Scale[1.5]{\qquad \quad {\rm Mean} \; {\rm Field} \qquad \;  {\rm Moran} \qquad  \qquad {\rm Voter}}$ \\
	\vspace{0.2cm}
	\raisebox{1.5cm}{$\Scale[1.5]{u < u_c}$} \quad
	\includegraphics[width=0.25\textwidth]{th-sample-equi-alpha-05-r}
	\includegraphics[width=0.25\textwidth]{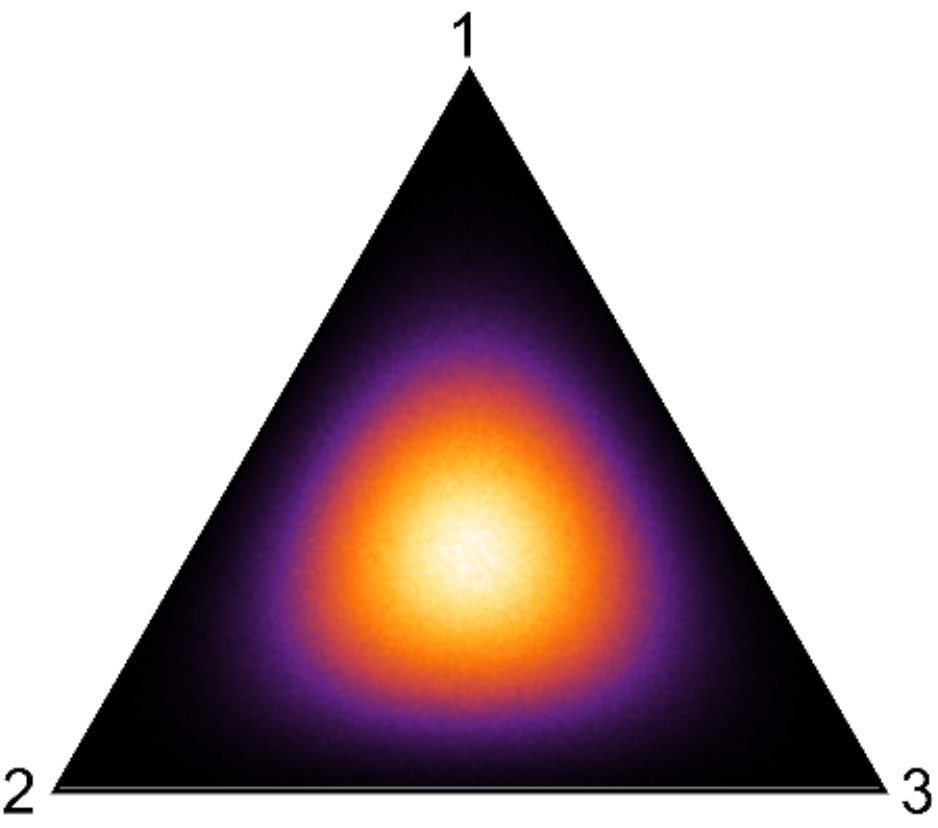}
	\includegraphics[width=0.25\textwidth]{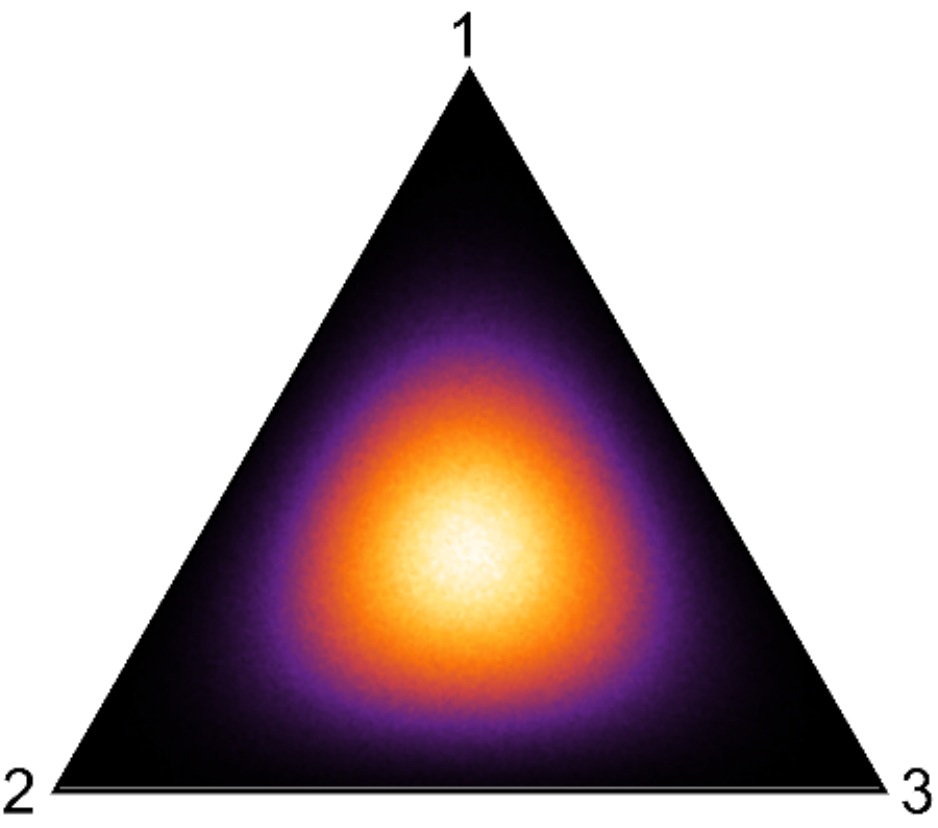}\\
	\raisebox{1.5cm}{$\Scale[1.5]{u = u_c}$} \quad
	\includegraphics[width=0.25\textwidth]{th-sample-equi-alpha-10-r}
	\includegraphics[width=0.25\textwidth]{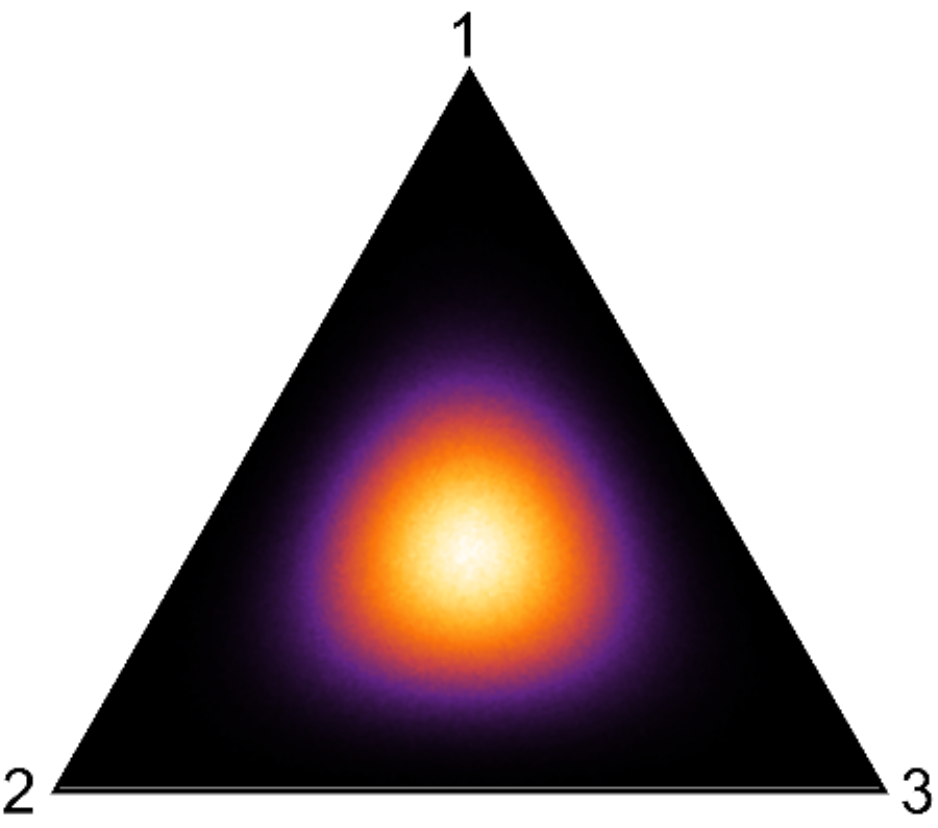}
	\includegraphics[width=0.25\textwidth]{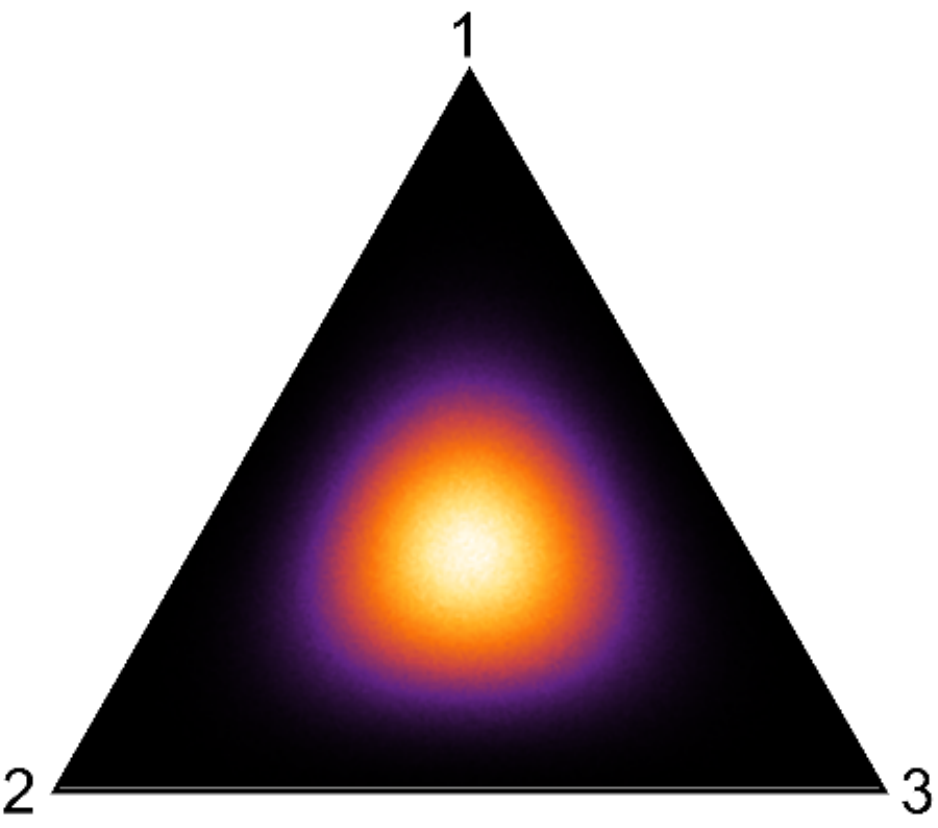}\\
	\raisebox{1.5cm}{$\Scale[1.5]{u > u_c}$} \quad
	\includegraphics[width=0.25\textwidth]{th-sample-equi-alpha-20-r}
	\includegraphics[width=0.25\textwidth]{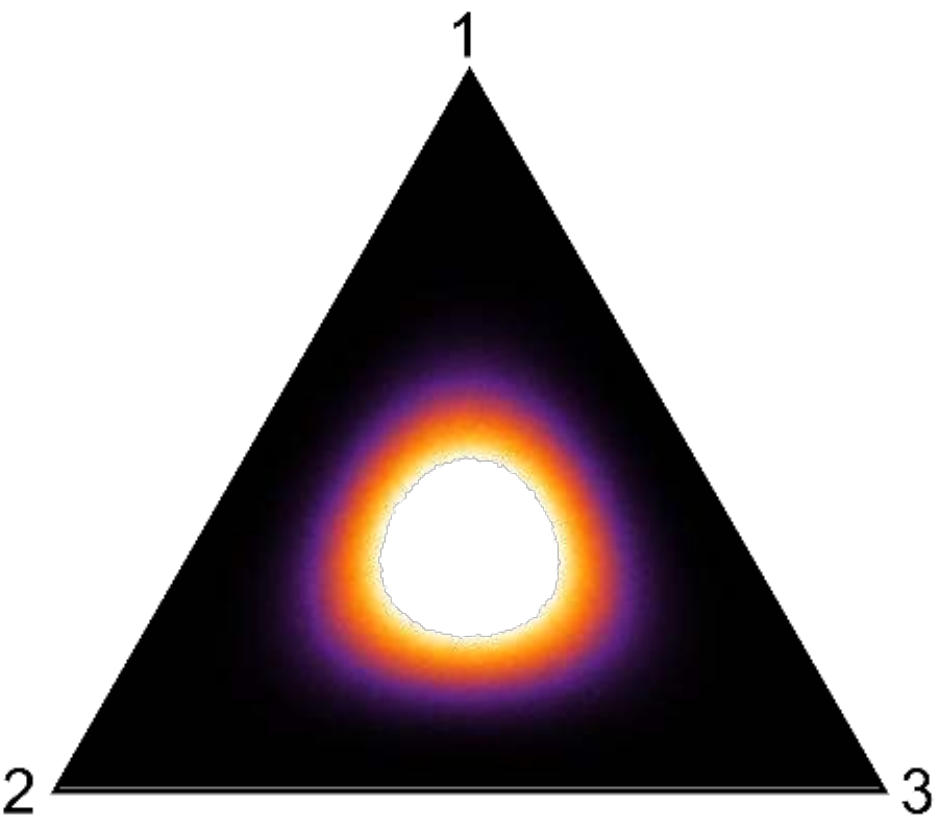}
	\includegraphics[width=0.25\textwidth]{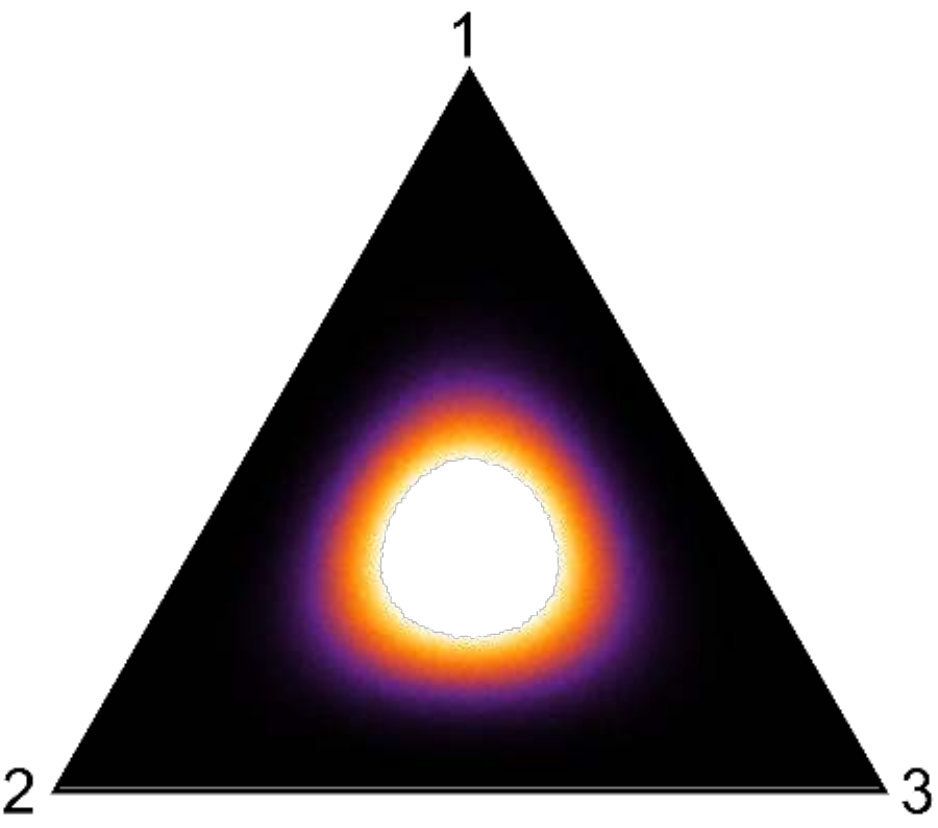}
	\caption{Simulations for 3 alleles on a simple ring graph, where each node is connected to the 2 nearest neighbors. The critical
		mutation probability is $u_c=1/(2n+1)$ (middle row) corresponding to $\alpha_c=1$. Top row: $u = 1/(4n-1) < u_c$ and
		$\alpha=0.5$; Bottom row: $u = 1/(n+2) > u_c$ and $\alpha=2.0$. Columns show the mean-field prediction, and the simulated stationary distributions for the Moran and voter models. Color bar is the same as in Fig.~\ref{fig:exact}.}
	\label{fig:ring1}
\end{figure}

\begin{figure}[htb!]
	\centering
	$\Scale[1.5]{\qquad \quad {\rm Mean} \; {\rm Field} \qquad \;  {\rm Moran} \qquad  \qquad \quad {\rm Voter}}$ \\
	\vspace{0.2cm}
	\raisebox{1.5cm}{$\Scale[1.5]{u < u_c}$} \quad
	\includegraphics[width=0.25\textwidth]{th-sample-equi-alpha-05-r}
	\includegraphics[width=0.25\textwidth]{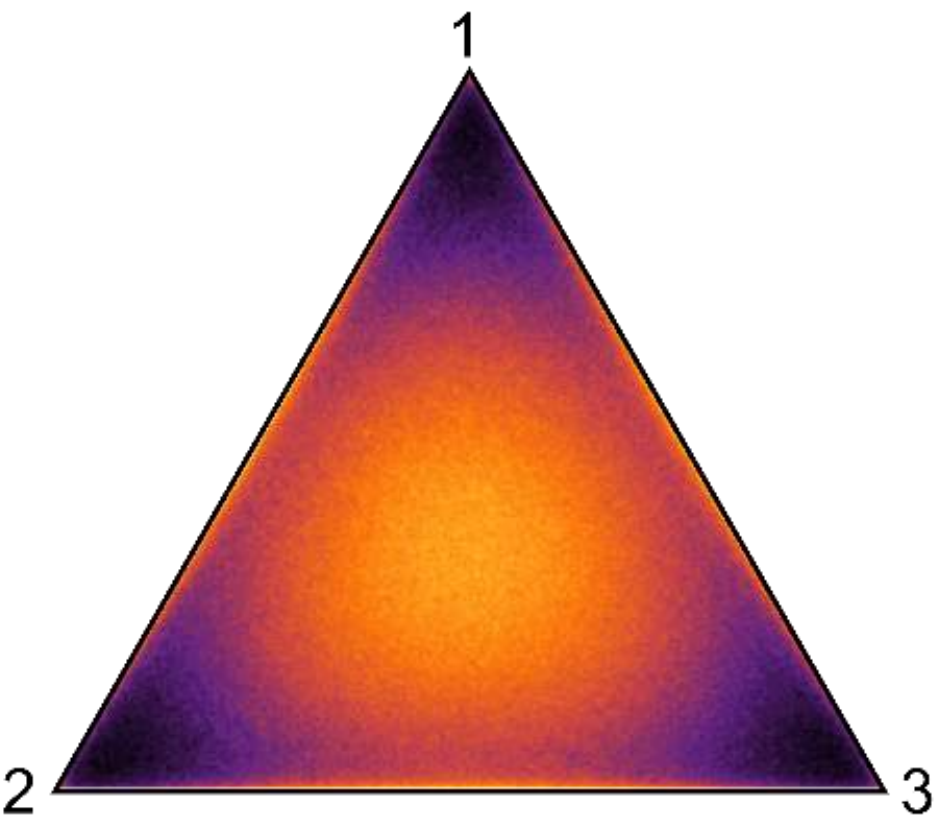}
	\includegraphics[width=0.25\textwidth]{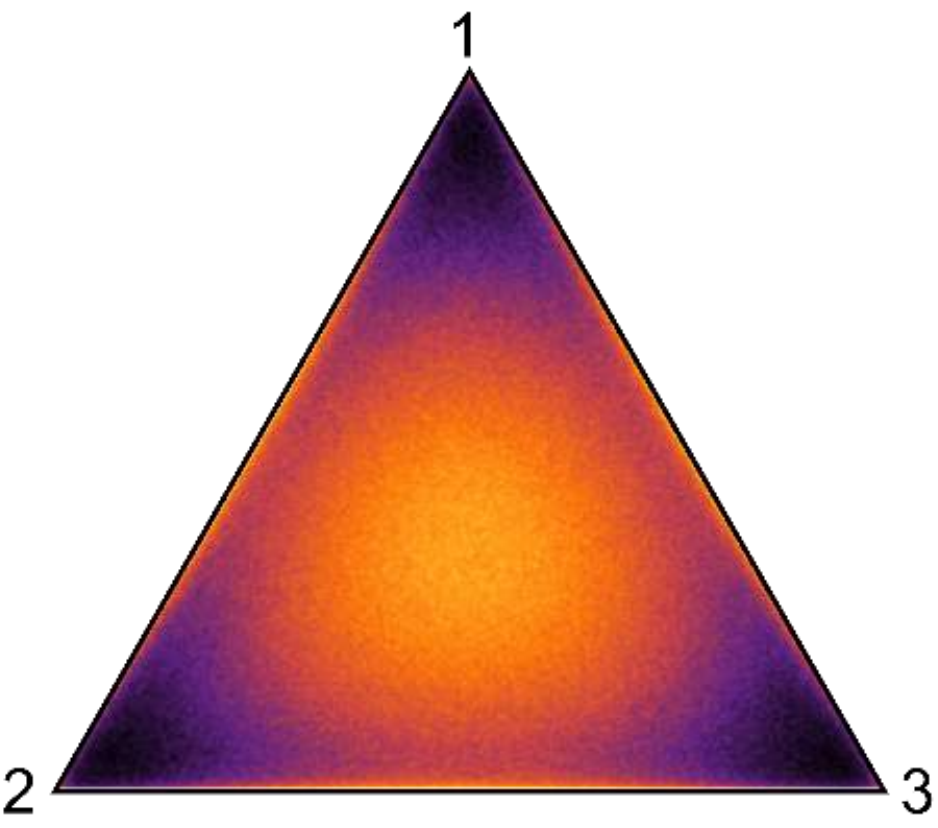}\\
	\raisebox{1.5cm}{$\Scale[1.5]{u = u_c}$} \quad
	\includegraphics[width=0.25\textwidth]{th-sample-equi-alpha-10-r}
	\includegraphics[width=0.25\textwidth]{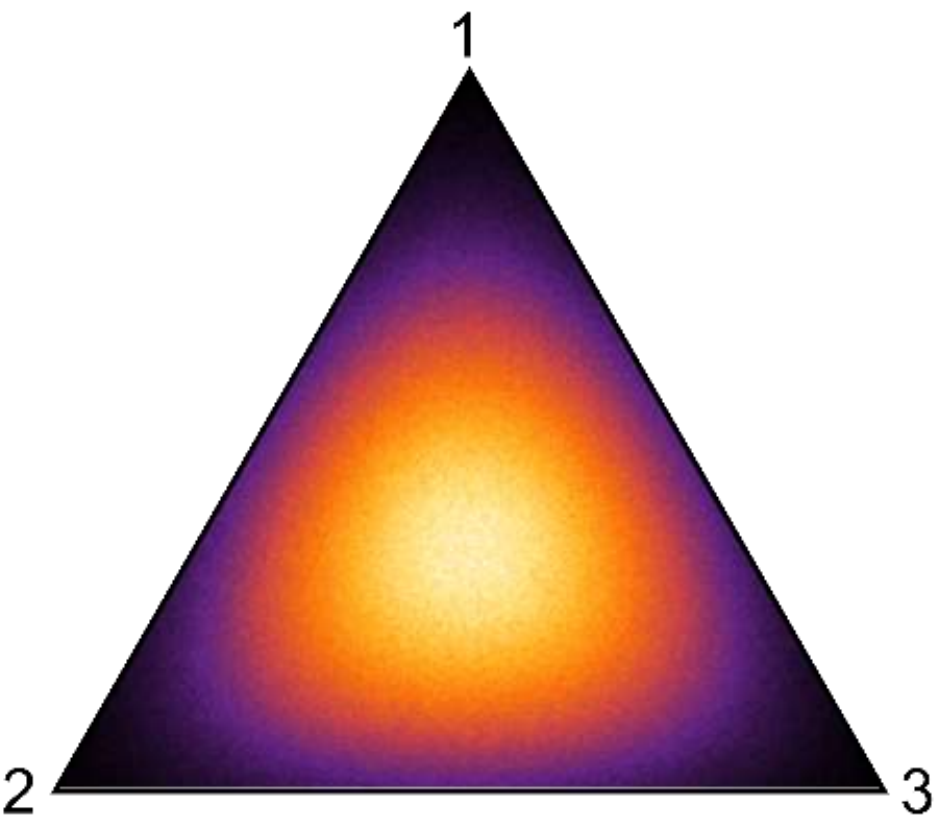}
	\includegraphics[width=0.25\textwidth]{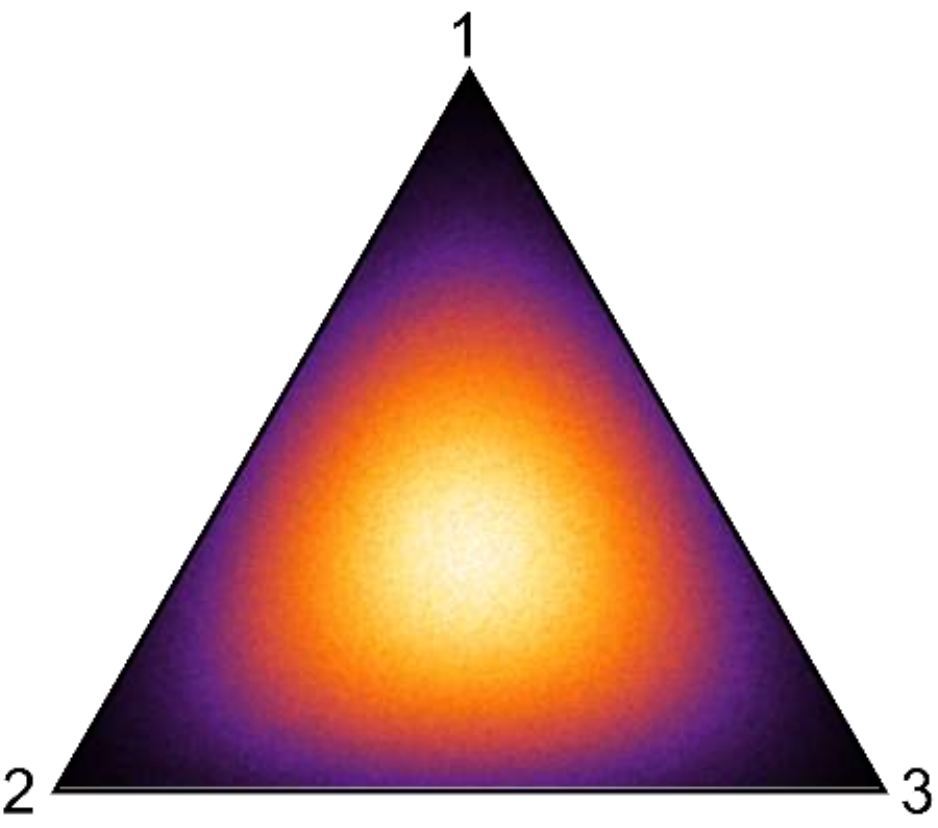}\\
	\raisebox{1.5cm}{$\Scale[1.5]{u > u_c}$} \quad
	\includegraphics[width=0.25\textwidth]{th-sample-equi-alpha-20-r}
	\includegraphics[width=0.25\textwidth]{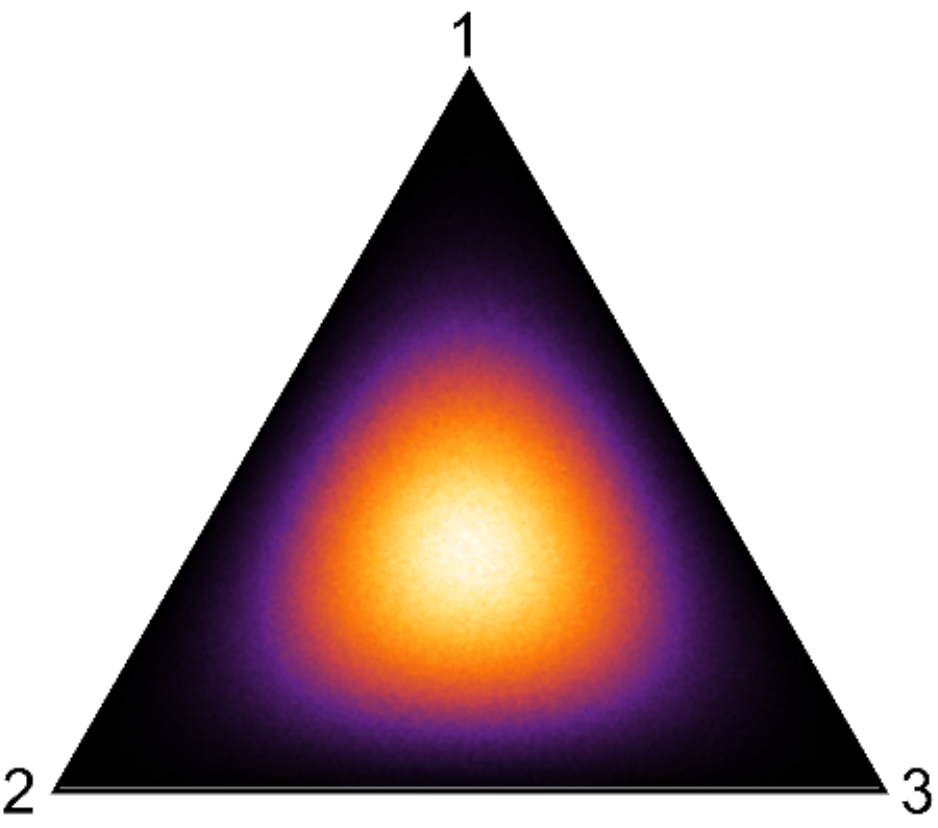}
	\includegraphics[width=0.25\textwidth]{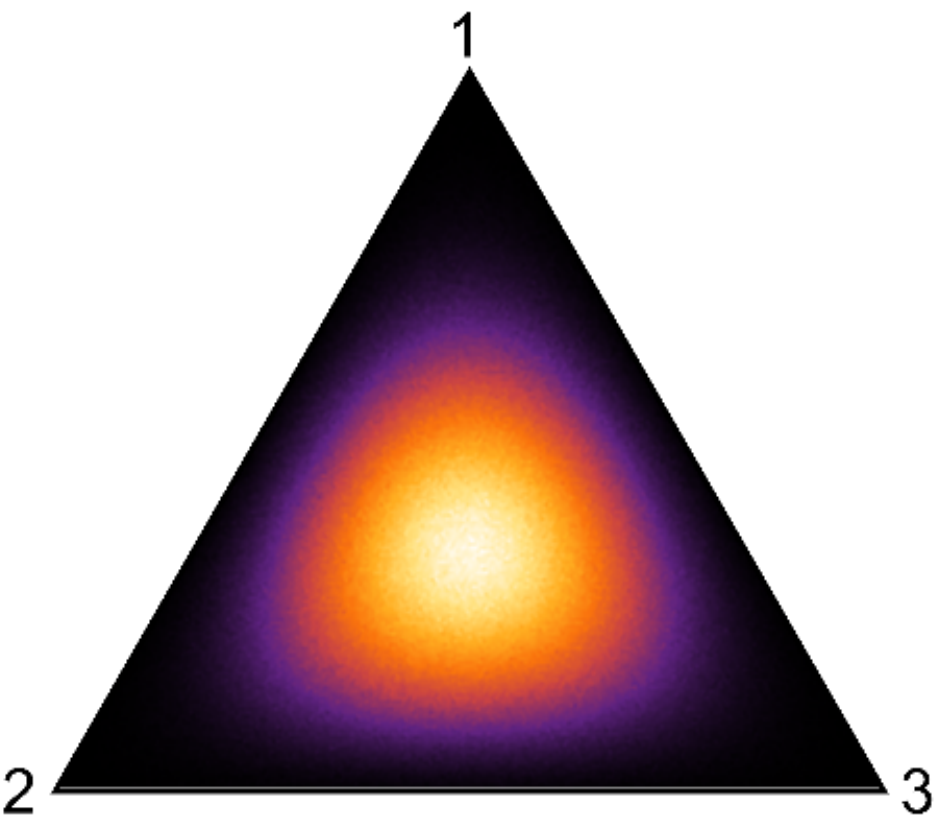}
	\caption{Simulations for 3 alleles on a ring graph, where each node is connected to the 10 nearest neighbors. The critical
		mutation probability is $u_c=1/(2n+1)$ (middle row) corresponding to $\alpha_c=1$. Top row: $u = 1/(4n-1) < u_c$ and
		$\alpha=0.5$; Bottom row: $u = 1/(n+2) > u_c$ and $\alpha=2.0$. Columns show the mean-field prediction, and the simulated stationary distributions for the Moran and voter models. Color bar is the same as in Fig.~\ref{fig:exact}.}
	\label{fig:ring5}
\end{figure}

\medskip
\noindent\textbf{Spatial structure: rings and small-world networks.}
We next probe the role of spatial structure by simulating on ring networks. Nodes are arranged uniformly on a circle, and each node is connected to the first $d/2$ neighbors on each side, so all nodes have the same degree $d$ but interactions are local. The network diameter (the largest shortest-path distance) is $(n-1)/d$ (for $n$ odd), so information propagates slowly through the network. Figure~\ref{fig:ring1} shows results for $n=201$ and $d=2$ with the same parameters as above. Here the effect of mutation is strongly enhanced: even for $u<u_c$ the stationary distribution is concentrated near the center of the simplex, indicating very high diversity. Thus the critical mutation is much lower than in fully mixed or ER populations, showing the opposite trend from scale-free networks. Note that the degree-fluctuation correction in Appendix~\ref{app:second-order-degree} does not apply here: ring neighborhoods are not exchangeable (see the discussion in Sec.~\ref{voter-random}), and the assumptions underlying the ER closure are violated, even though the degree variance is zero. Figure~\ref{fig:ring5} shows similar simulations for a ring with $d=10$. Spatial effects are weaker, but a distinct feature remains for low mutation: fixation is unlikely, yet two-allele states can still have substantial probability mass, visible as brighter regions along the simplex edges.

\begin{figure}[htb!]
    \centering
    $\Scale[1.4]{\qquad \quad p_r=0.01 \qquad \quad  p_r=0.05 \qquad   \quad p_r=0.10 \qquad p_r=0.20}$ \\
  	\vspace{0.2cm}  
    \raisebox{1.1cm}{$\Scale[1.5]{u < u_c}$} \quad
\includegraphics[width=0.17\textwidth]{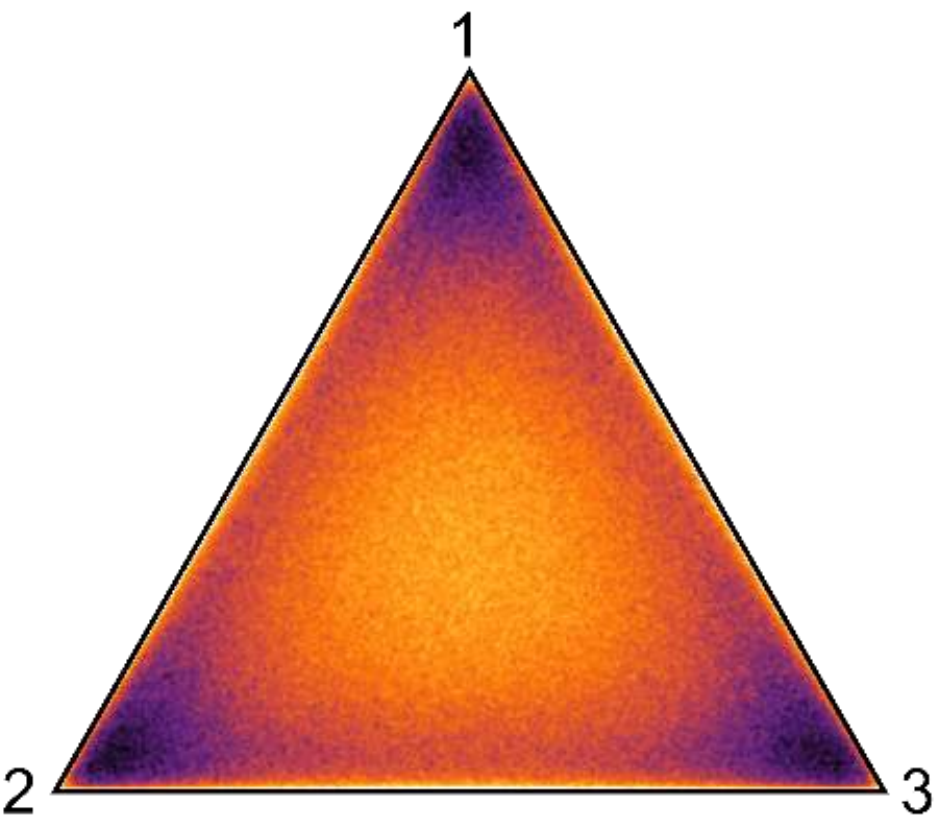} \qquad
\includegraphics[width=0.17\textwidth]{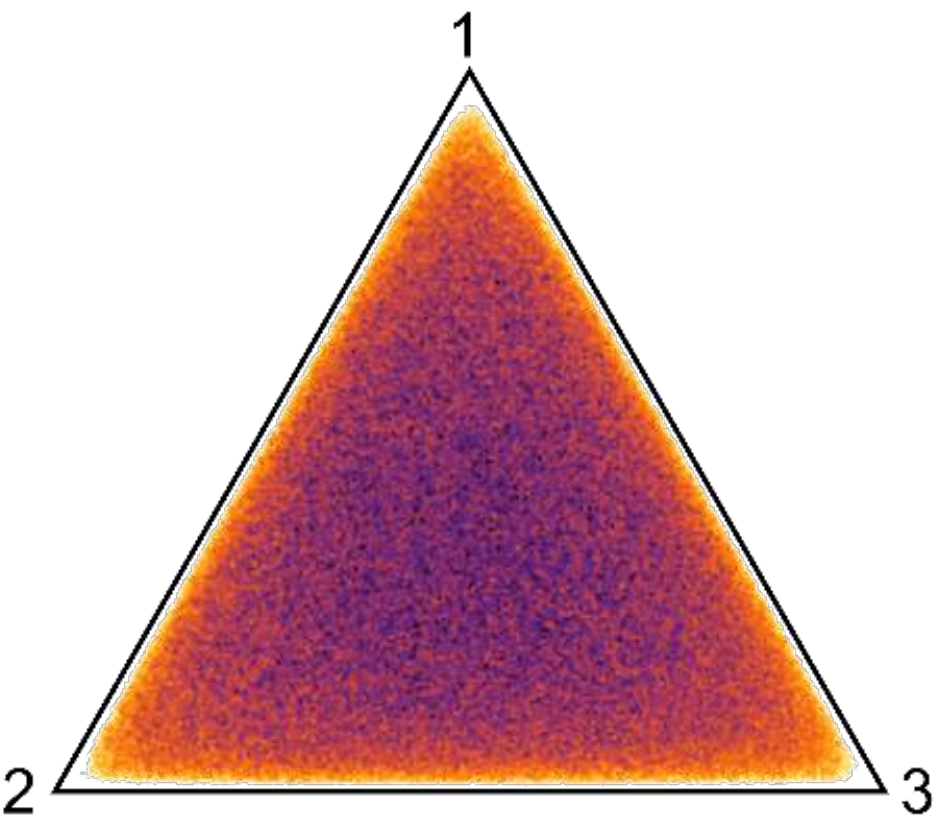}\qquad
\includegraphics[width=0.17\textwidth]{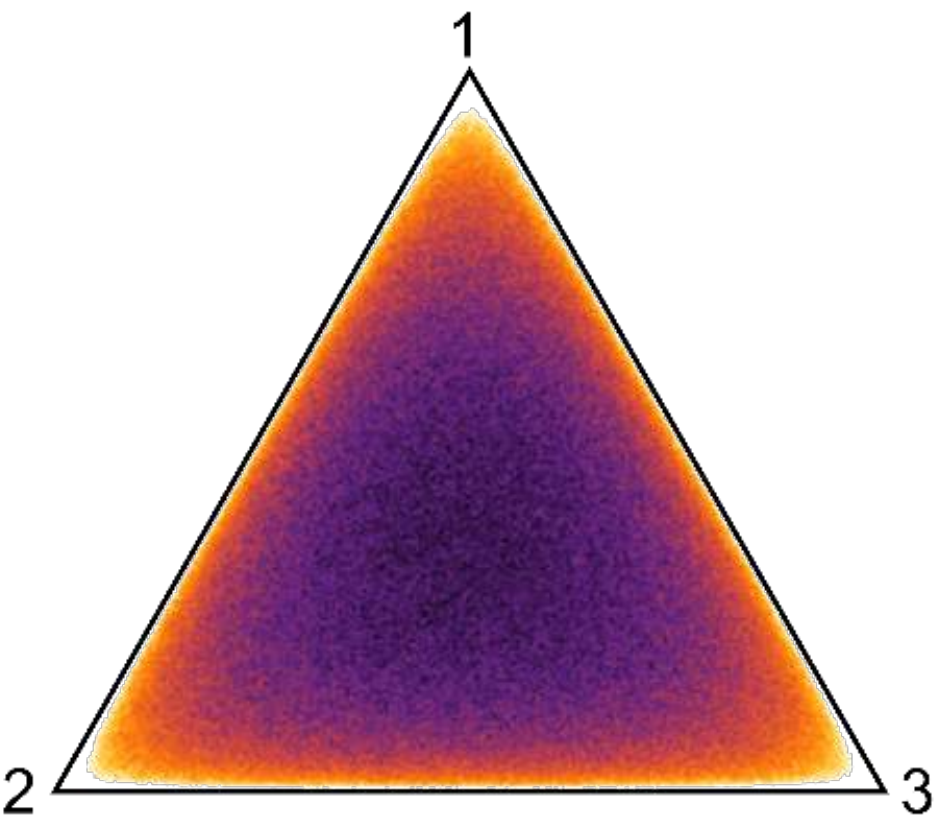} \qquad
\includegraphics[width=0.17\textwidth]{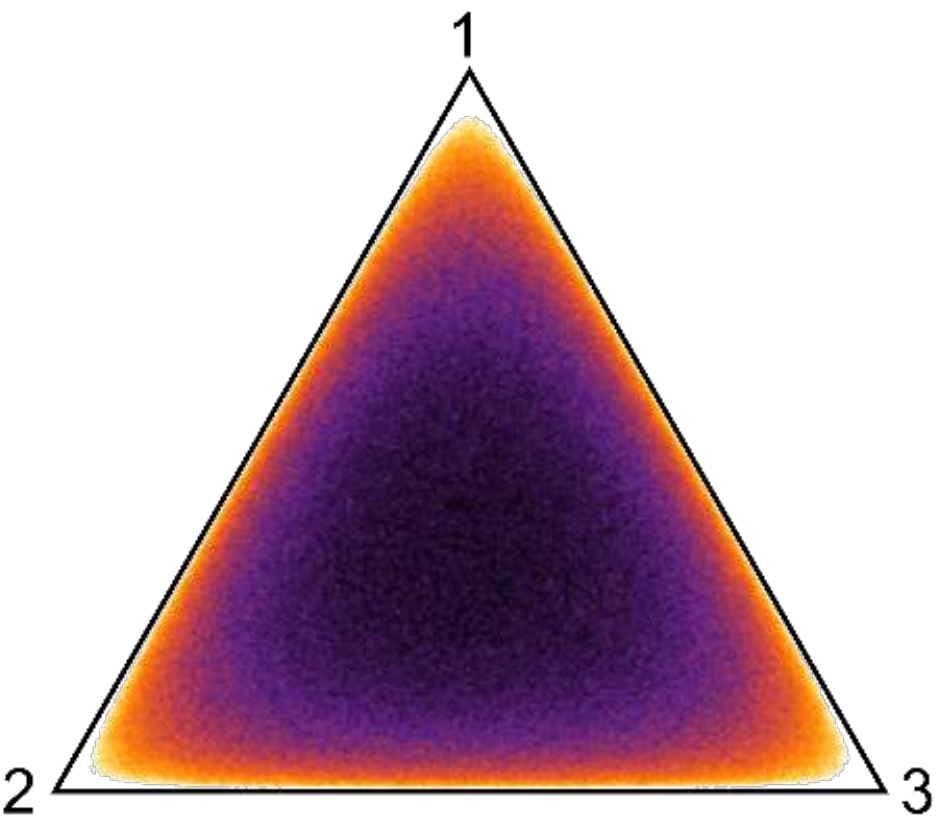} \\
    \raisebox{1.1cm}{$\Scale[1.5]{u = u_c}$} \quad
\includegraphics[width=0.17\textwidth]{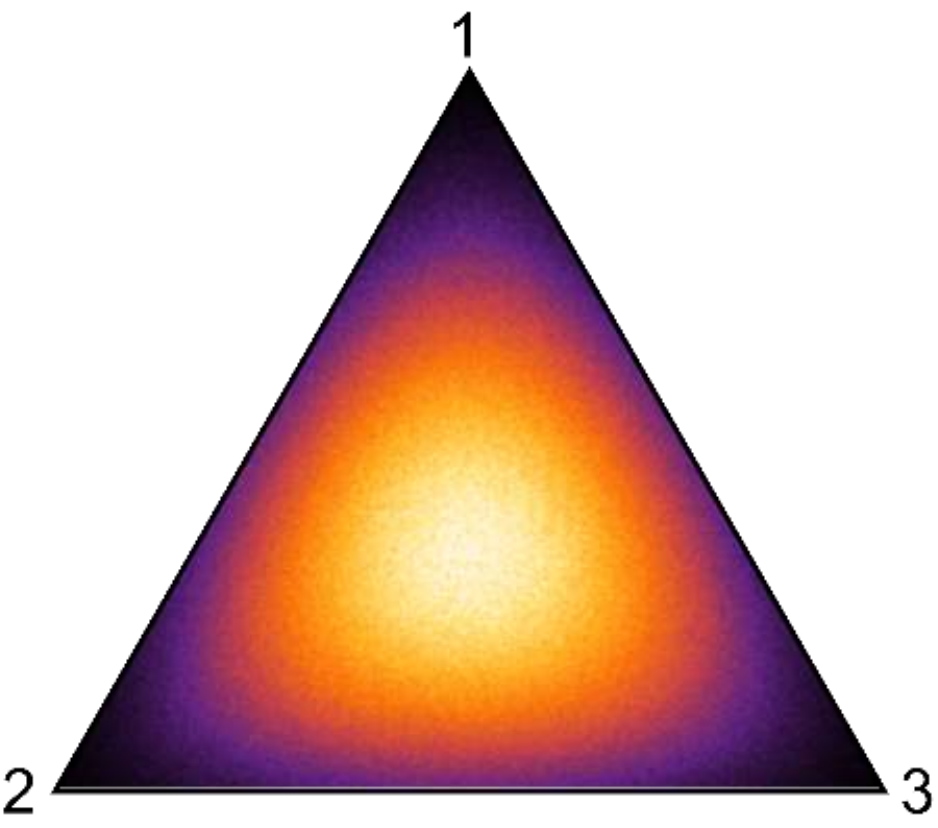} \qquad
\includegraphics[width=0.17\textwidth]{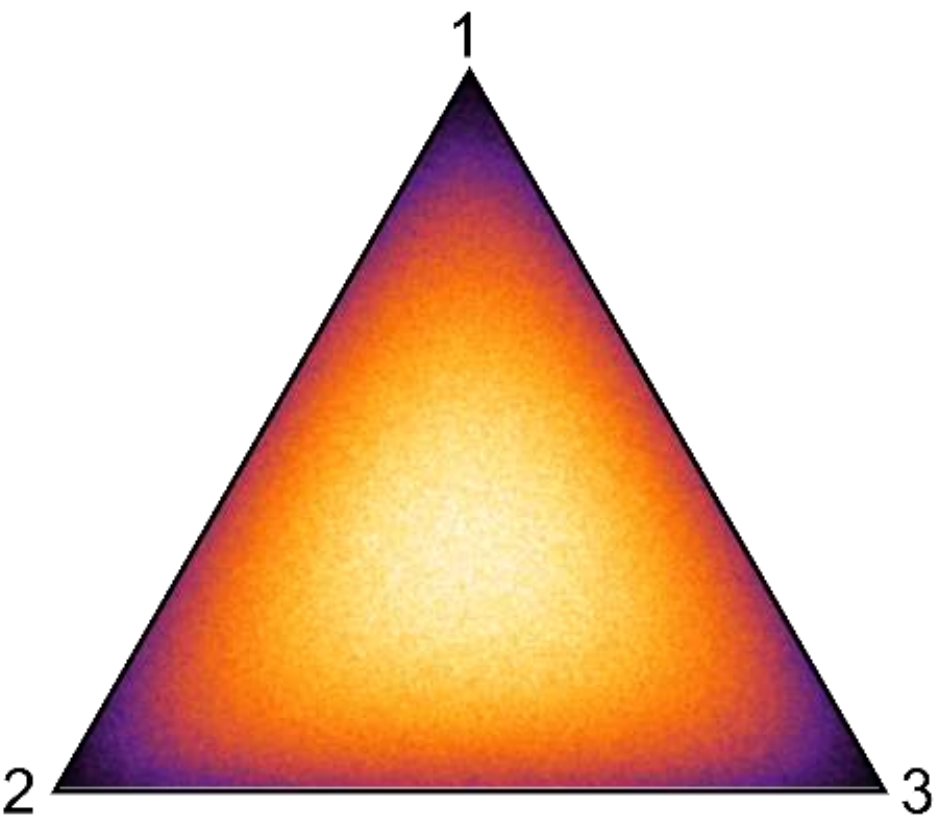}\qquad
\includegraphics[width=0.17\textwidth]{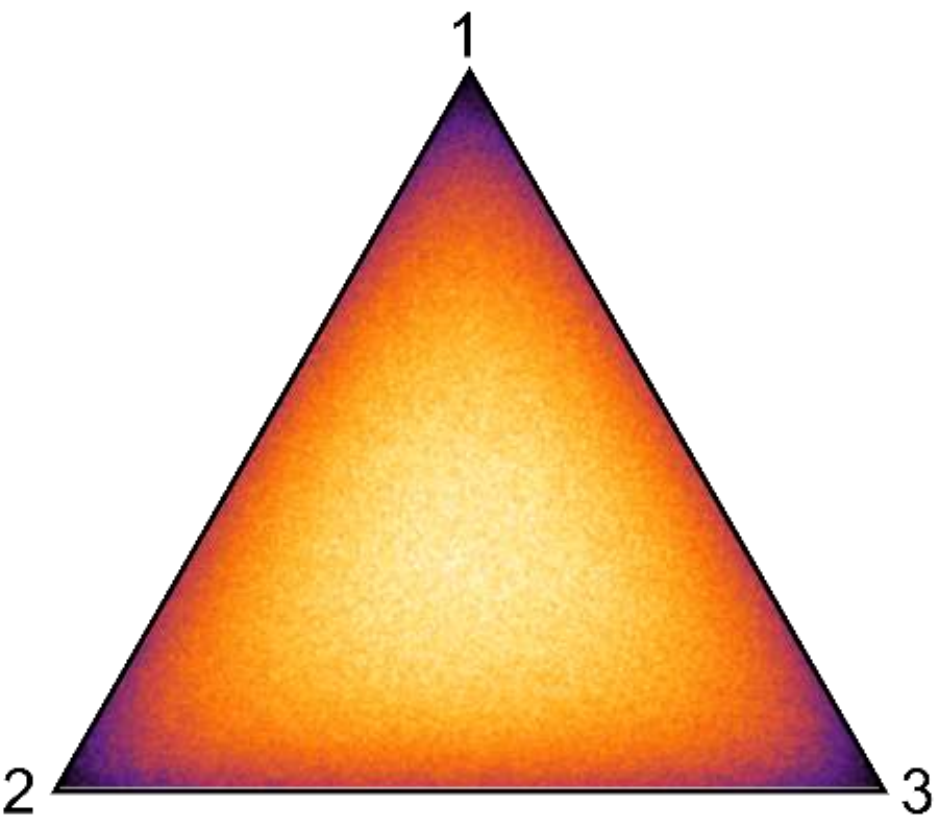} \qquad
\includegraphics[width=0.17\textwidth]{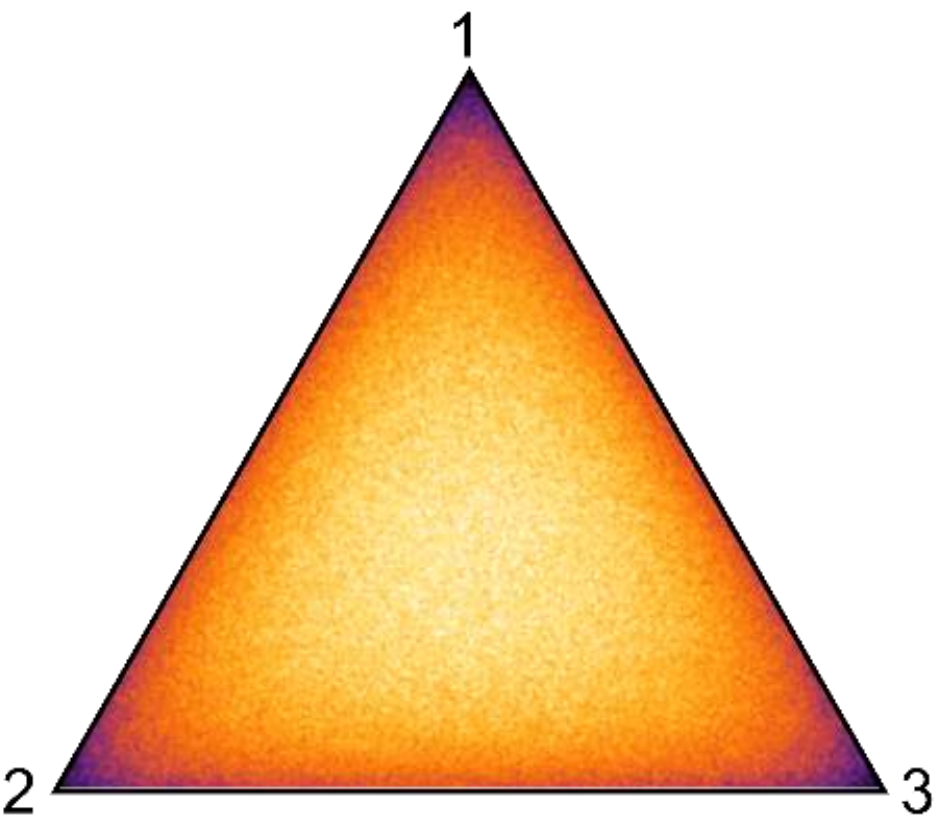} \\
    \raisebox{1.1cm}{$\Scale[1.5]{u > u_c}$} \quad
\includegraphics[width=0.17\textwidth]{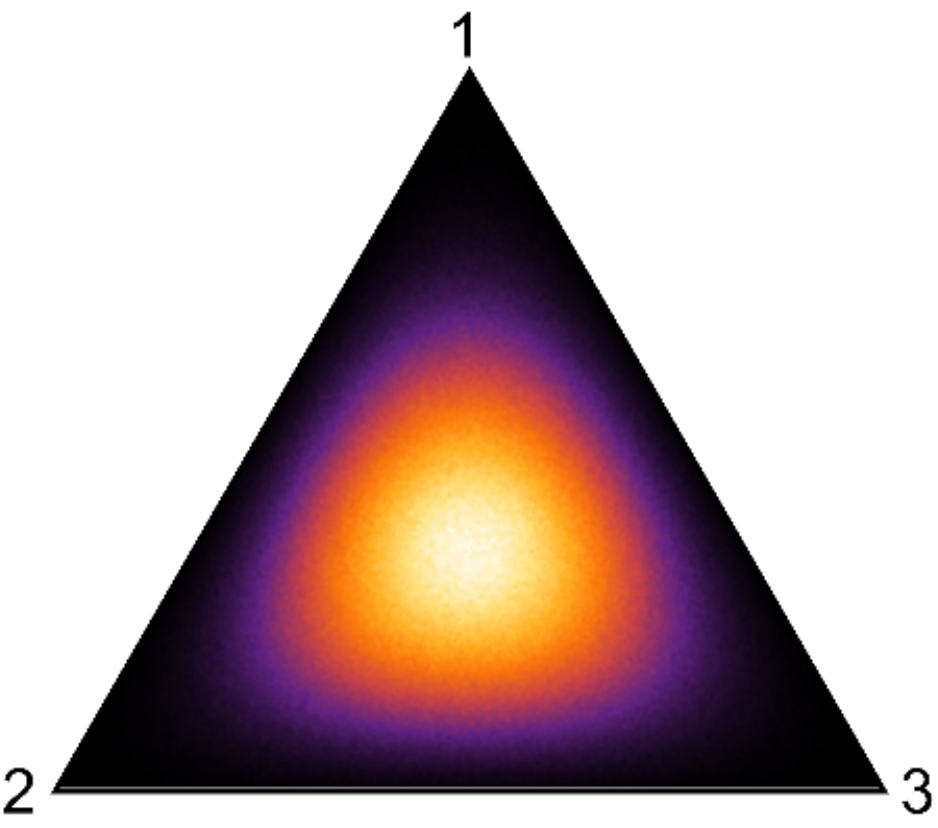} \qquad
\includegraphics[width=0.17\textwidth]{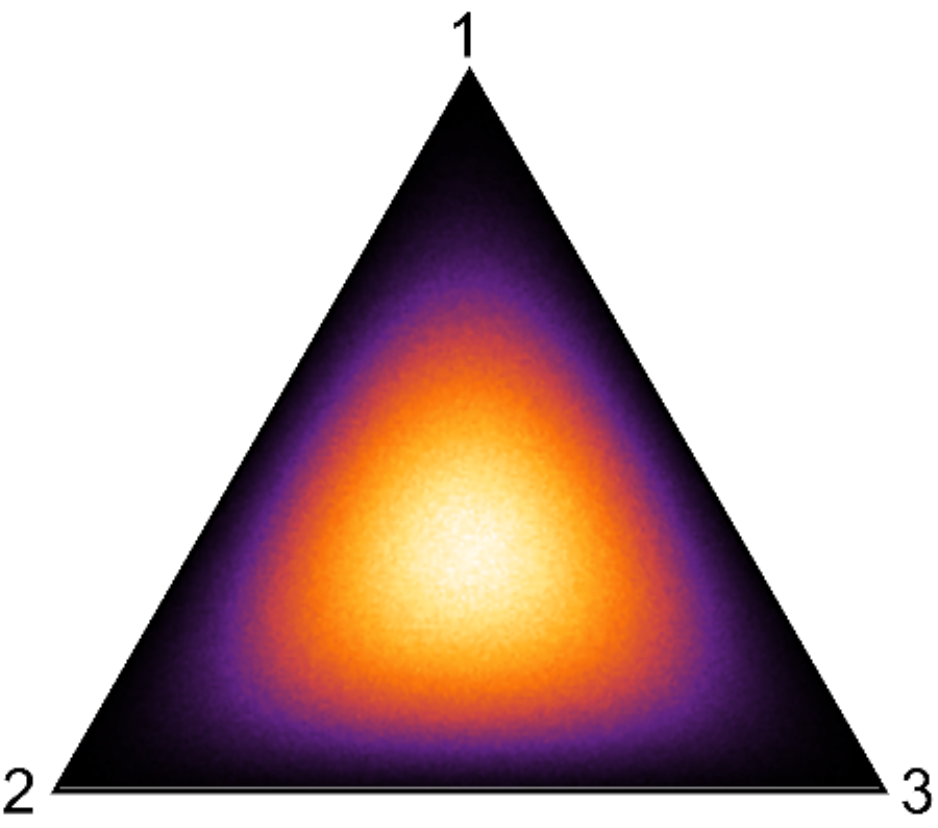}\qquad
\includegraphics[width=0.17\textwidth]{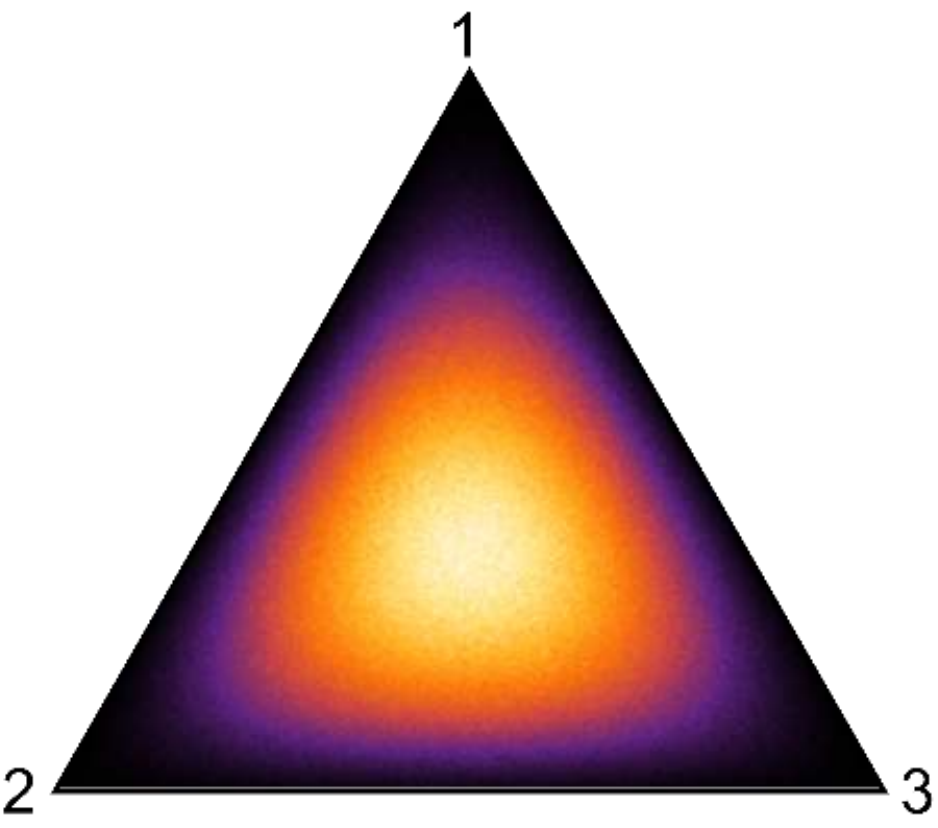} \qquad
\includegraphics[width=0.17\textwidth]{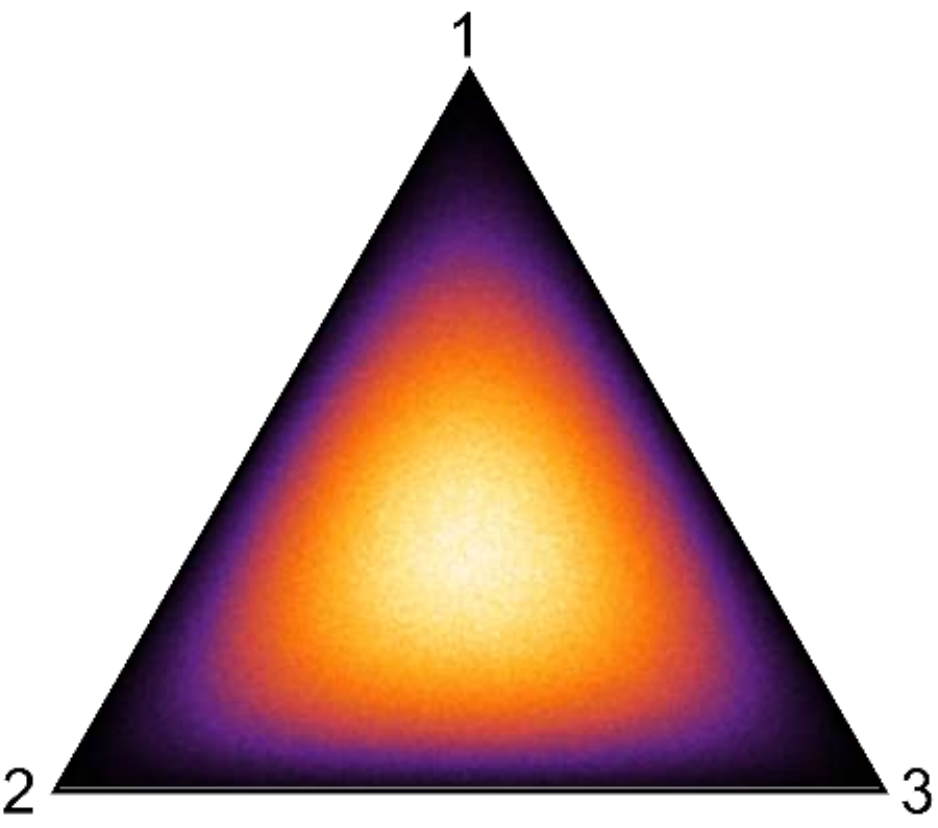}
    \caption{Simulations on small world graphs, starting from a ring with $d=10$. From left to right rewiring probabilities are 0.01, 0.05, 0.10 and 0.20.  Top: $u < u_c$ (corresponding to $\alpha=0.5$). Middle: critical mutation, corresponding to $\alpha=1$.  Bottom: $u > u_c$ (corresponding to $\alpha=2.0$). Color bar is the same as in Fig.\ref{fig:exact}.}
    \label{fig:ring5sw}
\end{figure}

Finally, Fig.~\ref{fig:ring5sw} shows a small-world interpolation \cite{watts1998collective} obtained by rewiring edges of the $d=10$ ring independently with probability $p_r$ (here $p_r=0.01,0.05,0.10,0.20$, with the same mutation rates as above). As $p_r$ increases the network becomes more random and approaches an ER-like structure, so the ER mean-field predictions become increasingly accurate. Indeed, for $p_r\gtrsim 0.1$ the distributions for $u<u_c$ and $u>u_c$ are already close to the ER mean-field baseline, whereas at $u=u_c$ substantially larger rewiring is needed to approach the expected uniform distribution; even at $p_r=0.2$ the distribution remains more concentrated near the simplex center than the uniform mean-field prediction.

\section{Discussion}
\label{discussion}

We derived the equilibrium (stationary) distribution of a neutral multi-allelic locus in a finite population. Our results assume \emph{parent-independent mutation} (PIM), meaning that the probability that an allele mutates to type~$i$ is independent of its parental type. While PIM is not the most general mutation scheme, it is the appropriate idealization for the constant mutation rates used in our simulations and is a standard approximation that yields analytically tractable and biologically interpretable models \cite{jukes1969evolution}, even though substitution rates between DNA bases are known to be asymmetric \cite{kimura1980simple}. Under PIM, we showed that the multi-allelic Moran model with mutation maps exactly to the multi-candidate voter model with zealots on fully mixed populations, and that the correspondence also holds (in a mean-field sense) for randomly connected populations.

These results are relevant to many biological and medical settings in which loci are highly polymorphic. Classic examples include the rabbit coat-color locus with four alleles \cite{snustad2015principles} and the ABO blood-group system. More broadly, medically important loci such as CFTR and the human leukocyte antigen (HLA) region exhibit extensive allelic variation. Even when selection and demographic history play important roles, the neutral stationary distribution provides a fundamental baseline: it clarifies how mutation alone partitions variation among allele types and helps separate mutational effects from selective and structural ones.

For fully mixed populations, the stationary distribution of allele counts is Dirichlet--multinomial (Eq.~\eqref{DR_voter}) with parameters
\[
\alpha_i=\frac{2(n-1)u_i}{1-\Lambda},
\qquad
\Lambda=\sum_{i=1}^m u_i.
\]
At criticality, the stationary distribution becomes uniform yielding the critical mutation rate
\[
u=u_c=\frac{1}{2n-2+m}.
\]
This explicit parametric form provides, in principle, a route to infer mutation parameters by fitting a Dirichlet--multinomial model to allele-count data under neutrality, and it furnishes a quantitative reference point for assessing departures from neutrality.

We also investigated non-fully mixed populations and considered four representative interaction structures:

\noindent\textbf{(i) Homogeneous random interactions.}
We modeled populations in which individuals interact with only $d<n$ potential partners, chosen at random, using Erd\H{o}s--R\'enyi graphs $G(n,p)$ with mean degree $d=p(n-1)$. Because these graphs are degree-homogeneous and connections are random, individuals are approximately exchangeable and the population state is still well summarized by the global allele-count vector. In mean-field, both the stationary distribution and the critical mutation rate are insensitive to~$d$ and coincide with the fully mixed case. This contrasts with the voter model, in which reducing the number of contacts amplifies the effective influence of zealots by a factor $(n-1)/d>1$. Curiously, this factor cancels in the Moran model. Intuitively, in an ER network a node's neighborhood is an (approximately) unbiased sample of the population, so the factor $d$ cancels between having $d$ potential neighbors and choosing one with probability $1/d$; mean degree affects mixing time, not the stationary distribution. These results are illustrated in Fig.~\ref{fig:er}.

\noindent\textbf{(ii) Heterogeneous interactions.}
To probe the effect of degree heterogeneity, we studied Barab\'asi--Albert scale-free networks, whose degree distribution satisfies $P(d)\sim d^{-\gamma}$ with $\gamma\approx 3$. Although our exact theory does not directly apply in this setting, simulations show that the Moran--voter correspondence remains nearly exact, while the resulting stationary distributions deviate from the fully mixed prediction. Relative to fully mixed and ER populations, degree-heterogeneous networks exhibit lower diversity at the same mutation rates and an upward shift in the apparent critical mutation threshold (Fig.~\ref{fig:sf}). Incorporating degree fluctuations as a second-order correction substantially improves agreement between theory and simulations.

\noindent\textbf{(iii) Spatially structured populations.}
To model local interactions in space, we considered ring networks in which each node has exactly $d$ neighbors (a regular, degree-homogeneous graph) but neighborhoods are spatially local rather than random. Rings have been analyzed previously in the two-allele case \cite{schneider2016mutation,franco2021shannon}. Here, degree regularity does not imply macroscopic exchangeability: the same count vector $k$ can correspond to dynamically inequivalent spatial configurations (e.g., contiguous domains versus interleaved patterns), and tracking only $k$ can miss these spatial effects. Consistent with this, our simulations show that rings substantially enhance diversity and reduce the apparent critical mutation rate (Figs.~\ref{fig:ring1} and~\ref{fig:ring5}). While a mean-field reduction is not quantitatively accurate on rings, the simulations suggest that the Moran--voter correspondence persists beyond Erd\H{o}s--R\'enyi graphs and can extend, at least qualitatively, to spatially structured populations.

\noindent\textbf{(iv) Small-world interactions.}
Finally, we examined small-world networks obtained by rewiring a fraction of edges in a ring. We found that a substantial amount of rewiring is required before the stationary distribution approaches the fully mixed prediction, including at the critical mutation rate.

Across all topologies considered, simulations indicate that the stationary distribution of alleles in the Moran model is nearly identical to the stationary distribution of votes in the corresponding voter model. This motivates the conjecture that the mapping developed here for fully mixed populations can be extended to more general graphs. At the same time, the stationary distribution on arbitrary graphs may be considerably more complex than a Dirichlet--multinomial form, and understanding when simple reduced descriptions are accurate remains an important direction for future work.

Beyond the Moran--voter correspondence, our simulations show how interaction structure shapes diversity: degree heterogeneity tends to suppress diversity, whereas spatial locality (here, rings) promotes it, consistent with prior observations \cite{schneider2016mutation}. These effects may be relevant for biological networks with heterogeneous connectivity, such as protein--protein interaction networks \cite{wuchty2001scale,nacher2009emergence}, even though the extent to which such networks are scale-free is debated \cite{de2005spectral,khanin2006scale,broido2019scale}. Conversely, spatially extended populations can sustain higher diversity through isolation by distance and can facilitate diversification and speciation \cite{de_aguiar_global_2009}.

More broadly, our work provides an exact, multi-allelic stationary distribution under neutrality together with a precise bridge between two classical stochastic processes, the Moran and voter models. This connection enables tools and intuition to move between population genetics, statistical physics, and network science, and it offers a principled baseline for interpreting allele-count data in structured populations. By clarifying which features of the stationary distribution are universal (robust to homogeneous random connectivity) and which are topology-dependent (sensitive to heterogeneity and spatial locality), the results help sharpen quantitative expectations for genetic diversity across realistic interaction structures.

\appendix
\section{Second-order correction for degree fluctuations in the ER mean-field reduction}
\label{app:second-order-degree}

This appendix develops a simple second-order correction to the ER mean-field reduction used in Sec.~\ref{voter-random}. The aim is to quantify how random degree fluctuations affect the
neighbor-selection probability
\[
\frac{N_{j,i}+\alpha^{\mathrm{ER}}_i}{d_j+\alpha^{\mathrm{ER}}_0},
\]
where $d_j$ is the free-degree of the updated node $j$, $N_{j,i}$ is the number of its free neighbors in
state $i$, and $\alpha^{\mathrm{ER}}_i$ and $\alpha^{\mathrm{ER}}_0:=\sum_{q=1}^m\alpha^{\mathrm{ER}}_q$ are the zealot
counts (recall that every free node is connected to all zealots). The second-order degree-fluctuation correction is not specific
to ER graphs: it applies to any network ensemble for which (i) neighbor states are approximately well mixed,
and (ii) the focal degree has finite mean and variance.

\subsection{Setup and an exact conditional reduction}
Fix a macroscopic free-voter configuration $k=(k_1,\dots,k_m)$ with $\sum_{i=1}^m k_i=n$.
Let $d_j$ denote the (random) free-degree of the updated node, and let $N_{j,i}$ denote the number of its free
neighbors in state $i$. For ER graphs, the random-neighborhood assumption asserts that conditional on $d_j$,
the $d_j$ free neighbors are approximately a uniform sample from the other $n-1$ free nodes. In particular,
excluding the updated node,
\begin{equation}
\mathbb{E}[N_{j,i}\mid k,d_j]=d_j\,p_i,
\qquad
p_i:=\frac{k_i}{n-1}.
\label{app:eq:ENgivenDj}
\end{equation}

A useful simplification is that the expectation of the zealot-augmented ratio admits an exact reduction under
\eqref{app:eq:ENgivenDj}. By the law of total expectation,
\begin{align}
\mathbb{E}\!\left[\frac{N_{j,i}+\alpha^{\mathrm{ER}}_i}{d_j+\alpha^{\mathrm{ER}}_0}\,\Big|\,k\right]
&=
\mathbb{E}\!\left[
\mathbb{E}\!\left[\frac{N_{j,i}+\alpha^{\mathrm{ER}}_i}{d_j+\alpha^{\mathrm{ER}}_0}\,\Big|\,k,d_j\right]
\,\Big|\,k\right] \nonumber\\
&=
\mathbb{E}\!\left[\frac{\mathbb{E}[N_{j,i}\mid k,d_j]+\alpha^{\mathrm{ER}}_i}{d_j+\alpha^{\mathrm{ER}}_0}\,\Big|\,k\right]
=
\mathbb{E}\!\left[\frac{d_j\,p_i+\alpha^{\mathrm{ER}}_i}{d_j+\alpha^{\mathrm{ER}}_0}\,\Big|\,k\right].
\label{app:eq:cond-reduction-1}
\end{align}
Writing $d_jp_i+\alpha^{\mathrm{ER}}_i=p_i(d_j+\alpha^{\mathrm{ER}}_0)+(\alpha^{\mathrm{ER}}_i-p_i\alpha^{\mathrm{ER}}_0)$
yields the identity
\begin{equation}
\mathbb{E}\!\left[\frac{N_{j,i}+\alpha^{\mathrm{ER}}_i}{d_j+\alpha^{\mathrm{ER}}_0}\,\Big|\,k\right]
=
p_i+(\alpha^{\mathrm{ER}}_i-p_i\alpha^{\mathrm{ER}}_0)\,
\mathbb{E}\!\left[\frac{1}{d_j+\alpha^{\mathrm{ER}}_0}\,\Big|\,k\right].
\label{app:eq:exact-reduction}
\end{equation}
Thus, beyond the baseline random-neighborhood closure \eqref{app:eq:ENgivenDj}, all remaining dependence on
degree heterogeneity enters through the single scalar
$\mathbb{E}\!\left[(d_j+\alpha^{\mathrm{ER}}_0)^{-1}\mid k\right]$.

\subsection{Second-order delta-method approximation}
Let $\mu:=\mathbb{E}[d_j]$ and $\sigma_d^2:=\operatorname{Var}(d_j)$. In the ER model with fixed $p$,
$\mu=d=p(n-1)$ and $\sigma_d^2=(n-1)p(1-p)$, but we keep the notation $(\mu,\sigma_d^2)$ for clarity. Consider
\[
g(x):=\frac{1}{x+\alpha^{\mathrm{ER}}_0}.
\]
A second-order Taylor expansion of $g(d_j)$ about $\mu$ gives
\[
g(d_j)\approx g(\mu)+g'(\mu)(d_j-\mu)+\frac{1}{2}g''(\mu)(d_j-\mu)^2,
\]
and taking expectations yields the standard second-order delta-method approximation
\begin{equation}
\mathbb{E}\!\left[\frac{1}{d_j+\alpha^{\mathrm{ER}}_0}\right]
\approx
\frac{1}{\mu+\alpha^{\mathrm{ER}}_0}+\frac{\sigma_d^2}{(\mu+\alpha^{\mathrm{ER}}_0)^3},
\label{app:eq:delta}
\end{equation}
since $g'(\mu)=-1/(\mu+\alpha^{\mathrm{ER}}_0)^2$ and $g''(\mu)=2/(\mu+\alpha^{\mathrm{ER}}_0)^3$.
Substituting \eqref{app:eq:delta} into \eqref{app:eq:exact-reduction} and setting $\mu=d$ gives the explicit
second-order approximation
\begin{equation}
\mathbb{E}\!\left[\frac{N_{j,i}+\alpha^{\mathrm{ER}}_i}{d_j+\alpha^{\mathrm{ER}}_0}\,\Big|\,k\right]
\approx
\frac{\tfrac{d}{n-1}k_i+\alpha^{\mathrm{ER}}_i}{d+\alpha^{\mathrm{ER}}_0}
+
\left(\alpha^{\mathrm{ER}}_i-\frac{k_i}{n-1}\alpha^{\mathrm{ER}}_0\right)\frac{\sigma_d^2}{(d+\alpha^{\mathrm{ER}}_0)^3}.
\label{app:eq:second-order-prob}
\end{equation}
The correction term scales as $\sigma_d^2/(d+\alpha^{\mathrm{ER}}_0)^3$ and disappears when the degree is deterministic, i.e., when $\sigma_d^2=0$.

\subsection{Second-order corrected copy-step kernel}
In Section~\ref{voter-random}, the copy step updates $k$ by $k\to k+e_i-e_j$ when a free $j$-voter is selected
and then copies state $i$. Using \eqref{app:eq:second-order-prob}, a degree-fluctuation corrected kernel is
\begin{equation}
Q^{\mathrm{ER},(2)}\big(k\to k+e_i-e_j\big)
\approx
\frac{k_j}{n}\left[
\frac{\tfrac{d}{n-1}k_i+\alpha^{\mathrm{ER}}_i}{d+\alpha^{\mathrm{ER}}_0}
+
\left(\alpha^{\mathrm{ER}}_i-\frac{k_i}{n-1}\alpha^{\mathrm{ER}}_0\right)\frac{\sigma_d^2}{(d+\alpha^{\mathrm{ER}}_0)^3}
\right].
\label{app:eq:Q-second-order}
\end{equation}
This expression makes explicit how ER degree heterogeneity perturbs the mean-field transition probabilities.
In the dense ER regime with fixed $p$ and $n\to\infty$, one has $d=p(n-1)\to\infty$ and
$\sqrt{\sigma_d^2}/d=O(d^{-1/2})$, so the correction term in \eqref{app:eq:Q-second-order} becomes small.

\subsection{Reparametrization via second-order effective zealot counts}
The right-hand side of \eqref{app:eq:Q-second-order} is affine in $k_i$, so it can be re-expressed in voter form
with corrected effective counts, up to second order. Specifically, one may write
\begin{equation}
Q^{\mathrm{ER},(2)}\big(k\to k+e_i-e_j\big)\approx
\frac{k_j}{n}\cdot\frac{k_i+\alpha_i^{\mathrm{eff},(2)}}{(n-1)+\alpha_0^{\mathrm{eff},(2)}},
\label{app:eq:Q-second-order-effective}
\end{equation}
where $\alpha_0^{\mathrm{eff},(2)}:=\sum_{i=1}^m \alpha_i^{\mathrm{eff},(2)}$. Equating the right-hand sides of
\eqref{app:eq:Q-second-order} and \eqref{app:eq:Q-second-order-effective} and solving for
$\alpha_i^{\mathrm{eff},(2)}$ yields the second-order effective zealot counts
\begin{equation}
\alpha_i^{\mathrm{eff},(2)}
=
(n-1)\alpha^{\mathrm{ER}}_i\,
\frac{(d+\alpha^{\mathrm{ER}}_0)^2+\sigma_d^2}{d(d+\alpha^{\mathrm{ER}}_0)^2-\alpha^{\mathrm{ER}}_0\sigma_d^2},
\qquad
\alpha_0^{\mathrm{eff},(2)}:=\sum_{i=1}^m \alpha_i^{\mathrm{eff},(2)}.
\label{app:eq:alpha-eff-2}
\end{equation}
When $\sigma_d^2=0$, \eqref{app:eq:alpha-eff-2} reduces to the mean-field effective counts
$\alpha_i^{\mathrm{eff}}=\frac{n-1}{d}\alpha^{\mathrm{ER}}_i$ from Eq.~(\ref{voterer2}).

\subsection{Remarks on isolates}
On ER graphs, $d_j$ can be zero. The ratio $(N_{j,i}+\alpha^{\mathrm{ER}}_i)/(d_j+\alpha^{\mathrm{ER}}_0)$ remains
well-defined because $\alpha^{\mathrm{ER}}_0>0$, so zealot links regularize the denominator in the global-zealot
setting considered in Section~\ref{voter-random}.

\bigskip

\noindent Acknowledgments: 

\noindent M.A.M.A. would like to thank FAPESP, grant 2021/14335-0, and CNPq, grant 303814/2023-3 for financial support.

\bigskip

\noindent Data availability statement:

\noindent The authors declare that the data supporting the findings of this study are available within the paper.

\clearpage

\begin{thebibliography}{38}
	\providecommand{\natexlab}[1]{#1}
	\providecommand{\url}[1]{\texttt{#1}}
	\expandafter\ifx\csname urlstyle\endcsname\relax
	\providecommand{\doi}[1]{doi: #1}\else
	\providecommand{\doi}{doi: \begingroup \urlstyle{rm}\Url}\fi
	
	\bibitem[Moran(1958)]{moran1958random}
	Patrick Alfred~Pierce Moran.
	\newblock Random processes in genetics.
	\newblock In \emph{Mathematical proceedings of the cambridge philosophical
		society}, volume~54, pages 60--71. Cambridge University Press, 1958.
	
	\bibitem[Gillespie(2004)]{gillespie_population_2004}
	John Gillespie.
	\newblock \emph{Population {Genetics}: {A} {Concise} {Guide}}.
	\newblock Johns Hopkins University Press, Baltimore, Md, 2nd edition edition,
	July 2004.
	\newblock ISBN 978-0-8018-8009-4.
	
	\bibitem[Nowak(2006)]{nowak_evolutionary_2006}
	Martin~A. Nowak.
	\newblock \emph{Evolutionary {Dynamics}: {Exploring} the {Equations} of
		{Life}}.
	\newblock Belknap Press, Cambridge, Mass, first edition edition edition,
	September 2006.
	\newblock ISBN 978-0-674-02338-3.
	
	\bibitem[Ewens(2004)]{ewens_mathematical_2004}
	Warren~J. Ewens.
	\newblock \emph{Mathematical {Population} {Genetics} 1: {Theoretical}
		{Introduction}}.
	\newblock Springer, New York, 2nd edition edition, January 2004.
	\newblock ISBN 978-0-387-20191-7.
	
	\bibitem[Burden(2018)]{burden2018population}
	Conrad~J Burden.
	\newblock Population genetics.
	\newblock In \emph{Encyclopedia of Bioinformatics and Computational Biology:
		ABC of Bioinformatics}, pages 759--788. Elsevier, 2018.
	
	\bibitem[Crow and Kimura(1970)]{crow_introduction_1970}
	James~F. Crow and Motoo Kimura.
	\newblock \emph{An {Introduction} to {Population} {Genetics} {Theory}}.
	\newblock The Blackburn Press, 1970.
	\newblock ISBN 1-932846-12-3.
	
	\bibitem[Brennan and Schrijver(2016)]{brennan2016cystic}
	Marie-Luise Brennan and Iris Schrijver.
	\newblock Cystic fibrosis: a review of associated phenotypes, use of molecular
	diagnostic approaches, genetic characteristics, progress, and dilemmas.
	\newblock \emph{The Journal of Molecular Diagnostics}, 18\penalty0
	(1):\penalty0 3--14, 2016.
	
	\bibitem[Snustad and Simmons(2015)]{snustad2015principles}
	D~Peter Snustad and Michael~J Simmons.
	\newblock \emph{Principles of genetics}.
	\newblock John Wiley \& Sons, 2015.
	
	\bibitem[Choo(2007)]{choo2007hla}
	Sung~Yoon Choo.
	\newblock The hla system: genetics, immunology, clinical testing, and clinical
	implications.
	\newblock \emph{Yonsei medical journal}, 48\penalty0 (1):\penalty0 11, 2007.
	
	\bibitem[Watterson(1961)]{watterson_markov_1961}
	G.~A. Watterson.
	\newblock Markov {Chains} with {Absorbing} {States}: {A} {Genetic} {Example}.
	\newblock \emph{The Annals of Mathematical Statistics}, 32\penalty0
	(3):\penalty0 716--729, September 1961.
	\newblock ISSN 0003-4851, 2168-8990.
	\newblock \doi{10.1214/aoms/1177704967}.
	\newblock URL \url{http://projecteuclid.org/euclid.aoms/1177704967}.
	
	\bibitem[Cannings(1974)]{cannings_latent_1974}
	C.~Cannings.
	\newblock The {Latent} {Roots} of {Certain} {Markov} {Chains} {Arising} in
	{Genetics}: {A} {New} {Approach}, {I}. {Haploid} {Models}.
	\newblock \emph{Advances in Applied Probability}, 6\penalty0 (2):\penalty0
	260--290, June 1974.
	\newblock ISSN 0001-8678.
	\newblock \doi{10.2307/1426293}.
	\newblock URL \url{http://www.jstor.org/stable/1426293}.
	
	\bibitem[Gladstien(1978)]{gladstien_characteristic_1978}
	K.~Gladstien.
	\newblock The {Characteristic} {Values} and {Vectors} for a {Class} of
	{Stochastic} {Matrices} {Arising} in {Genetics}.
	\newblock \emph{SIAM Journal on Applied Mathematics}, 34\penalty0 (4):\penalty0
	630--642, June 1978.
	\newblock ISSN 0036-1399.
	\newblock \doi{10.1137/0134050}.
	\newblock URL \url{http://epubs.siam.org/doi/abs/10.1137/0134050}.
	
	\bibitem[Kirman(1993)]{kirman_ants_1993}
	Alan Kirman.
	\newblock Ants, {Rationality}, and {Recruitment}.
	\newblock \emph{The Quarterly Journal of Economics}, 108\penalty0 (1):\penalty0
	137--156, February 1993.
	\newblock ISSN 0033-5533, 1531-4650.
	\newblock \doi{10.2307/2118498}.
	\newblock URL \url{http://qje.oxfordjournals.org/content/108/1/137}.
	
	\bibitem[Durrett and Moseley(2015)]{durrett_spatial_2015}
	Richard Durrett and Stephen Moseley.
	\newblock Spatial {Moran} models {I}. {Stochastic} tunneling in the neutral
	case.
	\newblock \emph{The Annals of Applied Probability}, 25\penalty0 (1):\penalty0
	104--115, February 2015.
	\newblock ISSN 1050-5164, 2168-8737.
	\newblock \doi{10.1214/13-AAP989}.
	\newblock URL \url{http://projecteuclid.org/euclid.aoap/1418740180}.
	
	\bibitem[D{\'e}barre et~al.(2014)D{\'e}barre, Hauert, and
	Doebeli]{debarre_social_2014}
	F.~D{\'e}barre, C.~Hauert, and M.~Doebeli.
	\newblock Social evolution in structured populations.
	\newblock \emph{Nature Communications}, 5, March 2014.
	\newblock \doi{10.1038/ncomms4409}.
	\newblock URL
	\url{http://www.nature.com/ncomms/2014/140306/ncomms4409/full/ncomms4409.html}.
	
	\bibitem[Harmon et~al.(2015)Harmon, Lagi, de~Aguiar, Chinellato, Braha,
	Epstein, and Bar-Yam]{harmon_anticipating_2015}
	Dion Harmon, Marco Lagi, Marcus A.~M. de~Aguiar, David~D. Chinellato, Dan
	Braha, Irving~R. Epstein, and Yaneer Bar-Yam.
	\newblock Anticipating {Economic} {Market} {Crises} {Using} {Measures} of
	{Collective} {Panic}.
	\newblock \emph{PLoS ONE}, 10\penalty0 (7):\penalty0 e0131871, July 2015.
	\newblock \doi{10.1371/journal.pone.0131871}.
	\newblock URL \url{http://dx.doi.org/10.1371/journal.pone.0131871}.
	
	\bibitem[Braha and De~Aguiar(2017)]{voting2017behavior}
	Dan Braha and Marcus A.~M. De~Aguiar.
	\newblock Voting contagion: Modeling and analysis of a century of us
	presidential elections.
	\newblock \emph{PloS one}, 12\penalty0 (5):\penalty0 e0177970, 2017.
	
	\bibitem[de~Aguiar and Bar-Yam(2011)]{de_aguiar_moran_2011}
	Marcus A.~M. de~Aguiar and Yaneer Bar-Yam.
	\newblock Moran model as a dynamical process on networks and its implications
	for neutral speciation.
	\newblock \emph{Physical Review E}, 84\penalty0 (3):\penalty0 031901, 2011.
	\newblock \doi{10.1103/PhysRevE.84.031901}.
	\newblock URL \url{http://link.aps.org/doi/10.1103/PhysRevE.84.031901}.
	
	\bibitem[Ramirez et~al.(2024)Ramirez, Vazquez, San~Miguel, and
	Galla]{ramirez2024ordering}
	Luc{\'\i}a~S Ramirez, Federico Vazquez, Maxi San~Miguel, and Tobias Galla.
	\newblock Ordering dynamics of nonlinear voter models.
	\newblock \emph{Physical Review E}, 109\penalty0 (3):\penalty0 034307, 2024.
	
	\bibitem[Jager and Amblard(2005)]{jager2005uniformity}
	Wander Jager and Fr{\'e}d{\'e}ric Amblard.
	\newblock Uniformity, bipolarization and pluriformity captured as generic
	stylized behavior with an agent-based simulation model of attitude change.
	\newblock \emph{Computational \& Mathematical Organization Theory}, 10\penalty0
	(4):\penalty0 295--303, 2005.
	
	\bibitem[Martins et~al.(2010)Martins, Pineda, and Toral]{martins2010mass}
	T~Vaz Martins, Miguel Pineda, and Raul Toral.
	\newblock Mass media and repulsive interactions in continuous-opinion dynamics.
	\newblock \emph{Europhysics Letters}, 91\penalty0 (4):\penalty0 48003, 2010.
	
	\bibitem[Mobilia et~al.(2007)Mobilia, Petersen, and Redner]{mobilia_role_2007}
	M~Mobilia, A~Petersen, and S~Redner.
	\newblock On the role of zealotry in the voter model.
	\newblock \emph{Journal of Statistical Mechanics: Theory and Experiment},
	2007\penalty0 (08):\penalty0 P08029--P08029, August 2007.
	\newblock ISSN 1742-5468.
	\newblock \doi{10.1088/1742-5468/2007/08/P08029}.
	\newblock URL
	\url{http://stacks.iop.org/1742-5468/2007/i=08/a=P08029?key=crossref.06d7f2bdcfcfb58cd82c138aa65a0871}.
	
	\bibitem[Chinellato et~al.(2015)Chinellato, Epstein, Braha, Bar-Yam, and
	Aguiar]{chinellato_dynamical_2015}
	David~D. Chinellato, Irving~R. Epstein, Dan Braha, Yaneer Bar-Yam, and Marcus
	A. M.~de Aguiar.
	\newblock Dynamical {Response} of {Networks} {Under} {External}
	{Perturbations}: {Exact} {Results}.
	\newblock \emph{Journal of Statistical Physics}, 159\penalty0 (2):\penalty0
	221--230, January 2015.
	\newblock ISSN 0022-4715, 1572-9613.
	\newblock \doi{10.1007/s10955-015-1189-x}.
	\newblock URL \url{http://link.springer.com/article/10.1007/s10955-015-1189-x}.
	
	\bibitem[Schneider et~al.(2016)Schneider, Martins, and
	de~Aguiar]{schneider2016mutation}
	David~M Schneider, Ayana~B Martins, and Marcus~AM de~Aguiar.
	\newblock The mutation--drift balance in spatially structured populations.
	\newblock \emph{Journal of theoretical biology}, 402:\penalty0 9--17, 2016.
	
	\bibitem[Franco et~al.(2021)Franco, Marquitti, Fernandes, Braha, and
	De~Aguiar]{franco2021shannon}
	Gabriella~Dantas Franco, Flavia Maria~Darcie Marquitti, Lucas~D Fernandes, Dan
	Braha, and Marcus A.~M. De~Aguiar.
	\newblock Shannon information criterion for low-high diversity transition in
	moran and voter models.
	\newblock \emph{Physical Review E}, 104\penalty0 (2):\penalty0 024315, 2021.
	
	\bibitem[Liggett(2012)]{liggett_interacting_2012}
	Thomas Liggett.
	\newblock \emph{Interacting {Particle} {Systems}}.
	\newblock Springer Science \& Business Media, December 2012.
	\newblock ISBN 978-1-4613-8542-4.
	
	\bibitem[Erd\H{o}s and R\'enyi(1959)]{erdds1959random}
	P~Erd\H{o}s and A~R\'enyi.
	\newblock On random graphs i.
	\newblock \emph{Publ. math. debrecen}, 6\penalty0 (290-297):\penalty0 18, 1959.
	
	\bibitem[Newman(2018)]{newman2018networks}
	Mark Newman.
	\newblock \emph{Networks}.
	\newblock Oxford university press, 2018.
	
	\bibitem[Albert and Barab{\'a}si(2002)]{albert2002statistical}
	R{\'e}ka Albert and Albert-L{\'a}szl{\'o} Barab{\'a}si.
	\newblock Statistical mechanics of complex networks.
	\newblock \emph{Reviews of modern physics}, 74\penalty0 (1):\penalty0 47, 2002.
	
	\bibitem[Watts and Strogatz(1998)]{watts1998collective}
	Duncan~J Watts and Steven~H Strogatz.
	\newblock Collective dynamics of ‘small-world’networks.
	\newblock \emph{nature}, 393\penalty0 (6684):\penalty0 440--442, 1998.
	
	\bibitem[Jukes et~al.(1969)Jukes, Cantor, et~al.]{jukes1969evolution}
	Thomas~H Jukes, Charles~R Cantor, et~al.
	\newblock Evolution of protein molecules.
	\newblock \emph{Mammalian protein metabolism}, 3\penalty0 (21):\penalty0 132,
	1969.
	
	\bibitem[Kimura(1980)]{kimura1980simple}
	Motoo Kimura.
	\newblock A simple method for estimating evolutionary rates of base
	substitutions through comparative studies of nucleotide sequences.
	\newblock \emph{Journal of molecular evolution}, 16\penalty0 (2):\penalty0
	111--120, 1980.
	
	\bibitem[Wuchty(2001)]{wuchty2001scale}
	Stefan Wuchty.
	\newblock Scale-free behavior in protein domain networks.
	\newblock \emph{Molecular biology and evolution}, 18\penalty0 (9):\penalty0
	1694--1702, 2001.
	
	\bibitem[Nacher et~al.(2009)Nacher, Hayashida, and Akutsu]{nacher2009emergence}
	Jose~C Nacher, Morihiro Hayashida, and Tatsuya Akutsu.
	\newblock Emergence of scale-free distribution in protein--protein interaction
	networks based on random selection of interacting domain pairs.
	\newblock \emph{Biosystems}, 95\penalty0 (2):\penalty0 155--159, 2009.
	
	\bibitem[de~Aguiar and Bar-Yam(2005)]{de2005spectral}
	Marcus A.~M. de~Aguiar and Yaneer Bar-Yam.
	\newblock Spectral analysis and the dynamic response of complex networks.
	\newblock \emph{Physical Review E}, 71\penalty0 (1):\penalty0 016106, 2005.
	
	\bibitem[Khanin and Wit(2006)]{khanin2006scale}
	Raya Khanin and Ernst Wit.
	\newblock How scale-free are biological networks.
	\newblock \emph{Journal of computational biology}, 13\penalty0 (3):\penalty0
	810--818, 2006.
	
	\bibitem[Broido and Clauset(2019)]{broido2019scale}
	Anna~D Broido and Aaron Clauset.
	\newblock Scale-free networks are rare.
	\newblock \emph{Nature communications}, 10\penalty0 (1):\penalty0 1017, 2019.
	
	\bibitem[de~Aguiar et~al.(2009)de~Aguiar, Baranger, Baptestini, Kaufman, and
	Bar-Yam]{de_aguiar_global_2009}
	M.~A.~M. de~Aguiar, M.~Baranger, E.~M. Baptestini, L.~Kaufman, and Y.~Bar-Yam.
	\newblock Global patterns of speciation and diversity.
	\newblock \emph{Nature}, 460\penalty0 (7253):\penalty0 384--387, July 2009.
	\newblock ISSN 0028-0836.
	\newblock \doi{10.1038/nature08168}.
	\newblock URL
	\url{http://www.nature.com/nature/journal/v460/n7253/full/nature08168.html}.
	
\end{thebibliography}

\end{document}